\documentclass[prx,twocolumn,showpacs,amsmath,amssymb,superscriptaddress,floatfix,reprint, nofootinbib]{revtex4-2}
\usepackage{macros}
\makeatletter
\def\l@subsection#1#2{}
\def\l@subsubsection#1#2{}
\makeatother

\begin{document}

\tolerance 10000
\title{From Symmetries to Commutant Algebras in Standard Hamiltonians
}
\author{Sanjay Moudgalya}
\email{sanjaym@caltech.edu}
\affiliation{Department of Physics and Institute for Quantum Information and Matter,
California Institute of Technology, Pasadena, California 91125, USA}
\affiliation{Walter Burke Institute for Theoretical Physics, California Institute of Technology, Pasadena, California 91125, USA}
\author{Olexei I. Motrunich}
\affiliation{Department of Physics and Institute for Quantum Information and Matter,
California Institute of Technology, Pasadena, California 91125, USA}
\begin{abstract}
In this work, we revisit several families of standard Hamiltonians that appear in the literature and discuss their symmetries and conserved quantities in the language of commutant algebras. 
In particular, we start with families of Hamiltonians defined by parts that are local, and study the algebra of operators that separately commute with each part.
The families of models we discuss include the spin-1/2 Heisenberg model and its deformations, several types of spinless and spinful free-fermion models, and the Hubbard model.
This language enables a decomposition of the Hilbert space into dynamically disconnected sectors that reduce to the conventional quantum number sectors for regular symmetries.
In addition, we find examples of non-standard conserved quantities even in some simple cases, which demonstrates the need to enlarge the usual definitions of symmetries and conserved quantities.
In the case of free-fermion models, this decomposition is related to the decompositions of Hilbert space via irreducible representations of certain Lie groups proposed in earlier works, while the algebra perspective applies more broadly, in particular also to arbitrary interacting models. 
Further, the von Neumann Double Commutant Theorem (DCT) enables a systematic construction of local operators with a given symmetry or commutant algebra, potentially eliminating the need for ``brute-force" numerical searches carried out in the literature, and we show examples of such applications of the DCT.
This paper paves the way for both systematic construction of families of models with exact scars and characterization of such families in terms of non-standard symmetries, pursued in a parallel paper~\cite{moudgalya2022exhaustive}.
\end{abstract}
\date{\today}
\maketitle
%

%%%%%%%%%%%%%%%%%%%%%%%%%%%%%%%%

\tableofcontents

\section{Introduction}
\label{sec:intro}
Symmetries play an important role in many parts of physics. 
In equilibrium quantum many-body physics, the study of symmetries of a system allows for a systematic understanding of phases of matter, their low-energy excitations, and transitions between them~\cite{sachdev2011quantum, fradkin2013field, zeng2019quantum}.
In non-equilibrium quantum many-body physics, symmetries and conserved quantities are crucial in the study of quantum dynamics and thermalization, e.g., the conserved quantities appear in the definitions of thermodynamic ensembles that the systems equilibriate to~\cite{d2016quantum, mori2018thermalization}.
The most natural symmetries that appear in the quantum many-body physics are examples of ``on-site" symmetries, which include usual Abelian symmetries such as $U(1)$ particle number conservation or non-Abelian ones such as $SU(2)$ spin conservation.  
An on-site symmetry is represented by a global unitary operator that is a tensor product of single-site unitary operators that form representations of a group that is referred to as the symmetry group of the system, and continuous on-site symmetries are accompanied by local conserved quantities that are sums of single-site terms. 
However, not all symmetries are of this type, and the exploration beyond ``conventional" global symmetries, including systems with non-on-site symmetries has only been recently initiated in various contexts~\cite{quella2020symmetry,lootens2021MPO, moudgalya2021hilbert, mcgreevy2022generalized}.
In the context of dynamics of isolated quantum many-body systems, focusing solely on the conventional on-site symmetries has also proven to be insufficient in many ways, as particularly evident from the study of phenomena of ``weak-ergodicity breaking" ~\cite{serbyn2020review, papic2021review, moudgalya2021review}.
In particular, the phenomenon of Hilbert space fragmentation~\cite{sala2020fragmentation, khemani2020localization, moudgalya2019thermalization, yang2019hilbertspace} was identified to be a violation of the conventional form of the Eigenstate Thermalization Hypothesis (ETH)~\cite{deutsch1991quantum, srednicki1994chaos, rigol2008thermalization, polkovnikov2011colloquium, d2016quantum, mori2018thermalization} that dictates the thermalization properties of non-integrable systems. % 
However, the conventional paradigm of thermalization only takes into account conventional on-site symmetries, whereas systems exhibiting Hilbert space fragmentation show evidence for a more restricted form of thermalization (dubbed Krylov-restricted thermalization)~\cite{moudgalya2019thermalization, yang2019hilbertspace, papic2021review}, which demands the consideration of additional non-local conserved quantities beyond the standard ones~\cite{moudgalya2021hilbert}. 
Surprisingly, a clear definition of a conserved quantity in an isolated quantum system is far from obvious.  
For example, allowing arbitrary operators that commute with the Hamiltonian to be conserved quantities is problematic, since exponentially many eigenstate projectors can be written down for such a system with a finite-dimensional Hilbert space, and restricting to local conserved quantities is not enough to capture the physics of certain phenomena such as weak ergodicity breaking.
In \cite{moudgalya2021hilbert}, we discussed a recipe to ameliorate this problem in the context of Hilbert space fragmentation by studying conserved quantities of \textit{families} of Hamiltonians (instead of  restricting the form of the conserved quantities of a particular Hamiltonian), and this naturally results in a precise definition.
In particular, a family of Hamiltonians is associated with an algebra of operators generated by the terms of the Hamiltonian, which we referred to as the ``bond algebra," and there is also algebra of operators that commute with each term of the Hamiltonian (i.e., the centralizer of the bond algebra), which we referred to as the ``commutant algebra," or more simply the ``commutant."
The commutant is the algebra of conserved quantities of the family of Hamiltonians, and since no restriction is imposed by hand on the conserved quantities in the commutant, this provides a more general definition of conserved quantities.
For example, the commutant can include non-on-site or even non-local operators, and it need not have any simple underlying group structure, as illustrated by examples in \cite{moudgalya2021hilbert}.
While our focus in \cite{moudgalya2021hilbert} was only on systems exhibiting Hilbert space fragmentation, the formalism of bond and commutant algebras is completely general and can be applied to a variety of systems. 
In fact, as briefly discussed there, conventional on-site symmetries can also be understood within this formalism, and the commutant algebra in this case is the algebra generated by the local conserved quantities. 
Correspondingly, there is a bond algebra associated with any on-site symmetry that is generated from a set of strictly local (few-site) operators.  
In this work, we slightly generalize the formalism to consider ``local algebras" instead of ``bond algebras," where we allow the generators to also consist of sums of strictly local operators, and revisit some of the ``standard models" of condensed matter physics, including free-fermion models and the Hubbard model, and discuss them in detail in the language of local and commutant algebras.
In addition to systematically recovering well-known results in the literature, we find this language to be illuminating in many ways, and we highlight a few of our main results below.
First, the generalized definition of conserved quantities that the commutant algebra language provides
is necessary even in some commonly-studied cases.
For example, this perspective clearly illustrates the difference between the on-site regular $SU(2)$ symmetry and the non-on-site dynamical $SU(2)$ symmetry~\cite{buca2019nonstationary, medenjak2020isolated} (also sometimes referred to as a Spectrum Generating Algebra~\cite{moudgalya2020eta, tang2020multi}), which exist in several standard systems such as the Hubbard model. 
Moreover, we find that the full commutant need not have a standard underlying group structure even in some classes of simple quadratic fermion models and explains the degeneracies between certain eigenstates that conventional symmetry groups miss. 
Second, this language provides a decomposition of the Hilbert space into dynamically disconnected subspaces that generalizes conventional symmetry quantum number sectors, as previously illustrated in the context of Hilbert space fragmentation~\cite{moudgalya2021hilbert}.
As we discuss in this work, this decomposition also explains the Hilbert space decompositions in terms of irreducible representations of certain Lie groups, recently also used by Pakrouski \textit{et al.}\ to understand Quantum Many-Body Scars as group-invariant states~\cite{pakrouski2020many, pakrouski2021group, sun2022majorana}.
Such decompositions were first noted in the context of tensor models~\cite{klebanov2018spectra}, which led to the discovery of rather mysterious Casimir identities~\cite{gaitan2020hagedorn} between Casimir elements of different Lie groups.
As we will show, the commutant language also provides a natural origin of these identities, and generalizes such decompositions to large classes of systems.
Third, perhaps most importantly, this formalism potentially provides a systematic way to construct local symmetric operators that commute with a given set of conserved quantities, due to a result known as the von Neumann Double Commutant Theorem (DCT)~\cite{landsman1998lecture}.
Given a commutant $\mC$ correspoding to a local algebra $\mA$, the DCT guarantees that all operators that commute with the conserved quantities in the commutant $\mC$ belong to the local algebra $\mA$.  
Since the local algebra $\mA$ is generated by a simple set of local operators, this property allows us to \textit{exhaustively} construct \textit{all} local Hamiltonians with a given set of conserved quantities, in principle eliminating the need for numerical methods~\cite{qi2019determininglocal, chertkov2020engineering} that perform ``brute-force" searches for operators of a fixed form or range (e.g., two-site, three-site, and so on) that commute with the given set of conserved quantities.
The DCT is applicable to many kinds of models, and in this work we discuss its implications in several well-known models.
In particular, for the Hubbard models, this approach provides a systematic approach to identify perturbations that preserve the dynamical symmetries/Spectrum Generating Algebra, previously achieved by guesswork or numerics in the literature~\cite{moudgalya2020eta, mark2020eta}. 
In addition to formally specifying all symmetric operators, the DCT also allows us to derive additional constraints due to locality (e.g., rule out the existence of local Hamiltonians that possess only certain conserved quantities) or show some general properties of the spectra of local Hamiltonians with a given set of conserved quantities (e.g., the appearance of equally spaced towers of states).  
This paper is organized as follows.
In Sec.~\ref{sec:commutants}, we review the concepts of bond, local, and commutant algebras, as well as discuss the Double Commutant Theorem (DCT) and Hilbert space decomposition that we use in the rest of this work.
In Sec.~\ref{sec:commutantexamples}, we study a few simple examples of spin systems with conventional symmetries such as $Z_2$, $U(1)$, or $SU(2)$, and also discuss the implications of the DCT in those cases.
In Secs.~\ref{sec:FFbondalgebras} and \ref{sec:Hubbard}, we study examples of bond, local, and commutant algebras that naturally appear in free-fermion systems and Hubbard models respectively, and we discuss implications of the DCT in those cases.
We conclude with open questions in Sec.~\ref{sec:conclusions}.
\section{Bond, Local, and Commutant Algebras}\label{sec:commutants}
We now review or introduce the concepts of bond and commutant algebras, and also introduce the terminology and notations used in the rest of this work.
The concept of locality of operators will play an important role throughout this work. 
In this work, we will use the phrase \textit{local operator} to either refer to \textit{strictly local operators}, with a support over a few sites in close vicinity, or \textit{extensive local operators} that are sums of strictly local terms throughout the system.\footnote{Note that strictly local and extensive local operators differ in the scaling of their Frobenius (Hilbert-Schmidt) norm with system size, where the norm is defined as $\norm{\hO}_{F} \defn \text{Tr} (\hO^\dagger \hO)/\Tr(\mathds{1})$. The norm of the former is independent of system size, whereas the norm of the latter scales linearly in system size.}
\subsection{Definition and Properties}
In quantum many-body physics, the commutant algebras naturally arise when studying symmetry algebras of \textit{families of} Hamiltonians~\cite{moudgalya2021hilbert}.
For concreteness, we focus on systems with a tensor product Hilbert space $\mH$ of local degrees of freedom on some lattice, and we are interested in Hamiltonians of the form 
\begin{equation}
    H = \sumal{\alpha}{}{J_{\alpha} \hH_{\alpha}}, 
\label{eq:genhamil}
\end{equation}
where $\{\hH_\alpha\}$ is some set of  local operators (in the above generalized sense), and $\{J_\alpha\}$ is an arbitrary set of coefficients.
The commutant algebra, denoted by $\mC$, is the symmetry algebra of Eq.~(\ref{eq:genhamil}), i.e., the algebra of operators that commute with the entire family of Hamiltonians of the form Eq.~(\ref{eq:genhamil}).
This in turn implies that any operator $\hO \in \mC$ separately commutes with \textit{each term} in $H$, i.e.,
\begin{equation}
    [\hH_\alpha, \hO] = 0\;\;\;\forall \alpha,     
\label{eq:eachtermcommute}
\end{equation}
which is a stronger condition than a regular symmetry for a particular instance of the Hamiltonian.
As we will also discuss in Sec.~\ref{sec:commutantexamples}, for families of Hamiltonians with only conventional symmetries, e.g., $U(1)$-symmetric Hamiltonians or $SU(2)$-symmetric Hamiltonians, the commutant $\mC$ is the associative algebra generated by the local conserved quantities~\cite{moudgalya2021hilbert}.
Due to Eq.~(\ref{eq:eachtermcommute}), any $\hO \in \mC$ also commutes with arbitrary products and linear combinations of the $\hH_\alpha$'s, or, in other words, $\hO$ commutes with the full algebra generated by the $\hH_\alpha$'s.
We denote this algebra by
\begin{equation}
    \mA = \lgen \{\hH_\alpha\} \rgen,
\label{eq:localalgebradefn}
\end{equation}
where we have introduced the notation $\lgen \cdots \rgen$ to denote the associative algebra generated by (linear combinations with complex coefficients and arbitrary products of) the enclosed elements and the identity operator $\mathds{1}$.\footnote{Although $\mathds{1}$ operator might not be generatable using the terms enclosed in $\lgen \cdots \rgen$ (e.g., from  the terms $\{\hH_{\alpha}\}$ in Eq.~(\ref{eq:localalgebradefn})), we will always implicitly include it in the algebras $\mA$ and $\mC$, since adding a constant to the Hamiltonian or its conserved quantities has no physical effect.}
In our previous work~\cite{moudgalya2021hilbert}, we referred to the algebra $\mA$ as the ``bond algebra"~\cite{nussinov2009bond, cobanera2010unified, cobenera2011bond}, since we were only interested in families of Hamiltonians where $\hH_\alpha$'s are the strictly local terms of the Hamiltonian with support on a few sites in close vicinity (usually associated with bonds on a lattice).
In this work, we will allow the $\hH_\alpha$ to either be strictly local or extensive local operators, hence we refer to algebras $\mA$ generated by such local operators more generally as ``locally-generated algebras" or simply ``local algebras."
Note that when all the $\hH_\alpha$'s are strictly local operators
(e.g., the strictly local terms in a Hamiltonian), we will continue to refer to $\mA$ as a ``bond algebra."
Further, in this work, we will be typically interested in systems where $\mA$ is non-Abelian (i.e., the case where the different $\hH_\alpha$'s do not all commute with each other).
Both the algebras $\mA$ and $\mC$ can be viewed as subalgebras of $\mL(\mH)$, the algebra of all linear operators on the Hilbert space $\mH$.\footnote{Throughout this work, we restrict ourselves to a fixed $D$-dimensional tensor product Hilbert space. Hence we do not attempt to distinguish between the algebra and its $D$-dimensional representation, and we always mean the latter when we say the former.}
We review three important properties of the local algebras $\mA$ and their commutant algebras $\mC$, which were also discussed in the context of bond and commutant algebras in \cite{moudgalya2021hilbert}.
First, $\mA$ and $\mC$ are both closed associative algebras and are generically non-commutative, i.e., for any $\hO_1, \hO_2 \in \mA/\mC$, $\alpha_1 \hO_1 + \alpha_2 \hO_2 \in \mA/\mC$ for any $\alpha_1, \alpha_2 \in \mathbb{C}$, and $\hO_1 \hO_2, \hO_2\hO_1 \in \mA/\mC$. 
Second, $\mA$ and $\mC$ are both \textit{unital} algebras, i.e., they contain the identity operator $\mathds{1}$. For $\mC$, this follows directly from Eq.~(\ref{eq:eachtermcommute}), and for $\mA$ we include it in the definition in Eq.~(\ref{eq:localalgebradefn}). 
Finally, $\mA$ and $\mC$ are both $\dagger$-algebras, i.e., they are closed under Hermitian conjugation ($\hO \in \mA/\mC \implies \hO^\dagger \in \mA/\mC$), and it is easy to see that this follows from the Hermiticity of the generators $\{\hH_{\alpha}\}$ of $\mA$ (w.l.o.g. required by the Hermiticity of Hamiltonians), and Eq.~(\ref{eq:eachtermcommute}). 
$\mA$ and $\mC$ are thus examples of von Neumann algebras~\cite{landsman1998lecture, harlow2017, kabernik2021reductions}, which refers to any algebra that satisfies these three properties above.
\subsection{Double Commutant Theorem}\label{subsec:DCT}
We now discuss the Double Commutant Theorem (DCT) for von Neumann algebras, also known as the double centralizer theorem or the bicommutant theorem, and its implications for the algebras $\mA$ and $\mC$.
Note that to avoid any confusion in the nomenclature, we will reserve the use of ``commutant" to refer to $\mC$ and use ``centralizer" to refer to the algebra that commutes with the given algebra.
As a direct consequence of Eqs.~(\ref{eq:eachtermcommute}) and (\ref{eq:localalgebradefn}), the commutant algebra $\mC$ is the centralizer of the algebra $\mA$ in the algebra of all operators on the Hilbert space, $\mL(\mH)$. 
Denoting the centralizer of $\mC$ in $\mL(\mH)$ by $\mD$, it is straightforward to show that $\mA \subseteq \mD$. 
However, since $\mA$ is a von Neumann algebra, $\mD = \mA$ as a consequence of the DCT.
\begin{restatable}[DCT]{thm}{dct}\label{thm:dct}
Given a unital $\dagger$-algebra $\mA$ and its centralizer $\mC$ in $\mL(\mH)$, the centralizer of $\mC$ in $\mL(\mH)$ is equal to $\mA$. 
\end{restatable}
While DCT can be proven in complete generality for all von Neumann algebras, including infinite-dimensional ones under appropriate operator topologies, we will only be working with finite-dimensional Hilbert spaces. 
For operator algebras on a finite-dimensional Hilbert space $\mH$, the proof of DCT is remarkably simple, and we reproduce it in App.~\ref{app:doublecommutant} for convenience.
We refer readers to \cite{landsman1998lecture, harlow2017, kabernik2021reductions} for clear and more detailed discussions of finite-dimensional von Neumann algebras and their properties.
The DCT has some deep implications when applied to local algebras and the corresponding commutant algebras.
When the commutant algebra $\mC$ is viewed as a \textit{symmetry algebra} (i.e., the algebra of some set of conserved quantities), DCT implies that $\mA$ is the algebra of operators that commute with operators in $\mC$, hence $\mA$ can be viewed as the algebra of all \textit{symmetric operators} in the Hilbert space.
Hence, if we know a family of Hamiltonians of the form $\sum_\alpha{J_\alpha\hH_\alpha}$ for which $\mC$ is the full symmetry algebra, the DCT states that \textit{all} operators that commute with $\mC$ should be a part of $\mA = \lgen \{\hH_\alpha\} \rgen$, i.e., any such operator should be expressible as some polynomial in terms of the local operators $\{\hH_\alpha\}$'s.
This provides a way of systematically determining (at least in principle) \textit{all} local operators that commute with some set of conserved quantities, starting from \textit{one} set of local operators (i.e., the $\hH_\alpha$'s) with \textit{only} those conserved quantities.
In Sec.~\ref{sec:commutantexamples}, we will discuss particular examples of this principle in action in the context of conventional symmetries. 
We highlight a few aspects and consequences of the DCT. 
First, the DCT makes it evident that conserved quantities or symmetries can be thought of in terms of a pair of algebras $(\mA, \mC)$ that are centralizers of each other in the algebra of all operators $\mL(\mH)$.
In the case of conventional symmetries, $\mA$ and $\mC$ are simply the algebra of symmetric operators and the algebra of conserved quantities respectively. 
Second, given two pairs of algebras $(\mA_1, \mC_1)$ and $(\mA_2, \mC_2)$, if $\mA_1$ is strictly contained within $\mA_2$ (i.e., $\mA_1 \subset \mA_2$), it is easy to show using DCT that $\mC_2$ is strictly contained within $\mC_1$ (i.e., $\mC_2 \subset \mC_1$).
Further, the converse is also true, i.e., if $\mC_2 \subset \mC_1$, the DCT implies that $\mA_1 \subset \mA_2$. 
Intuitively, enlarging the family of Hamiltonians (i.e., going from $\mA_1$ to $\mA_2$) leads to a smaller set of conserved quantities (i.e., $\mC_1$ gets reduced to $\mC_2$) and vice versa.
We will use this intuition in several places in the rest of this work.
Finally, in quantum matter applications, we often impose further locality conditions on the Hamiltonians.
For example, we may require that terms in a Hamiltonian have at most fixed range $r_\text{max}$.
Hence we are usually interested in cases where $\mA$ is generated by some set of strictly local operators that are distributed homogeneously on a regular lattice, and the possible commutants $\mC$ one can obtain with that constraint.
Here, we quickly realize that such a set of local Hamiltonians is a linear space but not an algebra, hence is only a subset of $\mA$, and additional considerations are needed when using results such as the DCT theorem.
As we will discuss with the help of examples in Sec.~\ref{sec:commutantexamples} and App.~\ref{app:localityDCT}, the combination of DCT and locality leads to interesting constraints on Hamiltonians that can be written down with particular conserved quantities.
We highlight some important results below.
Consider the case of on-site symmetries, when a commutant algebra $\mC$ is fully generated by (a family of) on-site unitary operators $\hU = \prod_j{\hu_j}$, and the bond algebra $\mA$ is generated by a homogeneous set of strictly local terms on a regular lattice. 
We obtain the following lemmas, proven in App.~\ref{subsec:DCTonsite}.
\begin{restatable}{lem}{strloc}\label{lem:strloc}
Any strictly local operator $h_R$ with support only within a contiguous region $R$ that is symmetric under an on-site unitary symmetry can be generated from the set of generators of the bond algebra $\mA$ restricted to a bounded region $R' \supseteq R$.
\end{restatable}
\begin{restatable}{lem}{typeI}\label{lem:typeI}
An extensive local operator $H$ that is symmetric under an on-site unitary symmetry can always be expressed as  $H = \sum_R{h_R}$, where each strictly local term $h_R$ has support in a bounded contiguous region $R$ and is symmetric under the same on-site unitary symmetry.
\end{restatable}
Such locality considerations, when combined with a DCT, allow for a systematic and exhaustive characterization of the local symmetric operators in a $\mA$.
We will also encounter other examples of local Hamiltonians with such range constraints that have additional features in their spectra such as fixed spacings of certain energy levels, e.g., the appearance of equally spaced towers of levels, which do not follow from the algebra/commutant considerations alone but allow similar systematic characterization, see Sec.~\ref{subsec:dynamicalSU2} and Lem.~\ref{lem:dynsym} below.
Finally, there are situations where entire bond algebras generated from terms with $r_{\max} \leq r_c$ for some ``critical range" $r_c$ asked to commute with certain conserved quantities necessarily have larger commutants than just the algebra generated by the requested conserved quantities.
One example of this is models with dipole conservation leading to Hilbert space fragmentation~\cite{sala2020fragmentation, khemani2020localization, moudgalya2019thermalization, moudgalya2021hilbert} for any finite $r_c$.
However, we will not be considering such situations in this paper.
\subsection{Hilbert Space Decomposition and Singlets}\label{subsec:Hilbertdecomp}
Given algebras $\mA$ and $\mC$ that are centralizers of each other in $\mL(\mH)$, their irreps can be used to construct a virtual bipartition~\cite{zanardi2001virtual, bartlett2007reference, lidar2014dfs, moudgalya2021hilbert} of the Hilbert space as follows~\cite{fulton2013representation}
\begin{equation}
    \mH = \bigoplus_{\lambda}{\left(\mH^{(\mA)}_\lambda \otimes \mH^{(\mC)}_\lambda\right)},
\label{eq:Hilbertdecomp}
\end{equation}
where $\mH^{(\mA)}_\lambda$ and $\mH^{(\mC)}_\lambda$ respectively denote $D_\lambda$- and $d_\lambda$-dimensional irreps of $\mA$ and $\mC$.
Eq.~(\ref{eq:Hilbertdecomp}) can be simply be viewed as a tensored basis in which all the operators in $\mA$ are simultaneously (maximally) block-diagonal.
That is, in the basis of Eq.~(\ref{eq:Hilbertdecomp}), any operator $\hh_\mA$ in $\mA$ and $\hh_{\mC}$ in $\mC$ have the matrix representations
\begin{equation}
    \hh_{\mA} = \bigoplus_{\lambda} (M^\lambda(\hh_\mA) \otimes \mathds{1}),\;\;\hh_{\mC} = \bigoplus_{\lambda} (\mathds{1} \otimes N^\lambda(\hh_\mC)),
\label{eq:ACopsmatrix}
\end{equation}   
where $M^\lambda(\hh_\mA)$ and $N^\lambda(\hh_\mC)$ are $D_\lambda$-dimensional and $d_\lambda$-dimensional matrices respectively.
Furthermore, $\mA$ and $\mC$ contain operators realizing arbitrary such matrices $M^\lambda$ and $N^\lambda$ respectively.
(In particular, when $\mC$ or $\mA$ is Abelian, we have all the $d_\lambda = 1$ or all the $D_\lambda = 1$ respectively.)
Another consequence of the DCT is that the centers of the algebras $\mA$ and $\mC$ (i.e., the subalgebra that commutes with all the elements in the algebra) coincide, and can be written as $\mZ = \mA \cap \mC$. (In particular, when $\mC$ or $\mA$ is Abelian, we obtain that $\mZ = \mC \subseteq \mA$ or $\mZ = \mA \subseteq \mC$ respectively.)
In the basis specified by Eq.~(\ref{eq:Hilbertdecomp}), any operator $\hh_{\mZ}$ in $\mZ$ has the matrix representation
\begin{equation}
 \hh_{\mZ} = \bigoplus_\lambda(c_\lambda(\hh_{\mZ}) \mathds{1} \otimes \mathds{1})
\label{eq:Zopsmatrix}
\end{equation}
where $c_\lambda(\hh_{\mZ})$ is a c-number; furthermore, arbitrary values of $c_\lambda$ are realized in the operators in $\mZ$.
Hence the different blocks labelled by $\lambda$'s in Eq.~(\ref{eq:Hilbertdecomp}) can be uniquely specified by  eigenvalues under elements in the center $\mZ$, in particular, under a minimal set of generators of $\mZ$.
Note that these blocks determine the unique partitioning of the Hilbert space if we demand that operators in $\lgen \mA \cup \mC \rgen$ act irreducibly within each block.
Since the Hamiltonians we study belong to the local algebra $\mA$, this decomposition can be used to precisely define dynamically disconnected ``Krylov subspaces" of the Hamiltonian~\cite{moudgalya2021hilbert}.
In particular, for each $\lambda$, Eq.~(\ref{eq:Hilbertdecomp}) implies the existence of $d_\lambda$ number of identical $D_\lambda$-dimensional Krylov subspaces, which, in systems with only conventional symmetries such as $U(1)$ or $SU(2)$, correspond to regular quantum number sectors.
These subspaces can be uniquely labelled by eigenvalues under a minimal set of generators of any maximal Abelian subalgebra of $\mC$~\cite{moudgalya2021hilbert}.
Note that any maximal Abelian subalgebra of $\mC$ partitions the Hilbert space into blocks such that operators in $\mA$ act irreducibly within each of the blocks.
In general any Abelian subalgebra of $\mC$ can be used to partition the Hilbert space into blocks that are closed under the actions of operators in $\mA$; although operators in $\mA$ do not act irreducibly within the blocks unless the subalgebra is a maximal Abelian subalgebra.
Further, unless the Abelian subalgebra of $\mC$ is the center $\mZ$, the partitions defined by the subalgebra are not closed under the action of operators in the $\mC$.
These properties will be important in Sec.~\ref{subsec:groupdecomp}, where we connect the decomposition of Eq.~(\ref{eq:Hilbertdecomp}) to group decompositions of earlier works~\cite{klebanov2018spectra,gaitan2020hagedorn,  pakrouski2020many, pakrouski2021group}.
Note that it is possible to have $D_\lambda = 1$ for some $\lambda$, which correspond to one-dimensional Krylov subspaces, i.e., simultaneous eigenstates of all the operators in the algebra $\mA$, including the family of Hamiltonians we are interested in. 
We refer to these eigenstates as ``singlets" of the algebra $\mA$, since they transform under one-dimensional representations of $\mA$.\footnote{Note that the usage of the term ``singlet" here differs from the conventional physics usage, which usually refers to simultaneous eigenstates of all operators in $\mC$, i.e., those that transform under one-dimensional representations of $\mC$.}
For every $\lambda$ such that $D_\lambda = 1$, there are $d_\lambda$ \textit{degenerate} singlets, i.e., all with the same eigenvalue under operators in $\mA$, and we represent these singlets by $\{\ket{\psi_{\lambda,\alpha}}\}$, where $1 \leq \alpha \leq d_\lambda$.
In general, $\mA$ could have many sets of singlets that are  \textit{non-degenerate} between the sets, e.g., when it has irreps such that $D_\lambda = D_{\lambda'} = 1$ for some $\lambda \neq \lambda'$, and the different sets of singlets differ by their eigenvalues under operators in $\mA$.
All singlets have a nice property that the projectors onto the singlet states are a part of the commutant algebra $\mC$, i.e., $\ketbra{\psi_{\lambda,\alpha}}{\psi_{\lambda,\alpha}}$ commutes with all the elements in $\mA$.
These are thus examples of eigenstate projectors that can be viewed as conserved quantities of the family of Hamiltonians we are interested in. 
More generally, this property also extends to any ``ket-bra" operator of the form $\ketbra{\psi_{\lambda,\alpha}}{\psi_{\lambda,\beta}}$ for degenerate singlets $\ket{\psi_{\lambda,\alpha}}$ and $\ket{\psi_{\lambda,\beta}}$ where $\alpha \neq \beta$.
\section{Conventional Examples}\label{sec:commutantexamples}
\begin{table*}[]
    \centering
    \renewcommand{\arraystretch}{1.5}
    \begin{tabular}{|c|c|c|c|c|c|c|c|}
        \hline
        \multirow{2}{*}{\#}&\multicolumn{4}{c|}{$\boldsymbol{\mA}$}&\multicolumn{3}{c|}{$\boldsymbol{\mC}$}
        \\
        \cline{2-8}
        & \multicolumn{2}{c|}{\bf Algebra} & {\bf Group} & {\bf Singlets}  & \multicolumn{2}{c|}{\bf Algebra} & {\bf Subgroups}  \\
        \hline
        \#1 & $\mA_{Z_2}$ & $\lgen \{S^x_j S^x_{j+1}\},  \{S^z_j\} \rgen$ & - & - & $\mC_{Z_2}$ & $\lgen \prod_j{S^z_j} \rgen$ & $Z_2$ \\
        \hline
        \#2 & $\mA_{Z_2 \times Z_2}$ & $\lgen \{S^x_j S^x_{j+1}\}, \{S^z_j S^z_{j+1}\} \rgen$ & - & -  & $\mC_{Z_2 \times Z_2}$ & $\lgen \prod_j{S^x_j}, \prod_j{S^z_j} \rgen$ & $2 \times Z_2$ \\
        \hline
        \#3 & $\mA_{U(1)}$ & $\lgen \{S^x_j S^x_{j+1} + S^y_j S^y_{j+1}\}, \{S^z_j\} \rgen$ & - & $\{\ket{F}\}, \{\sket{\bar{F}}\}$ & $\mC_{U(1)}$ & $\lgen S^z_{\tot}\rgen$ & $U(1)$ \\
        \hline
        \#4 & $\mA_{U(1) \times Z_2}$ & $\lgen \{S^x_j S^x_{j+1} + S^y_j S^y_{j+1}\}, \{S^z_j S^z_{j+1}\} \rgen$ & - & $\{\ket{F}$, $\sket{\bar{F}}\}$ & $\mC_{U(1) \times Z_2}$ & $\lgen \prod_j{S^x_j}, S^z_{\tot}\rgen$ & $U(1)$,  $Z_2$ \\
        \hline
        \#5 & $\mA_{SU(2)}$ &$\lgen \{\vec{S}_j\cdot\vec{S}_{j+1}\} \rgen$ & $S_L$ & $\{(S^-_{\tot})^n \ket{F}\}$ & $\mC_{SU(2)}$ &$\lgen S^x_{\tot},  S^y_{\tot}, S^z_{\tot} \rgen$ & $SU(2)$\\
        \hline
        \#6 & $\mA_{\dyn SU(2)}$ & $\lgen \{\vec{S}_j \cdot \vec{S}_{j+1}\},   S^z_{\tot} \rgen$ & - & $\{\{(S^-_{\tot})^n \ket{F}\}\}$ & $\mC_{\dyn SU(2)}$ & $\lgen \vec{S}_{\tot}^2, S^z_{\tot} \rgen$ & $U(1)$ \\
        \hline
    \end{tabular}
    \caption{
    Some algebras and their commutants that naturally occur in the study of one-dimensional spin-1/2 (or hard-core boson) models.
    The algebras are specified in terms of their generators, and although the full commutants $\mC$ need not have a conventional group interpretation with on-site unitary actions, they can have many such subgroups.
    Singlets of these algebra $\mA$ in all these cases can either be degenerate or non-degenerate, and all the singlets within $\{\cdot\}$ are degenerate.
    The cases $\#1-\#5$ are usually associated with conventional ``on-site" symmetry groups, and the algebras $\mA$ are examples of bond algebras.
    In cases \#1 and \#2, the bond algebras $\mA$ and their commutants $\mC$ are examples of Pauli string algebras discussed in Sec.~\ref{subsec:paulistring}.
    Case \#6 is an example that corresponds to a dynamical $SU(2)$ symmetry, where the commutant $\mC$ does not have any obvious group interpretation and the algebra $\mA$ is not a bond algebra.
    Note that in all these cases we consider the system size $L$ to be finite but large ($L \gg 1$), since very small $L$ can lead to additional ``accidental" symmetries.
    The proofs for the commutant algebras are rather straightforward in \#1-\#4, the proof for \#5 is discussed in App.~\ref{subsec:regularsu2proof}, and the proof for \#6 then follows straightforwardly.
    }    \label{tab:conventionalexamples}
\end{table*}
We now discuss some examples of local and commutant algebras, especially with regard to singlets and the application of the double commutant theorem.
We refer readers to \cite{moudgalya2021hilbert} for a more detailed discussion that focuses on the decomposition of Eq.~(\ref{eq:Hilbertdecomp}) in various systems.
For concreteness, we restrict ourselves to spin-1/2 systems with $L$ spins, and we represent the three on-site spin matrices on a site $j$ by $\{S^x_j, S^y_j, S^z_j\}$.
A few of the examples we discuss are summarized in Tab.~\ref{tab:conventionalexamples}.
Note that although we only discuss one-dimensional lattices here, all these cases generalize straightforwardly to higher dimensions.
\subsection{Regular SU(2)}\label{subsec:regularSU2}
We start with systems with conventional symmetries, for example spin-1/2 systems with the regular $SU(2)$ symmetry.
A family of $SU(2)$-symmetric systems are the Heisenberg models
\begin{equation}
    H_{SU(2)} = \sumal{j}{}{J_j (\vec{S}_j \cdot \vec{S}_{j+1})},
\label{eq:HSU2sym}
\end{equation}
where $\vec{S}_j \defn (S^x_j, S^y_j, S^z_j)$.
Denoting the total spin operators as
\begin{equation}
    S^\alpha_{\tot} \defn \sumal{j}{}{S^\alpha_j},\;\; \alpha \in \{x, y, z, +, -\}, 
\end{equation}
the local algebra $\mA$ and the corresponding commutant $\mC$ for this family models are given by (see \#5 in Tab.~\ref{tab:conventionalexamples})
\begin{equation}
    \mA_{SU(2)} = \lgen \{\vec{S}_j \cdot \vec{S}_{j+1} \}\rgen,\;\;\; \mC_{SU(2)} = \lgen S^x_{\tot}, S^y_{\tot}, S^z_{\tot} \rgen. 
\label{eq:SU2commutant}
\end{equation}
$\mC_{SU(2)}$ is simply the associative algebra generated by the generators of the Lie algebra $\mathfrak{su}(2)$, usually referred to as the ``Universal Enveloping Algebra" (UEA) of the Lie algebra and denoted by $\mU(\mathfrak{su}(2))$.
Since $\vec{S}_j \cdot \vec{S}_{j+1} = \frac{1}{2} P_{j,j+1} - \frac{1}{4}$, where $P_{j,j+1}$ is the permutation operator between the states of the spins on sites $j$ and $j+1$, we obtain $\mA_{SU(2)} = \lgen \{\vec{S}_j \cdot \vec{S}_{j+1}\}\rgen = \lgen \{P_{j,j+1}\}\rgen$.
Hence $\mA_{SU(2)} = \mathbb{C}[S_L]$, the group algebra of the symmetric group $S_L$ with complex coefficients. 
Note that since $\mA_{SU(2)}$ is generated by strictly local nearest-neighbor terms, it is an example of a bond algebra. 
The Hilbert space decomposition of Eq.~(\ref{eq:Hilbertdecomp}) in this case links the representations of the symmetric group $S_L$ and the Lie group $SU(2)$, a result known as the Schur-Weyl duality~\cite{fulton2013representation}.
The common center of the algebras $\mA_{SU(2)}$ and $\mC_{SU(2)}$ is given by $\mZ_{SU(2)} = \lgen \vec{S}^2_\tot \rgen$, where $\vec{S}^2_\tot$ is the quadratic Casimir of $SU(2)$.
Hence the $\lambda$ in Eq.~(\ref{eq:Hilbertdecomp}) runs over different values of the total spin angular momentum $S_\tot$ corresponding to eigenvalues $S_\tot (S_\tot + 1)$ of $\vec{S}^2_\tot$.
The singlets of the algebra $\mA_{SU(2)}$ are simply given by the ferromagnetic multiplet of states $\{(S^-_{\tot})^n \ket{F}\}$, where $\ket{F}$ is the ferromagnetic state fully polarized in the $\hat{z}$ direction, defined as
\begin{equation}
    \ket{F} \defn \ket{\uparrow \cdots \uparrow},
\end{equation}
and $S^-_{\tot}$ is the total $SU(2)$ spin lowering operator.
These are examples of degenerate singlets, and it is easy to verify that they satisfy $(\vec{S}_{j}\cdot\vec{S}_{j+1})(S^-_{\tot})^n\ket{F} = \frac{1}{4}(S^-_{\tot})^n\ket{F}$ for all $j$ and $n$.
Indeed, these singlets belong to a block in Eq.~(\ref{eq:Hilbertdecomp}) that is labelled by eigenvalue $\frac{L}{2}(\frac{L}{2}+1)$ under the generator $\vec{S}^2_\tot$ of the center $\mZ$, and hence $(D_\lambda, d_\lambda) = (1, L + 1)$.
Furthermore, for $L>2$ there are no further singlets.
(Note that here and below we use the name ``singlet" in the sense of the bond algebra singlet defined above, $D_\lambda=1$.  This is different from the common usage in physics that refers to $S_\tot=0$,  which is referring to the singlets of $\mC$, i.e., states that transform under one-dimensional representations of $\mC$ with $d_\lambda = 1$.)
Note that while the singlet projectors are part of the commutant $\mC_{SU(2)}$, they are included within $\mU(\mathfrak{su}(2))$, since they can all be expressed in terms of the operators there [e.g., $\ketbra{F}{F} = \prod_{m=-L/2}^{L/2-1}(S^z_{\tot} - m)/(L/2 - m)$, and all other ``ket-bra" operators of singlets can be expressed by including the action of raising and lowering operators $S^+_{\tot}, S^-_{\tot} \in \mC$.]
Given the pair of algebras $(\mA_{SU(2)}, \mC_{SU(2)})$ of Eq.~(\ref{eq:SU2commutant}) that are centralizers of each other, we now discuss the implication of the DCT of Thm~\ref{thm:dct}.
The DCT, when applied to this case, states that any operator $\hO$ that commutes with operators in $\mC_{SU(2)}$, i.e., is SU(2)-symmetric, is a part of the algebra $\mA_{SU(2)}$.
In other words, DCT is the statement that all $SU(2)$-symmetric operators can be generated from the operators $\{\vec{S}_j\cdot\vec{S}_{j+1}\}$, or, equivalently, from the permutation operators $\{P_{j,j+1}\}$.
Since any permutation between arbitrary sites $l$ and $m$ ($l \neq m$) can be generated from nearest-neighbor permutations (e.g., $P_{j,j+2} = P_{j,j+1} P_{j+1, j+2} P_{j,j+1}$), we deduce that $\vec{S}_l\cdot\vec{S}_{m}$ can be expressed in terms of $\{\vec{S}_j \cdot \vec{S}_{j+1}\}$.
Finally, since all $SU(2)$-symmetric Hamiltonians can be expressed in terms of some $\{\vec{S}_l\cdot\vec{S}_m\}$, they can all be written in terms of $\{\vec{S}_j\cdot\vec{S}_{j+1}\}$, and hence are a part of $\mA_{SU(2)}$.
Hence, as a consequence of the DCT, we are able to recover all the $SU(2)$-symmetric Hamiltonians starting from one family of Hamiltonians with \textit{only} $SU(2)$ symmetry. 
Moreover, since the commutant $\mC$ corresponds to an on-site unitary symmetry, Lems.~\ref{lem:strloc} and \ref{lem:typeI} apply.
In particular, we can show that symmetric strictly local operators residing inside any contiguous local region $R$ on the lattice can be generated by $\{\vec{S}_j \cdot \vec{S}_{j+1}\}$ terms restricted to the region $R$, and all symmetric extensive local operators are linear combinations of strictly local operators.
\subsection{Dynamical SU(2)}\label{subsec:dynamicalSU2}
We now turn to certain systems that break regular $SU(2)$-symmetry but nevertheless preserve a part of it, e.g., Heisenberg Hamiltonians in a uniform magnetic field given by 
\begin{equation}
    H_{\dyn SU(2)} = \sumal{j}{}{J_j (\vec{S}_j \cdot \vec{S}_{j+1})} + B \sumal{j}{}{S^z_j}.
\label{eq:HSU2Bsym}
\end{equation}
This Hamiltonian satisfies the relations $[H_{\dyn SU(2)}, S^\pm_{\tot}] = \pm B S^\pm_{\tot}$, where $S^\pm_{\tot}$ are the total spin $SU(2)$ raising and lowering operators, and is an example of a ``dynamical symmetry"~\cite{buca2019nonstationary, medenjak2020isolated} or a Spectrum Generating Algebra~\cite{yang1989eta,moudgalya2020eta}.
Note that these commutation relations reduce to those of regular $SU(2)$ symmetry when $B = 0$. 
The local algebra and its corresponding commutant relevant for studying the conserved quantities of the family of systems of Eq.~(\ref{eq:HSU2Bsym}) are (see \#6 in Tab.~\ref{tab:conventionalexamples})
\begin{equation}
    \mA_{\dyn SU(2)} = \lgen \{\vec{S}_j \cdot \vec{S}_{j+1} \}, S^z_{\textrm{tot}}\rgen,\;\;\; \mC_{\dyn SU(2)} = \lgen \vec{S}^2_{\tot}, S^z_{\tot} \rgen. 
\label{eq:SU2Bcommutant}
\end{equation}
Note that $\mA_{\dyn SU(2)}$ is not a bond algebra since $S^z_{\tot}$ is not a strictly local operator, but is nevertheless a local algebra. 
Note that $\mC_{\dyn SU(2)}$ in Eq.~(\ref{eq:SU2Bcommutant}) is Abelian (in fact, it is a maximal Abelian subalgebra of $\mC_{SU(2)}$), hence the center of the two algebras is simply $\mZ_{\dyn SU(2)} = \mC_{\dyn SU(2)}$. 
The singlets of $\mA_{\dyn SU(2)}$ are the same as that of $\mA_{SU(2)}$, and are given by $\{(S^-_{\tot})^n\ket{F}\}$.
However, unlike in the $SU(2)$ case, these are non-degenerate, i.e., they belong to different blocks in Eq.~(\ref{eq:Hilbertdecomp}) since they differ under the eigenvalues of $S^z_{\tot}$, which is now part of the center $\mZ_{\dyn SU(2)}$. 
Note that singlet projectors such as $\ketbra{F}{F}$ are part of the commutant $\mC_{\dyn SU(2)}$, since they can be expressed in terms of $\vec{S}^2_{\tot}$, $S^z_{\tot}$, and $\mathds{1}$.
Further, the DCT applied to this case states that any operator in $\mC_{\dyn SU(2)}$, i.e., that commutes with $\vec{S}^2_{\tot}$ and $S^z_{\tot}$, is a part of $\mA_{\dyn SU(2)}$, i.e., can be expressed in terms of $\{\vec{S}_j\cdot\vec{S}_{j+1}\}$ and $S^z_{\tot}$.
Adding locality restrictions to DCT on Hamiltonians is more interesting in the dynamical-$SU(2)$ case than for the regular $SU(2)$ symmetry.
Since the dynamical $SU(2)$ is not an on-site symmetry, Lems.~\ref{lem:strloc} and \ref{lem:typeI} do not apply directly.
Nevertheless, we are able to constrain the structure of local operators, as we discuss in App.~\ref{app:localityDCT}.
Most importantly, we prove the following Lemma.
\begin{restatable}{lem}{dynsym}\label{lem:dynsym}
Any extensive local Hamiltonian with dynamical $SU(2)$ symmetry (i.e., that commutes with $\vec{S}^2_{\tot}$ and $S^z_{\tot}$) is necessarily a \textit{linear combination} of strictly local $SU(2)$-symmetric terms and a uniform Zeeman field term $S^z_{\tot}$. 
As a consequence, its spectrum necessarily has multiple towers of equally spaced levels in its spectrum.
\end{restatable}
Note that equally spaced towers appear when the levels degenerate due to regular $SU(2)$ are lifted by $S^z_{\tot}$, hence the states in the tower differ in their eigenvalues under $S^z_{\tot}$.
While this equal spacing might be interpreted as an analog of the exact degeneracy that occurs in the presence of regular $SU(2)$ symmetry, there is an important difference -- the degeneracy in the regular $SU(2)$ occurs irrespective of the locality of the Hamiltonian, whereas the equal spacing in the case of the dynamical $SU(2)$ only occurs when finite-range locality is imposed.
On a different note, if we consider strictly local operators and require them to commute with $\vec{S}_\tot^2$, it turns out that they necessarily commute with $\vec{S}_\tot$, i.e., they have the regular $SU(2)$ spin symmetry (in particular, they will be generated from local $\vec{S}_l \cdot \vec{S}_m$ terms).
This shows an example where the algebra generated by a conserved quantity -- here $\lgen \vec{S}_\tot^2 \rgen$ -- cannot be realized as a commutant of a bond algebra whose generators are strictly local operators and have range bounded by a finite number: any such symmetric bond algebra necessarily has a larger commutant.
Even if we consider extensive local operators commuting with $\vec{S}_\tot^2$, these must consist of $SU(2)$-symmetric strictly local terms plus a uniform Zeeman field term, say $S^z_{\tot}$, and no other terms; hence the resulting commutant for such Hamiltonian instances is effectively $\lgen \vec{S}^2_{\tot}, S^z_{\tot} \rgen$, which is larger than the requested $\lgen \vec{S}^2_{\tot} \rgen$.\footnote{Note that it is not the case that one cannot write down a local algebra that is the centralizer of $\mC = \lgen \vec{S}_{\tot}^2 \rgen$ -- in fact such a local algebra can be shown to be $\mA = \lgen \{\vec{S}_j \cdot \vec{S}_{j+1}\}, S^z_{\tot}, S^x_{\tot}\rgen$. 
However, any $r$-local Hamiltonian we choose from this local algebra $\mA$ always turns out to be a part of a smaller local algebra of the form $\mA' = \lgen \{\vec{S}_j\cdot \vec{S}_{j+1}\}, S^\alpha_{\tot} \rgen$, where $S^\alpha_{\tot}$ is the total spin operator in some direction, which then has a larger commutant than requested.
That is, the commutant of $\mA'$ is $\mC' = \lgen \vec{S}_{\tot}^2, S^\alpha_{\tot} \rgen$, which is strictly larger than $\mC$.
}
\subsection{Pauli String Algebras}\label{subsec:paulistring}
For additional illustrations, we discuss some examples and properties of so-called Pauli string algebras.
For example, consider the family of transverse-field Ising models given by
\begin{equation}
    H_{Z_2} = \sumal{j}{}{\big(J_j X_j X_{j+1} + h_j Z_j\big)},
\label{eq:HZ2sym}
\end{equation}
where $\{X_j, Y_j, Z_j\}$ denote the three Pauli matrices on site $j$.
Since $Z_2$ is the only symmetry of the family of Hamiltonians, the relevant bond and commutant algebras are given by (see \#1 in Tab.~\ref{tab:conventionalexamples})
\begin{equation}
    \mA_{Z_2} = \lgen \{X_j X_{j+1}\}, \{Z_j\}\rgen,\;\;\; \mC_{Z_2} = \lgen \prodal{j}{}{Z_j} \rgen. 
\label{eq:U1commutantapp}
\end{equation}
Another example appears in the context of families of spin-1/2 XYZ models, given by 
\begin{equation}
    H_{\text{XYZ}} = \sum_j{(J^x_j X_j X_{j+1} + J^y_j Y_j Y_{j+1} + J^z_j Z_j Z_{j+1})} ~.
\label{eq:XYZmodels}
\end{equation}
This family of models possesses two $Z_2$ symmetries, and the corresponding bond and commutant algebras are given by (see \#2 in Tab.~\ref{tab:conventionalexamples})
\begin{align}
    \mA_{Z_2 \times Z_2} &= \lgen \{X_j X_{j+1}\}, \{Z_j Z_{j+1}\} \rgen, \nn \\
    \mC_{Z_2 \times Z_2} &= \lgen \prodal{j}{}{X_j}, \prodal{j}{}{Z_j}\rgen.
\label{eq:ACZ2Z2}
\end{align}
Note that the terms $\{Y_j Y_{j+1}\}$ need not be explicitly included in the generators of $\mA_{Z_2 \times Z_2}$ since they can be generated from $\{X_j X_{j+1}\}$ and $\{Z_j Z_{j+1}\}$. 
Further, note that even though the symmetry of the system is referred to as $Z_2 \times Z_2$, the two $Z_2$ symmetries do not commute with each other unless the system size is even.
Since the algebras $\mA_{Z_2}$ and $\mA_{Z_2 \times Z_2}$ (and their commutants) are all generated by Pauli strings, we refer to them as \textit{Pauli string algebras}.
Abelian Pauli string algebras are ubiquitous in the study of stabilizer code models such as the toric code~\cite{kitaev2003fault, kitaev2010topological} or certain fracton models~\cite{fractonreview, pretko2020fracton}.
However, such models are completely  solvable~\cite{moudgalya2021hilbert} and are not relevant for the study of non-integrable systems. 
On the other hand, non-integrable Hamiltonians can be constructed using non-Abelian Pauli string algebras, examples of which can be found in a variety of models in the literature.
For example, such algebras naturally appear in the study of subsystem codes and operator quantum error correction~\cite{poulin2005stabilizer, kribs2006oqec, bacon2006oqec}, and also in some systems with exotic subsystem symmetries~\cite{xu2004strong, you2018subsystem, zhou2021fractal, wildeboer2021symmetryprotected}.
We mention in passing a few general observations on Pauli string algebras.
First, the commutant of a Pauli string algebra $\mA$ is also a Pauli string algebra.\footnote{A simple proof is as follows.
Suppose $\mA = \lgen \{\hS_\mu, \mu \in M_\mA \} \rgen$,  generated by Pauli strings with labels in some set $M_\mA$.
Consider an operator $\hO \in \mC$ expanded over the basis of all Pauli strings, $\hO = \sum_\nu{c_\nu \hS_\nu}$.
We require $[\hO, \hS_\mu] = \sum_\nu{c_\nu [\hS_\nu, \hS_\mu]} = \sum_{\nu, [\hS_\nu, \hS_\mu] \neq 0}{c_\nu 2 \hS_\nu \hS_\mu} = 0$ for all $\mu \in M_\mA$, where the second equality follows from the fact that every pair of Pauli strings either commute or anticommute.
Since $\hS_\mu$ is invertible, we then obtain $\sum_{\nu, [\hS_\nu, \hS_\mu] \neq 0}{c_\nu \hS_\nu} = 0$ for each $\mu \in M_{\mA}$.
Hence $c_\nu = 0$ for every $\hS_\nu$ that anticommutes with at least one of the $\hS_\mu$ from $\mu \in M_\mA$, while $c_\nu$ can be arbitrary for $\hS_\nu$ that commutes with all $\hS_\mu$, $\mu \in M_\mA$, showing that the commutant $\mC$ is spanned by (hence generated by) Pauli strings.  
}
As a consequence, it is sufficient to work with the groups generated by the Pauli strings in $\mA$ and $\mC$, instead of their (group) algebras.
Second, non-Abelian Pauli string algebras do not admit any singlets.\footnote{To see this, any singlet $\ket{\psi}$ is a simultaneous eigenstate of all the Pauli strings $\hS_\alpha \in \mA$, i.e., $\hS_\alpha\ket{\psi} = s_\alpha\ket{\psi}$.
If $\mA$ is non-Abelian, there is at least one pair of anticommuting Pauli strings in $\mA$, say $\hS_\alpha$ and $\hS_\beta$ with $[\hS_\alpha, \hS_\beta] = 2\hS_\alpha \hS_\beta$, and it is easy to see that either $s_\alpha = 0$ or $s_\beta = 0$.
However, $0$ cannot be an eigenvalue of any Pauli string (since they are invertible), hence this is a contradiction.}
Finally, since Pauli strings are on-site unitary operators, Lems.~\ref{lem:strloc} and \ref{lem:typeI} apply for symmetric local operators in Pauli string algebras.
Moreover, all this discussion can also be straightforwardly generalized to Majorana string algebras, i.e., algebras generated by strings of Majorana fermion operators~\cite{chertkov2020engineering}.
\subsection{Other examples}\label{subsec:otherexamples}
Similar to $SU(2)$-symmetric systems, we could also consider families of Hamiltonians with other symmetries, for example $U(1)$-symmetric systems.
In this case, the corresponding bond and commutant algebras are given by (see \#3 in Tab.~\ref{tab:conventionalexamples})
\begin{equation}
    \mA_{U(1)} = \lgen \{S^x_j S^x_{j+1} + S^y_j S^y_{j+1}\}, \{S^z_j\}\rgen,\;\;\; \mC_{U(1)} = \lgen S^z_{\tot} \rgen. 
\label{eq:U1commutant}
\end{equation}
For example, the spin-1/2 XX or XXZ model in the presence of a magnetic field is a part of this bond algebra $\mA_{U(1)}$.
Since $\mC_{U(1)}$ is Abelian, it is also the center of $\mA_{U(1)}$, and the $\lambda$'s in Eq.~(\ref{eq:Hilbertdecomp}) are labelled by the $L + 1$ eigenvalues of $S^z_{\tot}$.
The algebra $\mA_{U(1)}$ possesses two singlets, given by the two ferromagnetic states $\ket{F} = \ket{\uparrow \cdots \uparrow}$ and $\sket{\bar{F}} = \ket{\downarrow \cdots \downarrow}$. 
Since $\ket{F}$ and $\sket{\bar{F}}$ are distinguished by $S^z_{\tot}$ that is part of the center (as well as by $S^z_j$'s that are part of $\mA_{U(1)}$), they are examples of non-degenerate singlets. 
Note that the projectors onto the singlets of course belong to $\mC_{U(1)}$, [e.g., $\ketbra{F}{F} = \prod_{m=-L/2}^{L/2-1}(S^z_{\tot} - m)/(L/2 - m)$].
An example of a system with both $U(1)$ and $Z_2$ symmetries occurs in families of XXZ models, where the corresponding bond and commutant algebras are given by (see \#4 in Tab.~\ref{tab:conventionalexamples})
\begin{gather}
    \mA_{U(1) \times Z_2} = \lgen \{S^x_j S^x_{j+1} + S^y_j S^y_{j+1}\}, \{S^z_j S^z_{j+1}\}\rgen,\nn \\
    \mC_{U(1) \times Z_2} = \lgen S^z_{\tot}, \prodal{j}{}{S^x_j}\rgen.
\label{eq:U1Z2commutant}
\end{gather}
Note that families of XXZ models can be expressed in terms of the generators of this bond algebra. 
Although the symmetry is sometimes referred to as $U(1) \times Z_2$, the generators of the symmetries do not commute, hence the commutant is non-Abelian. 
The common center of the bond and commutant algebra is thus given by $\mZ_{U(1) \times Z_2} = \lgen (S^z_{\tot})^2 \rgen$, which can be verified to be a part of both $\mA_{U(1) \times Z_2}$ and $\mC_{U(1) \times Z_2}$.
The singlets of $\mA_{U(1) \times Z_2}$ are the same as those of $\mA_{U(1)}$, i.e., $\ket{F}$ and $\sket{\bar{F}}$, although they are now examples of degenerate singlets.
Many more examples of bond algebras in the literature were discussed in \cite{moudgalya2021hilbert}.
These capture families of systems with quantum group symmetries such as $SU(2)_q$, or even those with Hilbert space fragmentation, where the dimension of the commutant grows exponentially with system size.
We note that in systems with Hilbert space fragmentation, the singlets of the bond algebras are referred to as ``frozen"~\cite{sala2020fragmentation, khemani2020localization, moudgalya2019thermalization} or ``jammed"~\cite{zadnik2021folded} states, since they are (typically product) states invariant by the action of the Hamiltonian.
Regarding the application of the DCT to these examples, Lems.~\ref{lem:strloc} and \ref{lem:typeI} apply when the commutants are generated by on-site unitary operators (e.g., for $\mC_{U(1)}$ and $\mC_{U(1) \times Z_2}$ discussed above). 
However, the commutants in systems with quantum group symmetries or fragmentation are not of this form, and we defer a complete analysis of the interplay between locality and DCT in general to future work.
\section{Free Fermion Bond Algebras}
\label{sec:FFbondalgebras}
We now turn to the discussion of systems of non-interacting fermions in the language of bond and commutant algebras.
More precisely, we study bond algebras generated by fermionic bilinear terms, and one application of the results for the commutants is the possibility of an exhaustive inventory of all fermionic Hamiltonians (including interacting ones) that have the specific commutants (i.e., the specific symmetries), in the same spirit as in Sec.~\ref{sec:commutantexamples} for spin models.
As we will see soon, since fermionic bilinears are closed under the commutator, the corresponding bond algebras can be viewed as enveloping algebras of Lie algebras and can be characterized in the language of Lie algebras or groups, and we will often refer to these algebras as free-fermion algebras.
As a consequence, the commutant algebra provides a different and systematic perspective on Casimir identities and Hilbert space decompositions studied in the context of tensor models~\cite{klebanov2018spectra, gaitan2020hagedorn, pakrouski2020many, pakrouski2021group}.
In addition, the study of simpler non-interacting models also sets the stage for the more interesting Hubbard algebra which we will discuss in Sec.~\ref{sec:Hubbard}, which captures the symmetries of the Hubbard model. 
A complete characterization of these free-fermion bond algebras and their singlets is useful for the study of quantum many-body scars, which we discuss in a parallel paper~\cite{moudgalya2022exhaustive}.
It is this application that explains some emphasis on the singlets of the bond algebras in our presentation of the results.
Consider a family of systems of $2N$ Majorana fermions with a Hamiltonian of the form
\begin{equation}
    H_{FF, 1} \defn \sumal{A}{}{J_A \hH_A},\;\;\;\hH_A \defn  \frac{1}{4}\sumal{ \alpha,\beta}{}{A_{\alpha,\beta} \gamma_\alpha \gamma_\beta},
\label{eq:FFfamily}
\end{equation}
where $A$ runs over some fixed set of $2N \times 2N$ antisymmetric purely imaginary (Hermitian) matrices (i.e., $A = -A^T = -A^\ast$), $J_A$'s are arbitrary real coefficients, and the $\gamma_\alpha$'s are real Majorana fermion operators (i.e., $\gamma_\alpha^\dagger = \gamma_\alpha$) that obey the usual anticommutation relations $\{\gamma_\alpha, \gamma_\beta\} = 2\delta_{\alpha,\beta}$. 
The local algebra $\mA_{FF, 1}$ in this case is the associative algebra generated by the quadratic fermion operators involved, 
\begin{equation}
    \mA_{FF, 1} \defn \lgen \{ \hH_A \} \rgen = \lgen \{\frac{1}{4}\sum_{\alpha,\beta}{}{A_{\alpha,\beta} \gamma_\alpha \gamma_\beta}\}\rgen.
\label{eq:FFbondalgebra}
\end{equation}
Associative algebras generated by quadratic fermion operators of the form of Eq.~(\ref{eq:FFbondalgebra}) have an additional structure in that they are enveloping algebras of Lie algebras with a relatively small number $\mathcal{O}(\poly(N))$ of generators.
This property is evident from the commutation relation of their generators, which reads $i [\hH_A, \hH_B] = \hH_{i[A,B]}$.
In particular, this relation shows that the Lie algebra generated by $\{\hH_{A}\}$ is (a representation of) the (Lie) algebra generated by the specified set of $2N \times 2N$ purely imaginary Hermitian matrices $\{A\}$, which is a Lie subalgebra of the Lie algebra $\mf{so}(2N, \mathbb{R})$ of all $2N \times 2N$ purely imaginary (or purely
real in an alternative representation) antisymmetric matrices.
Further, the terms $\{\hH_A\}$ can then be interpreted as the generators of a Lie subgroup of $SO(2N, \mathbb{R}) = SO(2N)$, and the associative algebra $\mA_{FF, 1}$ is the UEA of the corresponding Lie algebra.
There are further simplifications in systems with particle number conservation.
In this case, we will be typically interested in Hamiltonians with $N$ fermions of the form 
\begin{equation}
    H_{FF, 2} \defn \sumal{A}{}{J_A \hT_A},\;\;\;\hT_A \defn  \sumal{\alpha, \beta}{}{A_{\alpha,\beta} \cd_\alpha c_\beta},
\label{eq:FFfamily2}
\end{equation}
where $A$ runs over some fixed set of $N \times N$ Hermitian matrices, $J_A$'s are arbitrary real coefficients, $\{\cd_\alpha\}$ and $\{c_\alpha\}$ are complex fermion creation and annihilation operators that obey the usual anticommutation relations $\{\cd_\alpha,\cd_\beta\} = 0$, $\{c_\alpha, \cd_\beta\} = \delta_{\alpha,\beta}$.  
Note that $\alpha$ and $\beta$ in Eq.~(\ref{eq:FFfamily}) can be composite indices that label the position and other properties such as the spin of the fermion.
The corresponding local algebra has the form
\begin{equation}
    \mA_{FF, 2} \defn \lgen \{\hT_A\} \rgen = \lgen \{\sumal{\alpha,\beta}{}{A_{\alpha,\beta}\cd_\alpha c_\beta}\}\rgen.
\label{eq:FFbondalgebra2}
\end{equation}
The terms $\hT_A$ obey the same commutation relations as $\hH_A$, i.e., they satisfy 
\begin{equation}
    i [\hT_A, \hT_B] = \hT_{i[A, B]},    
\label{eq:TAcommutation}
\end{equation}
showing direct correspondence between Lie products of the quadratic fermion operators in the full Fock Hilbert space and the $N \times N$ Hermitian matrices~\cite{pakrouski2020many}.
Hence, the Lie algebra generated by $\{\hT_A\}$ is a subalgebra of $\fu{(N)}$, the terms $\{\hT_A\}$ can be interepreted as the generators of a Lie subgroup of $U(N)$~\cite{pakrouski2020many}, and $\mA_{FF, 2}$ is the UEA of the corresponding Lie algebra.
In the following, we discuss examples of such algebras that arise starting with quadratic spinless or spinful fermion terms.
For the sake of simplicity, we also restrict ourselves to systems with fermion number conservation, and only focus on nearest-neighbor bond algebras, where the fermions are arranged in a lattice, and the matrices $A$ in the generators (e.g., in Eqs.~(\ref{eq:FFfamily}) and (\ref{eq:FFfamily2})) ``couple" fermions that are on neighboring sites.  
Finally, we note that although the bond algebras $\mA_{FF}$ arise from free-fermion terms, the algebras $\mA_{FF}$ also contain interaction terms that are obtained by linear combinations of products of the quadratic terms (e.g., four-fermion terms), and everything we discuss (in particular, the corresponding commutant algebras and singlets) also holds for families of systems with such interaction terms. 
For example, the bond algebra corresponding to a family of Hamiltonians $\sumal{j}{}{J_j (i \gamma_j \gamma_{j+1})}$ is given by $\lgen \{i \gamma_j \gamma_{j+1}\}\rgen$, and it also contains the interaction term $\gamma_j \gamma_{j+1}\gamma_{j+2} \gamma_{j+3}$. 
Hence the conserved quantities in the commutant of this algebra are also conserved quantities of families of Hamiltonians such as $\sumal{j}{}{\left[J_j (i \gamma_j \gamma_{j+1}) + J'_j \gamma_j \gamma_{j+1} \gamma_{j+2} \gamma_{j+3}\right]}$.
\begin{table*}[]
    \centering
    \renewcommand{\arraystretch}{1.5}
    \begin{tabular}{|c|c|c|c|c|c|c|c|}
        \hline
        \multirow{2}{*}{\#} & \multicolumn{4}{c|}{$\boldsymbol{\mA}$}&\multicolumn{3}{c|}{$\boldsymbol{\mC}$}
        \\
        \cline{2-8}
        & \multicolumn{2}{c|}{\bf Algebra} & {\bf Group} & {\bf Singlets}  & \multicolumn{2}{c|}{\bf Algebra} & {\bf Subgroups}  \\
        \hline
        \#1 & $\mA_{c,\mu}$ & $\lgen \{T^{(\ast)}_{j,k}\}_{\tnn}, \{n_j\}\rgen$ & $U(N)$ & $\{\ket{\Omega}\}$, $\{\sket{\bar{\Omega}}\}$ & $\mC_{c,\mu}$ & $\lgen N_{\textrm{tot}}\rgen$ & $U(1)$ \\
        \hline
        \multirow{2}{*}{\#2} & \multirow{2}{*}{$\mA_c$} &$\lgen \{T^{(r)}_{j,k}\}_{\tnn},  \{T^{(i)}_{j,k}\}_{\tnn} \rgen$ & \multirow{2}{*}{$SU(N)$} & \multirow{4}{*}{$\{\ket{\Omega}$, $\sket{\bar{\Omega}}\}$} & \multirow{2}{*}{$\mC_c$} &\multirow{2}{*}{$\lgen N_{\textrm{tot}}, \ketbra{\bar{\Omega}}{\Omega}\rgen$} & \multirow{2}{*}{$U(1)$} \\
        & & $\lgen \{T^{(r)}_{j,k}\}_{\tnn}\rgen$ non-bipartite &&&&&\\ 
        \cline{1-4}\cline{6-8}
        \#3a & $\mA_i$ &$\lgen \{T^{(i)}_{j,k}\}_{\tnn} \rgen$ & \multirow{2}{*}{$SO(N)$} &  & $\mC_i$ &$\lgen N_{\tot}, Q_X \rgen$ & \multirow{2}{*}{$U(1)$, $Z_2$}\\
        \cline{1-3}\cline{6-7}
        \#3b & $\mA_r$ &$\lgen \{T^{(r)}_{j,k}\}_{\tnn} \rgen$, bipartite & & & $\mC_r$ & $\lgen N_{\tot}, \widetilde{Q}_X \rgen$ & \\
        \hline
    \end{tabular}
    \caption{Natural bond algebras generated by spinless free fermion terms with number conservation, their singlets, and commutant algebras.
    The bond algebras are generated by natural subsets of elementary terms of Eq.~(\ref{eq:spinlessgens}), and are associated with subgroups of $U(N)$ (see Sec.~\ref{subsec:spinlessfermions}). 
    Multiple subsets of such elementary terms can generate the same bond algebra, and different bond algebras can be isomorphic.
    Singlets of the bond algebras can either be degenerate or non-degenerate, and all the singlets within $\{\cdot\}$ are degenerate.
    The commutant algebras are specified in terms of their generators, and although the full commutants need not have a conventional group interpretation with on-site unitary actions, they can have many such subgroups.
    A detailed discussion of all the algebras and their centers is provided in App.~\ref{app:spinlessfermion}.
    Note that even though the generators of the bond algebras are free fermion terms, the algebras also contain interacting terms that can be constructed by linear combinations of products of free-fermion terms.
    The commutants for all of these cases can be proven rigorously, following the methods discussed in App.~\ref{app:commutantexhaustion}.
    }
    \label{tab:spinlessbondalgebra}
\end{table*}
\subsection{Spinless Fermions}\label{subsec:spinlessfermions}
We start with bond algebras of spinless complex fermions on a lattice with $N$ sites.
We will be interested in algebras of the form of Eq.~(\ref{eq:FFbondalgebra2}), where $\alpha, \beta$ denote sites on a lattice, i.e., 
\begin{equation}
    \hT_A = \sumal{j,k}{}{A_{j,k} \cd_j c_k},\;\;A = A^\dagger.
\label{eq:TAdefnspinless}
\end{equation}
All the $\hT_A$'s in Eq.~(\ref{eq:TAdefnspinless}) can be expressed as linear combinations (with real coefficients) of the ``elementary" terms 
\begin{gather}
    T^{(r)}_{j,k} \defn \cd_j c_k + \cd_k c_j
    ,\;\;T^{(i)}_{j,k} \defn i (\cd_j c_k - \cd_k c_j),\;\; n_j \defn \cd_j c_j,
\label{eq:spinlessgens}
\end{gather}
where $T^{(r)}_{j,k}$ and $T^{(i)}_{j,k}$ correspond to real and imaginary hoppings respectively, and $n_j$ is the on-site number operator (also referred to as a ``chemical potential" term).
Here, we will consider bond algebras generated by some natural subset of these elementary terms on sites/bonds of a lattice, and discuss the associated commutant algebras and singlets.
We always start with elementary terms between \textit{all} nearest-neighboring sites on a lattice, and we denote such sets with the subscript ``$\tnn$", e.g., $\{T^{(r)}_{j,k}\}_{\tnn}$ or $\{T^{(i)}_{j,k}\}_{\tnn}$.
Note that in many cases, we could in principle reduce the number of generators even further (i.e., some elementary terms can be generated from a set of other elementary terms); however, since we are interested in generic local many-body systems, it is natural to include all relevant local terms from the outset.
Further, we use the following shorthand notations to denote sets of generators
\begin{align}
    \{T^{(c)}_{j,k}\}_{\tnn} &\defn \{T^{(r)}_{j,k}\}_{\tnn} \cup \{T^{(i)}_{j,k}\}_{\tnn},\nn \\
    \{T^{(\ast)}_{j,k}\}_{\tnn} &\defn \{T^{(r)}_{j,k}\}_{\tnn}\ \text{or}\ \{T^{(i)}_{j,k}\}_{\tnn}\ \text{or}\ \{T^{(c)}_{j,k}\}_{\tnn},
\label{eq:shorthand}
\end{align}
which correspond to cases where both real and imaginary hopping terms are included in the set of generators or if either of them can be included.
A summary of the results is provided in Table~\ref{tab:spinlessbondalgebra}, and each of these cases is discussed in detail in App.~\ref{app:spinlessfermion}.
We discuss four examples of algebras generated by natural subsets of the elementary terms, and we label the distinct algebras by distinct case numbers. 
Note that distinct choices of elementary terms as generators can result in the same bond algebra (e.g., in case \#2), and distinct bond algebras can be isomorphic, which we denote as subcases (e.g., cases \#3a and \#3b).
The commutants of these algebras can be compactly expressed in terms of their generators, and we find examples of both Abelian (case \#1) and non-Abelian commutants (cases \#2 - \#3b). 
Further, the commutants we find do not necessarily have obvious group interpretations, i.e., they need not be generated by unitary operators with on-site actions, although there are such natural subgroups within the commutants.
For example, due to the fermion number conservation in the terms of Eq.~(\ref{eq:spinlessgens}), all the commutants we discuss contain a $U(1)$ subgroup generated by 
\begin{equation}
    N_{\tot} \defn \sum_j{n_j},
\end{equation}
which has an on-site action.
Further, the cases \#3a and \#3b generated by purely imaginary and real-bipartite hoppings respectively also possess extra discrete symmetries $Q_X$ and $\tQ_X$ respectively, defined as
\begin{align}
    Q_X &\defn (-1)^{\frac{N(N-1)}{4}}
    \prodal{j}{}{(\cd_j + c_j)},\nn \\
    \tQ_X &\sim \prodal{j \in \text{I}}{}{(\cd_j - c_j)} \prodal{j \in \text{II}}{}{(\cd_j + c_j)}, 
\label{eq:QXQXtdefns}
\end{align}
where I and II in the definition of $\tQ_X$ denote the two sublattices on a bipartite lattice.\footnote{Note that the overall factor for $Q_X$ is included to ensure $Q_X^2 = 1$, and the precise overall phase factor for $\tQ_X$ that ensures $\tQ_X^2 = 1$ depends on the number of sites on each sublattice but is not important here.}
These are operators that interchange particles and holes and have interpretations as $Z_2$ symmetries with on-site actions. 
On the other hand, in case \#2 generated by complex hoppings, there are non-local symmetries in the commutant that do not have a conventional unitary symmetry interpretation, but they nevertheless explain features such as degeneracies in the spectra of Hamiltonians.
In all the considered cases, the only singlets of the bond algebra are the vacuum $\ket{\Omega}$ and the anti-vacuum $\sket{\bar{\Omega}}$, defined as
\begin{equation}
    \ket{\Omega} \defn \ket{0\ 0\ \cdots 0},\;\;\sket{\bar{\Omega}} \defn \prodal{j}{}{\cd_j}\ket{\Omega},
\label{eq:Omegadefns}
\end{equation}
where $0$ denotes an empty site.
They are non-degenerate in the presence of the chemical potential terms (case \#1) but are degenerate in the other cases that exhibit non-Abelian commutants (cases \#2, \#3a-b) -- and the degeneracy can be traced to the symmetries $Q_X$ and $\tQ_X$ in the cases \#3a-b respectively, and to the non-local symmetry in the case \#2.
We now discuss the application of the DCT of Thm.~\ref{thm:dct} to the cases in Tab.~\ref{tab:spinlessbondalgebra}.
In particular, we wish to construct extensive local Hamiltonians within the bond algebras shown.
Note that the commutants in cases \#1 and \#3a-b can be generated from purely on-site unitary operators (corresponding to a $U(1)$ symmetry in case \#1 and $U(1)$ and $Z_2$ symmetries in cases \#3a-b, see App.~\ref{app:spinlessfermion}), hence Lems.~\ref{lem:strloc} and \ref{lem:typeI} directly apply.
On the other hand, in case \#2, the full commutant cannot be understood in terms of an on-site symmetry. 
Nevertheless, we are able to use the fact that $\ket{\Omega}$ and $\sket{\bar{\Omega}}$ are product states and singlets of the bond algebra $\mA_{c,\mu}$ to extend some of the results to case \#2, see App.~\ref{subsubsec:DCTcomplexhopping}.
In particular, while the results on symmetric strictly local operators hold, we need to restrict ourselves to symmetric \textit{translation-invariant} extensive local operators for the similar results to hold (see App.~\ref{subsubsec:DCTcomplexhopping} for details).
These results reveal the possible structures in symmetric local operators (i.e., local operators in the bond algebra), and potentially provide a route for their systematic construction.
\subsection{Spinful Fermions}\label{subsec:spinfulfermions}
\begin{table*}[]
    \centering
    \renewcommand{\arraystretch}{1.5}
    \begin{tabular}{|c|c|c|c|c|c|c|c|}
        \hline
        \multirow{2}{*}{\#} & \multicolumn{4}{c|}{$\boldsymbol{\mA}$}&\multicolumn{3}{c|}{$\boldsymbol{\mC}$}\\
        \cline{2-8}
        & \multicolumn{2}{c|}{\bf Algebra} &  {\bf Group} & {\bf Singlets} & \multicolumn{2}{c|}{\bf Algebra} & {\bf Subgroups} \\
        \hline
        \#1a & $\mA_{c,\mu}$ & $\lgen \{T^{(\ast)}_{j,k}\}_{\tnn}, \{K_j\} \rgen$ &  \multirow{3}{*}{$U(N)$} & $\{(S^-_{\tot})^n\ket{F}\}, \{\ket{\Omega}\}, \{\sket{\bar{\Omega}}\}$ & $\mC_{c,\mu}$ & $\lgen \{S^\alpha_{\tot}\}, N_\tot \rgen$ & \multirow{3}{*}{$SU(2) \times U(1)$} \\
        \cline{1-3}\cline{5-7}
        \#1b & $\mA_{i,h}$ & $\lgen \{T^{(i)}_{j,k}\}_{\tnn}, \{M_i\}\rgen$ &  & $\{\ket{F}\}, \{\sket{\bar{F}}\}, \{(\ed_0)^n\ket{\Omega}\}$ & $\mC_{i,h}$ & $\lgen S^z_{\tot}, \{\eta^\alpha_0\} \rgen$ & \\
        \cline{1-3}\cline{5-7}
        \#1c & $\mA_{r,h}$ & $\lgen \{T^{(r)}_{j,k}\}_{\tnn}, \{M_i\}\rgen$ bipartite & & $\{\ket{F}\}, \{\sket{\bar{F}}\}, \{(\ed_\pi)^n\ket{\Omega}\}$ & $\mC_{r,h}$ & $\lgen S^z_{\tot}, \{\eta^\alpha_\pi\} \rgen$ & \\
        \hline
        \multirow{2}{*}{\#2} & \multirow{2}{*}{$\mA_c$} & $\lgen \{T^{(r)}_{j,k}\}_{\tnn}, \{T^{(i)}_{j,k}\}_{\tnn}\rgen$ &  \multirow{2}{*}{$SU(N)$} & \multirow{2}{*}{$\{(S^-_{\tot})^n\ket{F}, \ket{\Omega}, \sket{\bar{\Omega}}\}$} & \multirow{2}{*}{$\mC_c$} & \multirow{2}{*}{\makecell{$\lgen \{S^\alpha_{\tot}\}, N_{\tot},$\\ $\ketbra{\bar{\Omega}}{\Omega}, \ketbra{F}{\Omega}\rgen$}} & \multirow{2}{*}{$SU(2) \times U(1)$} \\
        &&$\lgen \{T^{(r)}_{j,k}\}_{\tnn}\rgen$ non-bipartite &&&&&\\
        \hline
        {\#3a} & $\mA_i$ & $\lgen \{T^{(i)}_{j,k}\}_{\tnn} \rgen$  & \multirow{2}{*}{$SO(N)$} & $\{(S^-_{\tot})^n\ket{F}, (\ed_0)^n \ket{\Omega}\}$ & $\mC_i$ & $\lgen \{S^\alpha_{\tot}\}, \{\eta^\alpha_0\}, \{Q^\sigma_X\} \rgen$ & $SU(2) \times SU(2)$ \\
        \cline{1-3}\cline{5-7}
        {\#3b} & $\mA_r$ & $\lgen \{T^{(r)}_{j,k}\}_{\tnn} \rgen$ bipartite  & & $\{(S^-_{\tot})^n\ket{F}, (\ed_\pi)^n \ket{\Omega}\}$ & $\mC_r$ & $\lgen \{S^\alpha_{\tot}\}, \{\eta^\alpha_\pi\}, \{\widetilde{Q}^\sigma_X\} \rgen$ & $2 \times Z_2$\\
        \hline
        \#4 & $\mA_{c,\mu,h}$ & $\lgen \{T^{(\ast)}_{j,k}\}_{\tnn}, \{K_j\}, \{M_j\}\rgen$ & $U(N) \times U(N)$ & $\{\ket{F}\}, \{\sket{\bar{F}}\}, \{\ket{\Omega}\}, \{\sket{\bar{\Omega}}\}$ & $\mC_{c,\mu,h}$ & $\lgen S^z_{\tot}, N_{\tot} \rgen$ & $U(1) \times U(1)$  \\
        \hline
        \multirow{2}{*}{\#5} & \multirow{2}{*}{$\mA_{c,h}$} & $\lgen \{T^{(r)}_{j,k}\}_{\tnn}, \{T^{(i)}_{j,k}\}_{\tnn}, \{M_j\}\rgen$ & \multirow{2}{*}{$\frac{U(N) \times U(N)}{U(1)}$}  & \multirow{2}{*}{$\{\ket{F}\}, \{\sket{\bar{F}}\}, \{\ket{\Omega}, \sket{\bar{\Omega}}\}$} & \multirow{2}{*}{$\mC_{c,h}$} &  \multirow{2}{*}{$\lgen S^z_\tot, N_\tot, \ketbra{\bar{\Omega}}{\Omega}\rgen$} & \multirow{2}{*}{$U(1) \times U(1)$}  \\
        &&$\lgen \{T^{(r)}_{j,k}\}_{\tnn}, \{M_j\}\rgen$ non-bipartite &&&&&\\
        \hline
    \end{tabular}
    \caption{Natural bond algebras generated by spin-1/2 free fermion terms with particle number and spin conservation, their singlets and commutant algebras.
    The bond algebras are generated by natural subsets of elementary terms of Eq.~(\ref{eq:spinfulgens}), and can also be associated with subgroups of $U(N) \times U(N)$ (see Sec.~\ref{subsec:spinfulfermions}). 
    Multiple subsets of such elementary terms can generate the same bond algebra, and different bond algebras can be isomorphic. 
    Singlets of the bond algebras can either be degenerate or non-degenerate, and all the singlets within $\{\cdot\}$ are degenerate.
    The commutant algebras are specified in terms of their generators, and although the full commutants need not have a conventional group interpretation with on-site unitary actions, they can have many such subgroups.
    A detailed discussion of all the algebras and their centers is provided in App.~\ref{app:spinfulfermion}.
    Note that even though the generators of the bond algebras are free fermion terms, the algebra also contains interacting terms that can be constructed by linear combinations of products of free-fermion terms.
    Although the expressions for the commutants in all these cases are well-motivated conjectures, we do not have rigorous proofs for them. 
    We believe that they can be proven using the techniques discussed in App.~\ref{app:commutantexhaustion} but with significantly more effort.
    }
    \label{tab:spinfulbondalgebra}
\end{table*}
We now discuss examples of bond algebras that naturally appear in systems of spinful fermions on a lattice with $N$ sites.
We will be interested in terms of the form of Eq.~(\ref{eq:FFbondalgebra2}), where $\alpha,\beta$ now label both the sites $\{j\}$ of a lattice as well as the spins $\sigma \in \{\uparrow, \downarrow\}$.
Further, we will only be interested in terms that conserve spin (more precisely, one spin component) in addition to particle number, i.e.,
\begin{equation}
    \hT_A = \sumal{j, j', \sigma}{}{A^\sigma_{j, j'}\cd_{j, \sigma} c_{j', \sigma}},\;\;\; (A^\sigma)^\dagger = A^\sigma,\;\;\sigma \in \{\uparrow, \downarrow\}, 
\label{eq:TAdefinitionspin}
\end{equation}
where $A$ can be viewed as a $2N \times 2N$ block-diagonal matrix with two $N \times N$ blocks $A^\uparrow$ and $A^\downarrow$.
Applying the relations of Eq.~(\ref{eq:TAcommutation}), it is easy to see that the Lie algebra of the terms of Eq.~(\ref{eq:TAdefinitionspin}) is a subalgebra of $\fu(N) \oplus \fu(N)$ (two copies of the Lie algebra of $N \times N$ Hermitian matrices), and these terms can thus be considered to be generators of a subgroup of $U(N) \times U(N)$.
All the $\hT_A$'s in Eq.~(\ref{eq:TAdefinitionspin}) can be expressed (generated) in terms of the following ``elementary" terms:
\begin{gather}
    T^{(r)}_{j,k} \defn \sumal{\sigma \in \{\uparrow, \downarrow\}}{}{(\cd_{j,\sigma} c_{k,\sigma} + \cd_{k,\sigma} c_{j,\sigma})},\nn \\
    T^{(i)}_{j,k} \defn \sumal{\sigma \in \{\uparrow, \downarrow\}}{}{i (\cd_{j,\sigma} c_{k,\sigma} - \cd_{k,\sigma} c_{j,\sigma})}, \nn\\
    K_j \defn n_{j,\uparrow} + n_{j,\downarrow}, \;\; 
    M_j \defn n_{j,\uparrow}  - n_{j,\downarrow},
\label{eq:spinfulgens}
\end{gather}
where $n_{j,\sigma} \defn \cd_{j,\sigma} c_{j,\sigma}$. 
$T^{(r)}_{j,k}$ and $T^{(i)}_{j,k}$ respectively denote real and imaginary hopping terms (that are symmetric between the two spins), $K_j$ and $M_j$ respectively denote the on-site particle number operator (``chemical potential" term) and the on-site spin operator (``magnetic field" term). 
In the following, we will consider bond algebras generated by some natural subsets of the elementary terms of Eq.~(\ref{eq:spinfulgens}), and discuss the associated commutants and singlets. 
We will stick to the same shorthand notations as in Eq.~(\ref{eq:shorthand}) for denoting sets of nearest-neighbor hopping terms.
We also sometimes consider a more restricted class of terms that are symmetric under the interchange of the spins $\uparrow$ and $\downarrow$, i.e., we impose the condition that $A^\uparrow = A^\downarrow$ in Eq.~(\ref{eq:TAdefinitionspin}). 
Such terms can be expressed as linear combinations of the elementary terms of Eq.~(\ref{eq:spinfulgens}) without the $M_j$'s, and the Lie algebra generated by such terms is a subalgebra of that of $N \times N$ Hermitian matrices $\fu(N)$, and they generate a subgroup of $U(N)$. 
The Lie algebras then turn out to be the same as the ones generated by analogous elementary operators of Eq.~(\ref{eq:spinlessgens}) in the case of spinless fermions, although the bond and commutant algebras are different due to the larger Hilbert space in the spinful case.
The cases \#1a, \#2, and \#3a-b in Table \ref{tab:spinfulbondalgebra} correspond to such examples, and we discuss them in detail in App.~\ref{subsec:spincomplexmu}-\ref{subsec:spinimaginary}.
A summary of the results is provided in Table~\ref{tab:spinfulbondalgebra}, and each of the cases is discussed in detail in App.~\ref{app:spinfulfermion}.
We discuss eight examples of algebras generated by natural subsets of the elementary terms, and we label the distinct algebras by distinct case numbers. 
Similar to the study of spinless fermions, distinct choices of elementary terms as generators can result in the same bond algebra (e.g., two sets of generators shown in each of the cases \#2 and \#5), and distinct bond algebras can be isomorphic, which we denote as subcases (e.g., cases \#1a-c and \#3a-b).
The commutants of these algebras are expressed in terms of their generators, and can either be Abelian (case \#4) or non-Abelian (cases \#1a-c, \#2, \#3a-b, \#5). 
Similar to spinless fermions, the commutants need not have obvious group interpretations in terms of unitary operators with on-site actions, although there can be many such subgroups.
For example, due to the fermion number and spin conservation in the terms of Eq.~(\ref{eq:spinfulgens}), all the commutants we discuss contain a $U(1) \times U(1)$ subgroup generated by $N_{\tot}$ and $S^z_{\tot}$, defined as 
\begin{equation}
    N_{\tot} \defn \sum_j{K_j},\;\;\;S^z_{\tot} \defn \sum_j{M_j}/2,
\end{equation}
which have on-site actions as groups.
Note that it is sometimes more convenient to use $\eta^z_0 \defn (N_{\tot} - N)/2$ instead of $N_{\tot}$.
In cases \#1a, \#2, and \#3a-b, the $U(1)$ generated by $S^z_{\tot}$ is enhanced to a ``spin" $SU(2)$ symmetry generated by $\{S^\alpha_{\tot}\}$, defined as
\begin{equation}
    S^\alpha_{\tot} \defn \frac{1}{2}\sum_{j, \tau, \tau'}{\cd_{j, \tau} \sigma^\alpha_{\tau, \tau'} c_{j, \tau'}},\;\;\alpha \in \{x, y, z\}, 
\label{eq:Salphadefn}
\end{equation}
where $\{\sigma^\alpha\}$ are the Pauli matrices in the usual spin basis.
The Casimir operator corresponding to this $SU(2)$ symmetry is defined as $\vec{S}^2_{\tot} \defn \sum_\alpha{(S^\alpha_{\tot})^2}$, and the raising and lowering operators are defined as $S^{\pm}_{\tot} \defn 
S^x_{\tot} \pm i S^y_{\tot}$.
In cases \#1b-c and \#3a-b, the $U(1)$ generated by $N_{\tot}$ is enhanced to a ``pseudospin" $SU(2)$  symmetry generated by $\{\eta^\alpha_{0/\pi}\}$ defined as
\begin{gather}
    \eta^x_{0/\pi} \defn \frac{1}{2}(\eta^\dagger_{0/\pi} + \eta_{0/\pi}),\;\;\eta^y_{0/\pi} \defn \frac{1}{2i}(\eta^\dagger_{0/\pi} - \eta_{0/\pi}),\nn \\
    \eta^z_{0/\pi} \defn -i[\eta^x_{0/\pi}, \eta^y_{0/\pi}] =  \frac{1}{2}\left( N_{\tot} - N\right),\nn \\
    \ed_{0/\pi} \defn \sumal{j}{}{\zeta_{j, 0/\pi}\cd_{j,\uparrow} \cd_{j, \downarrow}},\;\;\eta_{0/\pi} = \sumal{j}{}{\zeta_{j, 0/\pi}c_{j, \downarrow}c_{j,\uparrow}},
\label{eq:etaalphadefns}
\end{gather}
where the subscripts $0$ or $\pi$ denote the ``momenta" of the operators with periodic boundary conditions, and $\zeta_{j, 0} \defn 1$ while $\zeta_{j, \pi} = (-1)^j$ on a bipartite lattice.
Note that $\eta_\pi^\dagger$ and $\eta_\pi$ are precisely the $\eta$-pairing operators that are studied in the context of the Hubbard model~\cite{yang1989eta, essler2005one, moudgalya2020eta, mark2020eta}.
The Casimir operator corresponding to the pseudospin $SU(2)$ symmetry can be defined as $\vec{\eta}_{0/\pi}^{2} \defn \sumal{\alpha}{}{(\eta^\alpha_{0/\pi})^2}$.
Further, the cases \#3a-b also possess extra discrete symmetries $Q^\sigma_X$ and $\tQ^\sigma_X$ respectively that interchange particles and holes of a particular spin; similar to Eq.~(\ref{eq:QXQXtdefns}) these are defined as
\begin{align}
    Q^\sigma_X &\defn (-1)^{\frac{N(N-1)}{4}} \prodal{j}{}{(\cd_{j,\sigma} + c_{j,\sigma})},\;\;\;\sigma \in \{\uparrow, \downarrow\},  \nn \\
    \tQ^\sigma_X &\sim \prodal{j \in \text{I}}{}{(\cd_{j,\sigma} - c_{j,\sigma})}\prodal{j \in \text{II}}{}{(\cd_{j,\sigma} + c_{j,\sigma})},\;\;\;\sigma \in \{\uparrow, \downarrow\}, 
\label{eq:QXQXtdefnsspin}
\end{align}
where I and II denote the two sublattices, and they have interpretations as two $Z_2$ symmetries with on-site actions. 
Note that $\tQ^\sigma_X$ is precisely the Shiba transformation operator defined in the context of the Hubbard model~\cite{essler2005one}.
On the other hand, in cases \#2 and \#5, there are non-local symmetries in the commutant that do not have a conventional unitary symmetry interpretation.
In all the considered cases, the vacuum $\ket{\Omega}$, anti-vacuum $\sket{\bar{\Omega}}$, and the two spin-polarized ferromagnetic states $\ket{F}$ and $\sket{\bar{F}}$ are singlets of the bond algebra; these are defined as
\begin{gather}
    \ket{\Omega} \defn \ket{0\ 0\ \cdots 0},\;\;\sket{\bar{\Omega}} \defn \prodal{j}{}{\cd_{j, \uparrow}\cd_{j, \downarrow}} \nn \\
    \ket{F} \defn \prodal{j}{}{\cd_{j, \uparrow}}\ket{\Omega},\;\;\sket{\bar{F}} \defn \prodal{j}{}{\cd_{j, \downarrow}}\ket{\Omega},
\label{eq:statespinsdefns}
\end{gather}
where 0 denotes an empty site.
In the cases where the $U(1)$ symmetries are enhanced to $SU(2)$, the singlets also contain ``towers" with extensive number of states -- either the ``eta-pairing" tower generated by $\eta^\dagger_{0/\pi}$ from $\ket{\Omega}$ to $\sket{\bar{\Omega}}$  and/or the ``ferromagnetic" tower generated by $S^-_{\tot}$ from $\ket{F}$ to $\sket{\bar{F}}$.
The singlets can either be degenerate or non-degenerate, and any degeneracies can be traced to some non-commuting (local or non-local) operators in the commutant.
The application of the DCT of Thm.~\ref{thm:dct} to the spinful cases in Tab.~\ref{tab:spinfulbondalgebra} closely follows the spinless cases discussed in the previous section.
In particular, note that the commutants in cases \#1a-c, \#3a-b, and \#4 can be generated from purely on-site unitary operators (see App.~\ref{app:spinlessfermion}), hence Lems.~\ref{lem:strloc} and \ref{lem:typeI} directly apply.
On the other hand, in cases \#2 and \#5, the commutant cannot be understood in terms of on-site symmetries.
Nevertheless, we are able to extend some of the results on the interplay between locality and DCT to these cases, see Apps.~\ref{subsubsec:DCTcomplexspin} and \ref{subsubsec:DCTspincomplexmag}. 
These results provide a potential route for a systematic construction of symmetric local operators starting from the generators of the bond algebra.
\subsection{Connection to Group Hilbert Space Decomposition, Casimir Relations, and Group Singlets}\label{subsec:groupdecomp}
The bond and commutant algebra language in the case of free-fermion systems explains the Hilbert space decompositions in terms of groups representations, and Casimir element relations that appear in earlier literature.
References \cite{klebanov2018spectra,gaitan2020hagedorn, pakrouski2020many, pakrouski2021group, sun2022majorana} identified several decompositions (partitionings) of the full fermionic Fock Hilbert space in a way such that the states within each partition transform under certain irreducible representations (irreps) of a group of the form $G_1 \times G_2$, where $G_1$ and $G_2$ are Lie groups.
In this case, each partition can be labelled by eigenvalues of all the (independent) Casimir elements  $\{C^{G_1}_\alpha\}$ and $\{C^{G_2}_\alpha\}$ of the Lie groups $G_1$ and $G_2$.\footnote{We remind the readers that a Lie group of rank $r$ has at most $r$ independent Casimir elements in any representation, where the rank of a Lie group is defined as the number of elements in the Cartan subalgebra (i.e., the maximal Abelian Lie subalgebra) of the corresponding Lie algebra. For example, $SU(n + 1)$, $U(n)$, $SO(2n)$, and $SO(2n+1)$ all have ranks $n$~\cite{georgi2000lie}.}
The representations of these Casimir elements are not independent, and \cite{klebanov2018spectra, gaitan2020hagedorn} derived relations between them that link the representations of $G_1$ to those of $G_2$ by means of certain Casimir relations.
For example, in some cases, states that are singlets of (i.e., transform under one-dimensional representations of) the group $G_1$ can be highest-weight states under group $G_2$. 
Such group decompositions of the Hilbert space and corresponding Casimir operators are useful in understanding spectra of certain tensor models~\cite{bulycheva2018spectra, klebanov2018spectra,pakrouski2019spectrum,  gaitan2020hagedorn}.
These models are extensively studied in the high energy physics literature, particularly in the context of toy models for black holes such as the Sachdev-Ye-Kitaev (SYK) model, where the truncation of the Hilbert space to the subspace spanned by singlets is important for the gauge-gravity correspondence~\cite{gurau2012colored, klebanov2018TASI, witten2019SYK, rosenhaus2019SYK}.
Here we understand the origin of the Hilbert space partitioning discussed in that literature in terms of Eq.~(\ref{eq:Hilbertdecomp}), where the blocks labelled by different $\lambda$'s constitute a partitioning of main interest to our problems.
States within the same block, labelled by a given $\lambda$ ,transform under $D_\lambda$-dimensional and $d_\lambda$-dimensional irreps of $\mA$ and $\mC$ respectively, and each such block can be also viewed as hosting a $(D_\lambda d_\lambda)$-dimensional irrep of an algebra $\lgen \mA \cup \mC \rgen$.
As discussed in Sec.~\ref{subsec:Hilbertdecomp}, the states within each block can be labelled by elements of (or just the generators of) $\mZ$, the common center of $\mA$ and $\mC$. 
We emphasize that Eq.~(\ref{eq:Hilbertdecomp}) is the unique partitioning of the Hilbert space if we demand that operators in $\mA$ and $\mC$ both (i.e., operators in $\lgen \mA \cup \mC \rgen$) act irreducibly within each partition. 
On the other hand, if we relax the latter requirement, there can be multiple inequivalent ways to partition the Hilbert space.
Given the unique partioning in terms of the bond and commutant algebras, we can understand situations in which this \textit{algebra partitioning} can be understood as \textit{group partitioning} in terms of irreps of Lie groups of the form $G_1 \times G_2$, and also situations where such an interpretation fails.
In App.~\ref{app:partitioning}, we discuss various subtleties of the relations between the group and algebra partitionings with the help of several examples from the spinless and spinful fermion algebras.
We assume $\mA$ is the UEA of the Lie algebra corresponding to the group $G_1$, and that the UEA of the Lie algebra corresponding to group $G_2$ is a subalgebra of the commutant $\mC$.
These properties always hold for the bond and commutant algebras discussed in Secs.~\ref{subsec:spinlessfermions}-\ref{subsec:spinfulfermions}. 
Given a Lie group $G$, we define its \textit{Casimir algebra} $\mZ_G$ as the algebra generated by the independent Casimir elements $\{C^G_\alpha\}$ of $G$, i.e., $\mZ_G \defn \lgen \{C^{G}_\alpha\} \rgen$.
Whether or not the group partitionings and Casimir relations apply depends on the relations between the Casimir algebras $\mZ_{G_1}$ and $\mZ_{G_2}$ with the center $\mZ \defn \mA \cap \mC$, and we analyze these relations case by case in App.~\ref{app:partitioning}.
In summary, these conditions demistify the origins of the Casimir relations studied in the previous literature.
Finally, we note that due to the group interpretation of the bond algebra $\mA$, (i.e., as the UEA of some Lie group $G$), the singlets of $\mA$ can also be understood as singlets of the $G$.
Indeed, since the singlets of $\mA$ are eigenstates of all the generators of the group $G$, they are also eigenstates of the group elements themselves (which can be expressed as exponentials of a linear combination of the generators), and they transform under one-dimensional representations of $G$. 
Hence these singlets were referred to as \textit{group invariant states} in \cite{pakrouski2020many, pakrouski2021group}.
However, strictly speaking, states that are referred to as ``group invariant" under a Lie group $G$ satisfy $\hat{g}\ket{\psi} = \ket{\psi}$ for any $\hat{g} \in G$ (in other words, the generators of $G$ are required to \textit{vanish} on the group invariant states).
In this work, we use the term singlets to refer to states $\ket{\psi}$ that satisfy $\hat{g}\ket{\psi} = \alpha(\hat{g})\ket{\psi}$, where $\alpha(\hat{g})$ is a phase [in other words, the singlets are eigenstates of all generators of $G$ with any (zero or non-zero) eigenvalues].
This general definition also incorporates the notion of degenerate and non-degenerate sets of singlets.
As we discuss in a parallel paper~\cite{moudgalya2022exhaustive}, when these algebras are used as stepping stones to constructions of models with exact scars, the knowledge about degeneracies of the various sets of singlets under the action of the bond algebra is necessary to understand the full set of scar states that can be embedded in those models.
\section{Hubbard Algebras}
\label{sec:Hubbard}
\begin{table*}[]
    \centering
    \renewcommand{\arraystretch}{1.5}
    \begin{tabular}{|c|c|c|c|c|c|c|}
        \hline
        \multirow{2}{*}{\#} & \multicolumn{3}{c|}{$\boldsymbol{\mA}$}&\multicolumn{3}{c|}{$\boldsymbol{\mC}$}
        \\
        \cline{2-7}
        & \multicolumn{2}{c|}{\bf Algebra} & {\bf Singlets}  & \multicolumn{2}{c|}{\bf Algebra} & {\bf Subgroups}  \\
        \hline
        \#1a & $\mA_{i,\hub}$ & $\lgen \{T^{(i)}_{j,k}\}_{\tnn}, \{V_j\}\rgen$ & $\{(S^-_{\tot})^n\ket{F}\}$, $\{(\ed_0)^n \ket{\Omega}\}$ & $\mC_{i, \hub}$ & $\lgen \{S^\alpha_{\tot}\}, \{\eta^\alpha_0\} \rgen$ & \multirow{2}{*}{$SU(2) \times SU(2)$} \\
        \cline{1-6}
        \#1b & $\mA_{r,\hub}$ & $\lgen \{T^{(r)}_{j,k}\}_{\tnn},  \{V_j\} \rgen$ bipartite & $\{(S^-_{\tot})^n\ket{F}\}$, $\{(\ed_\pi)^n \ket{\Omega}\}$ & $\mC_{r,\hub}$ & $\lgen \{S^\alpha_{\tot}\}, \{\eta^\alpha_\pi\} \rgen$ & \\
        \hline
        \multirow{2}{*}{\#2} & \multirow{2}{*}{$\mA_{c,\hub}$} & $\lgen \{T^{(r)}_{j,k}\}_{\tnn}, \{T^{(i)}_{j,k}\}_{\tnn}, \{V_{j}\} \rgen$ & \multirow{2}{*}{$\{(S^-_{\tot})^n\ket{F}\}$, $\{\ket{\Omega}, \sket{\bar{\Omega}}\}$} & \multirow{2}{*}{$\mC_{c,\hub}$} &  \multirow{2}{*}{$\lgen \{S^\alpha_{\tot}\}, N_{\tot}, \ketbra{\bar{\Omega}}{\Omega}\rgen$} & \multirow{2}{*}{$SU(2) \times U(1)$} \\
        &&$\lgen \{T^{(r)}_{j,k}\}_{\tnn}, \{V_j\}\rgen$ non-bipartite &&&& \\
        \hline
        \#3a & $\mA^{(\dyn \eta)}_{i,\hub}$ & $\lgen \{T^{(i)}_{j,k}\}_{\tnn}, \{V_j\}, N_{\tot}\rgen$ & $\{(S^-_{\tot})^n\ket{F}\}$, $\{\{(\ed_0)^n \ket{\Omega}\}\}$ & $\mC^{(\dyn \eta)}_{i, \hub}$ & $\lgen  \{S^\alpha_{\tot}\}, \vec{\eta}^2_0, \eta^z_0 \rgen$ & \multirow{4}{*}{$U(1) \times SU(2)$}\\
        \cline{1-6}
        \#3b & $\mA^{(\dyn \eta)}_{r,\hub}$ & $\lgen \{T^{(r)}_{j,k}\}_{\tnn}, \{V_j\}, N_{\tot}\rgen$\;\;\text{bipartite} & $\{(S^-_{\tot})^n\ket{F}\}$, $\{\{(\ed_\pi)^n \ket{\Omega}\}\}$ & $\mC^{(\dyn \eta)}_{r, \hub}$ & $\lgen \{S^\alpha_{\tot}\}, \vec{\eta}^2_\pi, \eta^z_\pi \rgen$ &\\
        \cline{1-6}
        \#3c & $\mA^{(\dyn S)}_{i,\hub}$ & $\lgen \{T^{(i)}_{j,k}\}_{\tnn}, \{V_j\}, S^z_{\tot}\rgen$ & $\{\{(S^-_{\tot})^n\ket{F}\}\}$, $\{(\ed_0)^n \ket{\Omega}\}$ & $\mC^{(\dyn S)}_{i, \hub}$ & $\lgen \vec{S}^2_{\tot}, S^z_{\tot}, \{\eta^\alpha_0\} \rgen$ &\\
        \cline{1-6}
        \#3d & $\mA^{(\dyn S)}_{r,\hub}$ & $\lgen \{T^{(r)}_{j,k}\}_{\tnn}, \{V_j\}, S^z_{\tot}\rgen$\;\;\text{bipartite} & $\{\{(S^-_{\tot})^n\ket{F}\}\}$, $\{(\ed_\pi)^n \ket{\Omega}\}$ & $\mC^{(\dyn S)}_{r, \hub}$ & $\lgen \vec{S}^2_{\tot}, S^z_{\tot}, \{\eta^\alpha_\pi\} \rgen$ &\\
        \hline
    \end{tabular}
    \caption{Some natural local algebras (``Hubbard algebras") generated by spin-1/2 free fermion hopping terms with particle number and spin conservation and interacting Hubbard terms, their singlets and commutant algebras.
    Singlets of the bond algebras can either be degenerate or non-degenerate, and all the singlets within $\{\cdot\}$ are degenerate.
    The commutant algebras are specified in terms of their generators, and although the full commutant need not have a conventional group interpretation with on-site unitary actions, they can have many such subgroups.
    The bond algebras (\#1a-b and \#2) are generated by natural subsets of elementary hopping terms defined in Eq.~(\ref{eq:spinfulgens}) and the on-site Hubbard terms defined in Eq.~(\ref{eq:Hubbardterms}).
    Unlike the bond algebras in Tabs.~\ref{tab:spinlessbondalgebra} and \ref{tab:spinfulbondalgebra}, these are generated by interacting terms and cannot be associated with any simple Lie group.
    The local algebras with dynamical symmetries (\#3a-d) are obtained by adding a uniform chemical potential or a uniform magnetic field to the bond algebras. 
    Note that we have not shown several algebras for the sake of brevity.
    A detailed discussion of all these cases is provided in App.~\ref{app:Hubbard}.
    Although the expressions for the commutants in all these cases are well-motivated conjectures, we do not have rigorous proofs for them.
    }
    \label{tab:hubbardalgebra}
\end{table*}
We now discuss other examples of fermionic bond algebras that can no longer be understood as being generated from any natural subset of elementary terms of Eq.~(\ref{eq:spinfulgens}). 
Specifically, we consider algebras obtained by adding on-site Hubbard-like terms to the generators.
The motivation for studying such algebras is two-fold. 
First, such an algebra appears in the context of the well-known Hubbard model ``at half-filling'' (i.e., at average density of one electron per site). 
Second, the Hubbard model is known to have spin and pseudospin $SU(2)$ symmetries, hence we expect the appropriate commutant to be generated by  $\{S^\alpha_{\tot}\}$ and $\{\eta^\alpha_{0/\pi}\}$ [see Eqs.~(\ref{eq:Salphadefn}) and (\ref{eq:etaalphadefns})].
As discussed in Sec.~\ref{subsec:DCT}, local and commutant algebras appear in pairs, and we can ask if there is a local algebra with this commutant.\footnote{Note that all the free-fermion bond algebras in Tab.~\ref{tab:spinfulbondalgebra} with spin and pseudospin $SU(2)$ symmetries in their respective commutants necessarily have additional symmetries.}
As we now discuss, there indeed exists such a local algebra that contains the Hamiltonian of the Hubbard model.
Moreover, we also find additional closely related algebras that are applicable to the Hubbard model under various conditions, e.g., with complex hopping terms or in the presence of a chemical potential or a magnetic field. 
We refer to these algebras as \textit{Hubbard algebras}.
We start with the families of Hubbard Hamiltonians of the form 
\begin{equation}
    H_{\Upsilon, \hub} = \sumal{\langle j, k \rangle}{}{t_{j,k} T^{(\Upsilon)}_{j,k}} + \sumal{j}{}{u_j V_j} - \mu \sumal{j}{}{K_j} - B\sumal{j}{}{M_j},
\label{eq:hubfamily}
\end{equation}
where $\langle j, k\rangle$ denotes nearest-neighboring sites on a lattice,  $T^{(\Upsilon)}_{j,k}$, $K_j$, and $M_j$ are the elementary terms defined in Eq.~(\ref{eq:spinfulgens}), $\Upsilon \in \{r, i, c\}$ corresponding to real, imaginary, or complex hoppings, and $V_j$ is the on-site Hubbard term defined as
\begin{gather}
     V_j \defn \left(n_{j,\uparrow} - \frac{1}{2}\right)\left(n_{j,\downarrow} - \frac{1}{2}\right) \nn\\
     =\frac{1}{2}\left[(K_j - 1)^2 - \frac{1}{2} \right] = \frac{1}{2}\left[\frac{1}{2} - M_j^2 \right].
\label{eq:Hubbardterms}
\end{gather}
Further, $\{t_{j,k}\}$ and $\{u_j\}$ in Eq.~(\ref{eq:hubfamily}) are arbitrary parameters, and $\mu$ and $B$ are the chemical potential and magnetic field strengths respectively that can either be arbitrary parameters or be set to zero to obtain subfamilies of Hamiltonians of Eq.~(\ref{eq:hubfamily}).
The symmetries of various families of Hamiltonians can be understood via the corresponding local algebras. 
As we summarize in Tab.~\ref{tab:hubbardalgebra}, we find many distinct Hubbard local algebras, depending on the choices of parameters and hoppings in Eq.~(\ref{eq:Ahubbard}).
If we focus on the family of Hubbard Hamiltonians with $\mu = B = 0$ in Eq.~(\ref{eq:hubfamily}), we find examples of Hubbard bond algebras and commutants with spin and/or pseudospin $SU(2)$ symmetries, each similar to the SU(2) symmetry discussed in Sec.~\ref{subsec:regularSU2} in the context of the spin-1/2 Heisenberg model.
On the other hand, if we are interested in symmetries of larger families of Hamiltonians allowing arbitrary non-zero values of $\mu$ and $B$, we find examples of dynamical symmetries similar to ones discussed in Sec.~\ref{subsec:dynamicalSU2}.
We summarize our results below and provide a more detailed discussion in App.~\ref{app:Hubbard}.
\subsection{Regular SU(2)}\label{subsec:Hubbardregular}
We start with the family of particle-hole symmetric on-site Hubbard terms, obtained by setting $\mu = B = 0$ in Eq.~(\ref{eq:hubfamily}).
We denote the corresponding bond algebras by $\mA_{\Upsilon, \hub}$, and they are obtained by adding the Hubbard terms $\{V_j\}$ to the free-fermion bond algebras generated by the appropriate hopping terms (cases \#3a-b and \#2 in Tab.~\ref{tab:spinfulbondalgebra}), i.e., 
\begin{equation}
    \mA_{\Upsilon, \hub} \defn \lgen \{T^{(\Upsilon)}_{j,k}\}_{\tnn}, \{V_j\}\rgen,\;\;\;\Upsilon \in \{r, i, c\}. 
\label{eq:Ahubbard}
\end{equation}
Since the on-site Hubbard terms $V_j$ can be expressed in terms of $K_j$ or $M_j$, as shown in Eq.~(\ref{eq:Hubbardterms}), it follows that $\mA_{\Upsilon, \hub}$ is a part of the free-fermion algebras $\mA_{c, \mu}$ and $\mA_{\Upsilon, h}$ (defined in Tab.~\ref{tab:spinfulbondalgebra}), and we conjecture using plausible arguments that $\mA_{\Upsilon,\hub} = \mA_{c,\mu} \cap \mA_{\Upsilon, h}$.\footnote{Note that it may seem strange that we invoke algebras with applied magnetic field $\mA_{\Upsilon,h}$ while the Hubbard algebras considered here, as we will discuss, are all spin-$SU(2)$-symmetric.
Such an algebra equality would not hold with just $\mA_\Upsilon$, since $\mA_{\Upsilon, \hub}$ is evidently larger than $\mA_{\Upsilon}$.
In the R.H.S., the intersection with $\mA_{c,\mu}$ guaranties spin-$SU(2)$-invariance while the intersection with $\mA_{\Upsilon,h}$ guaranties that hoppings are $\Upsilon$-type-generated and fixes the form of the Hubbard term; and this writing allows to reuse free-fermion results.
Alternatively, one could start with $\mA_{\Upsilon} \subseteq \mA_{\Upsilon,\hub} \subseteq \mA_{c,\mu}$ and think through some relatively simple analysis/plausible conjectures afresh.}
While it is not a priori clear if the Hubbard algebras are distinct from other free-fermion algebras in Tab.~\ref{tab:spinfulbondalgebra}, i.e., if on-site terms such as $K_j$ and $M_j$ can be generated from the generators of the Hubbard algebra $\mA_{\Upsilon, \hub}$, the differences between the algebras are already evident from (the degeneracies of) their singlets, which are listed in Tab.~\ref{tab:spinfulbondalgebra}. 
We discuss the properties of all these Hubbard algebras, their singlets, and commutants in more detail in App.~\ref{app:Hubbard}.
We are primarily interested in the commonly-studied case with only real hopping terms~\cite{essler2005one, moudgalya2020eta, mark2020unified}. 
On a bipartite lattice the Hubbard bond algebra is isomorphic to the bond algebra generated with purely imaginary hoppings (i.e., $\mA_{i,\hub}$), see App.~\ref{subsec:Hubbardreal}.
The commutants $\mC_{i,\hub}$ and $\mC_{r,\hub}$ in these cases possess a spin-$SU(2)$ symmetry, generated by $\{S^\alpha_{\tot}\}$ defined in Eq.~(\ref{eq:Salphadefn}), as well as a ``pseudospin" $SU(2)$ symmetry generated by operators $\{\eta^\alpha_{\pi}\}$ or $\{\eta^\alpha_0\}$ defined in Eq.~(\ref{eq:etaalphadefns}). 
For reasons we discuss in Apps.~\ref{subsec:Hubbardimag} and \ref{subsec:Hubbardreal}, we conjecture that the commutants $\mC_{r,\hub}$ and $\mC_{i,\hub}$ are completely generated by these two $SU(2)$ symmetries. 
On a non-bipartite lattice, the Hubbard algebra generated with real hoppings is equal to the Hubbard algebra generated with complex hoppings (i.e., $\mA_{c,\hub}$).
While the commutant $\mC_{c,\hub}$ in this case possesses the spin-$SU(2)$ symmetry generated by $\{S^\alpha_{\tot}\}$, the full pseudospin $SU(2)$ generated by $\{\eta^\alpha_{0/\pi}\}$ is no longer present.  
A complete discussion of this algebra can be found in App.~\ref{subsec:Hubbardcomplex}.
\subsection{Dynamical SU(2)}\label{subsec:Hubbarddynamical}
We now discuss examples of dynamical $SU(2)$ symmetries that can occur in the family of Hubbard Hamiltonians of Eq.~(\ref{eq:hubfamily}).
We start by setting $B = 0$ while allowing $\mu$ to be an arbitrary parameter, which corresponds to the addition of a uniform chemical potential term $\sum_j{K_j} = N_{\tot}$ to the bond algebra $\mA_{\Upsilon, \hub}$, and motivates the study of the local algebra
\begin{equation}
    \mA^{(\dyn \eta)}_{\Upsilon, \hub} = \lgen \{T^{(\Upsilon)}_{j,k}\}, \{V_j\}, N_{\tot} \rgen. 
\label{eq:dynhubeta}
\end{equation}
In fact, this algebra contains the most standard form of the Hubbard model, which is recovered by setting $\Upsilon = r$ and $u_j = U$, $t_{j,k} = -t$ in Eq.~(\ref{eq:hubfamily}).
With real hoppings on a bipartite lattice or imaginary hoppings (i.e., $\Upsilon \in \{r, i\}$), the commutant $\mC^{(\dyn \eta)}_{\Upsilon, \hub}$ contains the regular spin $SU(2)$ symmetry along with a dynamical pseudospin $SU(2)$ symmetry.
This modification of the regular pseudospin $SU(2)$ symmetry to a dynamical pseudospin $SU(2)$ symmetry upon the addition of a uniform chemical potential is analogous to the modification of a regular $SU(2)$ symmetry to a dynamical $SU(2)$ symmetry discussed in the context of the Heisenberg model in a uniform magnetic field in Sec.~\ref{subsec:dynamicalSU2}.   
Alternately, we could set $\mu = 0$ and allow $B$ to be an arbitrary parameter in Eq.~(\ref{eq:hubfamily}), which corresponds to the addition of a uniform magnetic field term $\sum_j{M_j} = S^z_{\tot}$ to the bond algebra $\mA_{\Upsilon, \hub}$, and motivates the study of the local algebra
\begin{equation}
    \mA^{(\dyn S)}_{\Upsilon, \hub} = \lgen \{T^{(\Upsilon)}_{j,k}\}, \{V_j\}, S^z_{\tot}\} \rgen. 
\label{eq:dynhubspin}
\end{equation}
Similar to the previous case, this results in the modification of the regular spin $SU(2)$ symmetry in the commutant to a dynamical spin $SU(2)$ symmetry, and this happens with any type of hopping (i.e., for $\Upsilon \in \{r, i, c\}$). 
Finally, we could allow both $\mu$ and $B$ to be arbitrary in Eq.~(\ref{eq:hubfamily}), which corresponds to the addition of a uniform chemical potential and a uniform magnetic field to the bond algebra $\mA_{\Upsilon, \hub}$. 
In this case, both the regular spin and pseudospin $SU(2)$ symmetries in the commutant $\mC_{\Upsilon, \hub}$ get modified into dynamical $SU(2)$ symmetries.
A complete discussion of these algebras can be found in App.~\ref{app:Hubbard}.
\subsection{Application of DCT}\label{subsec:hubbardDCT}
We now discuss the application of the DCT to the Hubbard algebras in Tab.~\ref{tab:hubbardalgebra}.
We restrict ourselves to the physically interesting case with real hoppings on a bipartite lattice or purely imaginary hoppings, where the commutants have spin and pseudospin (regular or dynamical) $SU(2)$ symmetries. 
We start with the discussion of $\mA_{i,\hub}$; all these results also hold for $\mA_{r,\hub}$ due to the isomorphism discussed in App.~\ref{subsec:Hubbardreal}.
As a consequence of DCT of Thm.~\ref{thm:dct}, any operator that commutes with spin and pseudospin $SU(2)$ symmetries generated by $\{S^\alpha_{\tot}\}$ and $\{\eta^\alpha_{0}\}$ should belong to the Hubbard algebra $\mA_{i,\hub}$.
Since the commutant $\mC_{i,\hub}$ can be understood as being completely generated by the family of on-site unitary operators that correspond to the spin and pseudospin $SU(2)$ symmetries, Lems.~\ref{lem:strloc} and \ref{lem:typeI} apply. 
A systematic numerical search for such one-site and two-site nearest neighbor symmetric operators was performed in \cite{mark2020eta}, and they found 10 linearly independent terms that commute with the spin and pseudospin symmetries, see Tab.~III, part A, there.\footnote{Note that the entries \#1, \#5, and \#6 there each correspond to two linearly independent operators. 
In addition, the identity as well as all one-site operators are included in the span of the operators \#1-\#7 listed there.}
Indeed, we find that the dimension of the two-site Hubbard algebra, i.e., $\lgen T^{(i)}_{j,j+1}, V_j, V_{j+1} \rgen$, is $10$, and we have verified that the 10 terms obtained in \cite{mark2020eta} span this algebra.
Although the expressions of the operators there in terms of the two-site generators $T^{(i)}_{j,j+1}$, $V_j$, $V_{j+1}$ can in general be complicated, some expressions are particularly simple. 
For example, the operator \#2 in Tab.~III of \cite{mark2020eta} can be expressed as $4(V_j + 1/4)(V_{j+1} + 1/4)$, the operator \#7 as $2(V_j + 1/4)(V_{j+1} + 1/4) \big(T_{j,j+1}^{(r)} \big)^2$ (remembering that the system in \cite{mark2020eta} has real bipartite hopping), etc.
The DCT now obviates the toils to extend the numerical search in \cite{mark2020eta} to three- and more-site symmetric operators since we know they can be produced from the basic bond algebra generators.
We then move on to the DCT in the context of the algebra $\mA^{(\dyn\eta)}_{i,\hub}$; the following results also hold for the algebras $\mA^{(\dyn\eta)}_{r,\hub}$, $\mA^{(\dyn S)}_{i,\hub}$, and $\mA^{(\dyn S)}_{r,\hub}$ due to their isomorphisms discussed in App.~\ref{app:Hubbard}.
Since the commutant $\mC^{(\dyn \eta)}_{i,\hub}$ consists of the full family of on-site unitaries of the spin $SU(2)$ symmetry, the Lems.~\ref{lem:strloc} and \ref{lem:typeI} apply w.r.t.\ this symmetry.
Further, since $\mC^{(\dyn \eta)}_{i,\hub}$ contains a dynamical pseudospin $SU(2)$ symmetry, Lem.~\ref{lem:dynsym} can also be applied w.r.t. this symmetry.
These arguments show that strictly local operators in $\mA^{(\dyn \eta)}_{i,\hub}$ possess both regular spin and pseudospin $SU(2)$ symmetries, hence are a part of $\mA_{i,\hub}$. 
Moreover, extensive local operators in $\mA^{(\dyn \eta)}_{i,\hub}$ are necessarily linear combinations of $N_{\tot}$ and other strictly local terms with regular spin and pseudospin $SU(2)$ symmetries (i.e., that are a part of $\mA_{i,\hub}$). 
Similar to the case discussed in Sec.~\ref{subsec:dynamicalSU2}, this shows that any local Hamiltonian in the algebra $\mA^{(\dyn \eta)}_{i,\hub}$ necessarily has towers of equally spaced levels in its spectrum, consisting of levels from quantum number sectors labelled by different eigenvalues under $N_{\tot}$.
This is in addition to degenerate eigenstates consisting of levels from quantum number sectors labelled by different eigenvalues under $S^z_{\tot}$. 
Note that a similar analysis can be carried out for the algebra $\mA^{(\dyn S, \dyn \eta)}_{i,\hub}$ (which is isomorphic to $\mA^{(\dyn S, \dyn \eta)}_{r,\hub}$), which possesses spin and pseudospin dynamical $SU(2)$ symmetries.
Using Lem.~\ref{lem:dynsym}, we can again show that extensive local operators in this algebra are linear combinations of $S^z_{\tot}$, $N_{\tot}$, and other terms with regular spin and pseudospin $SU(2)$ symmetries.
This shows that any local Hamiltonian in the algebra $\mA^{(\dyn S, \dyn \eta)}_{i,\hub}$ necessarily has two towers of equally spaced levels in its spectrum, each consisting of levels from quantum number sectors labelled by different eigenvalues under $S^z_{\tot}$ or $N_{\tot}$.
\section{Conclusions and Outlook}\label{sec:conclusions}
In this work, we expanded on the framework of commutant algebras introduced in \cite{moudgalya2021hilbert}, and illustrated its application to understand the conserved quantities in several standard models including the spin-1/2 Heisenberg model, various non-interacting fermionic models, and the electronic Hubbard model.
These Hamiltonians can all be understood as being parts of a larger family of local Hamiltonians, and the commutant algebra framework provides a systematic way to define conserved quantities solely based on this family.
In particular, the commutant algebra is the centralizer of the algebra generated by individual local terms that define the family, which is referred to as a bond algebra if all the individual terms are strictly local, or more generally as a local algebra if the individual terms can include extensive local operators.
The commutant algebra framework hence provides a systematic way to identifying conserved quantities without imposing restrictions on their forms, in contrast to the conventional approach of restricting to conserved quantities that are on-site unitary operators or sums of local operators.
Without such restrictions, the commutant language captures features that are not explained by conventional symmetries such as the existence of non-obvious dynamically disconnected subspaces in fragmented systems~\cite{moudgalya2021hilbert} or unexpected degeneracies in the spectrum in free-fermion models (with some examples presented in this work, see the spinless and spinful bond algebra $\mA_c$ in Tabs.~\ref{tab:spinlessbondalgebra} and \ref{tab:spinfulbondalgebra}).
At the same time, when applied to well-known models with only on-site unitary symmetries, the commutant language is equivalent to the conventional symmetry language, and the commutant algebra can be understood as being fully generated by the family of on-site unitary operators.
Among the models we study, the commutants in the spin-1/2 Heisenberg model, several of the non-interacting fermionic models, and the Hubbard model are all generated by on-site unitary symmetry operators, whereas the commutants in certain non-interacting fermionic models contain ``unconventional" non-local conserved quantities.
Understanding conserved quantities in terms of the commutant has several additional benefits. 
For example, the association of the commutant algebras to families of Hamiltonians directly explains several features that are not immediately clear a priori, e.g., the persistence of $\eta$-pairing in the Hubbard model with some types of disorder~\cite{yang1992remarks,moudgalya2020eta, mark2020eta}.
Further, in the free-fermion cases, the commutant provides a different perspective on the Hilbert space decomposition in terms of irreducible representations of Lie groups mostly introduced in the high-energy physics literature~\cite{klebanov2018spectra, gaitan2020hagedorn, pakrouski2020many, pakrouski2021group, sun2022majorana}, and motivates the associated ``Casimir relations" (see Sec.~\ref{subsec:groupdecomp}).
The commutant language is also useful when applied to conserved quantities that do not have an obvious underlying group structure or when identifying ``symmetry groups" in conserved quantities is ambiguous (e.g., $SO(4)$ versus $SU(2) \times SU(2)$ symmetry in the Hubbard model), since it avoid invoking groups by focusing directly on a minimal set of conserved quantities that generate all operators that commute with that Hamiltonian.   
Focusing on the associative algebra structure of the conserved quantities, as done in this work, instead of the Lie algebra or Lie group structures, e.g., as studied in previous works~\cite{kraus2007quantum, nussinov2009bond, cobenera2011bond, zeier2011symmetry, zimboras2015symmetry, pakrouski2020many}, is more insightful particularly due to the Double Commutant Theorem (DCT).
The DCT provides a way to generate all operators that have some desired symmetry; for example, all spin-1/2 $SU(2)$ symmetric operators can be generated from (i.e., can be expressed in terms of products and linear combinations of) the nearest-neighbor Heisenberg terms.
Along with some locality considerations, this provides a systematic way to construct all strictly local or extensive local operators with the desired symmetry, allowing us to provide an ``exhaustive'' description of families of such symmetric Hamiltonians.
For example, for commutants corresponding to on-site unitary symmetries, we showed that extensive local symmetric operators can be written as a sum of strictly local symmetric operators (see Lem.~\ref{lem:typeI}).
In a parallel work~\cite{moudgalya2022exhaustive}, we show that such a statement is not true for more general non-standard commutants; in particular there we construct ``Type II" symmetric Hamiltonians that cannot be expressed as sums of strictly local symmetric operators, which can be contrasted with ``Type I" symmetric Hamiltonians that can be expressed so.
Further, in the case of dynamical symmetries, we showed that all extensive local symmetric Hamiltonians necessarily have equally spaced eigenvalues in their spectra (see Lem.~\ref{lem:dynsym}).
Such applications of the DCT potentially eliminate need of brute-force searches for constructing operators with a desired symmetry~\cite{chertkov2020engineering, qi2019determininglocal}, as we demonstrated in the case of the Hubbard model, where we analytically recovered nearest-neighbor operators with the spin and pseudospin symmetries previously discovered numerically~\cite{mark2020eta}.
Looking forward, it would be interesting to extend this kind of analysis to place constraints on the nature or structure of symmetric unitary or Floquet operators.
Some of the results discussed in this work are also helpful in understanding other phenomena involving ``unconventional conserved quantities" within the commutant framework. 
As we will discuss in a parallel work~\cite{moudgalya2022exhaustive}, the singlets of the algebras we studied here also play an important role in understanding models with exact Quantum Many-Body Scars (QMBS)~\cite{serbyn2020review, moudgalya2021review, papic2021review}, and in particular we show that local algebras can be constructed such that the projectors onto the QMBS eigenstates can themselves be viewed as non-local conserved quantities exhausting the commutant.
In addition, in another upcoming work~\cite{szminprep}, we show that some examples of strong zero modes~\cite{alicea2016topological, fendley2016strong} can also be understood within the commutant framework.
It is also clear from examples studied in this work that proving the exhaustions of the commutants, i.e., showing that the \textit{entire} commutant is generated from a given set of simple operators, can be tedious even in simple cases, and we found it necessary to resort to numerical checks on small system sizes.
In another work~\cite{moudgalya2022numerical}, we elaborate on these numerical techniques and demonstrate their application on some of the examples studied in this work.
We hope that all these results, along with results for Hilbert space fragmentation~\cite{moudgalya2021hilbert}, might provide a way towards a unified and exhaustive understanding of both conventional and unconventional conserved quantities that appear in local Hamiltonians. 
Finally, this framework also demonstrates parallels between symmetries in quantum systems and decoherence-free subspaces or noiseless subsystems studied in the quantum information literature~\cite{lidar2003decoherencereview,  holbrook2003noiseless}.  
In particular,  when the terms of the Hamiltonians,  or the generators of the local algebra $\mA$, are interpreted as sources of ``noise" in the system,  information in the subspaces that transform under irreducible representations of $\mC$,  i.e., the $\{\mH^{(\mC)}_{\lambda}\}$ in Eq.~(\ref{eq:Hilbertdecomp}), are ``protected" against such a noise. 
While decoherence-free subspaces have mostly been explored for few-qubit systems, it might be interesting to use this framework to systematically explore many-body decoherence-free subspaces and their applications for realistic noise sources.
\section*{Acknowledgements}
We thank Arpit Dua, Igor Klebanov, Nick O'Dea, Daniel Mark, and Paolo Zanardi for useful discussions.
This work was supported by the Walter Burke Institute for Theoretical Physics at Caltech; the Institute for Quantum Information and Matter, an NSF Physics Frontiers Center (NSF Grant PHY-1733907); and the National Science Foundation through grant DMR-2001186.
A part of this work was done at the Aspen Center for Physics, which is supported by National Science Foundation grant PHY-1607611.
This work was partially supported by a grant from the Simons Foundation.
S.M. acknowledges the hospitality of the Centro de Ciencias de Benasque Pedro Pascual, where a part of this work was completed. 
\bibliography{refs}

\begin{thebibliography}{78}%
\makeatletter
\providecommand \@ifxundefined [1]{%
 \@ifx{#1\undefined}
}%
\providecommand \@ifnum [1]{%
 \ifnum #1\expandafter \@firstoftwo
 \else \expandafter \@secondoftwo
 \fi
}%
\providecommand \@ifx [1]{%
 \ifx #1\expandafter \@firstoftwo
 \else \expandafter \@secondoftwo
 \fi
}%
\providecommand \natexlab [1]{#1}%
\providecommand \enquote  [1]{``#1''}%
\providecommand \bibnamefont  [1]{#1}%
\providecommand \bibfnamefont [1]{#1}%
\providecommand \citenamefont [1]{#1}%
\providecommand \href@noop [0]{\@secondoftwo}%
\providecommand \href [0]{\begingroup \@sanitize@url \@href}%
\providecommand \@href[1]{\@@startlink{#1}\@@href}%
\providecommand \@@href[1]{\endgroup#1\@@endlink}%
\providecommand \@sanitize@url [0]{\catcode `\\12\catcode `\$12\catcode
  `\&12\catcode `\#12\catcode `\^12\catcode `\_12\catcode `\%12\relax}%
\providecommand \@@startlink[1]{}%
\providecommand \@@endlink[0]{}%
\providecommand \url  [0]{\begingroup\@sanitize@url \@url }%
\providecommand \@url [1]{\endgroup\@href {#1}{\urlprefix }}%
\providecommand \urlprefix  [0]{URL }%
\providecommand \Eprint [0]{\href }%
\providecommand \doibase [0]{https://doi.org/}%
\providecommand \selectlanguage [0]{\@gobble}%
\providecommand \bibinfo  [0]{\@secondoftwo}%
\providecommand \bibfield  [0]{\@secondoftwo}%
\providecommand \translation [1]{[#1]}%
\providecommand \BibitemOpen [0]{}%
\providecommand \bibitemStop [0]{}%
\providecommand \bibitemNoStop [0]{.\EOS\space}%
\providecommand \EOS [0]{\spacefactor3000\relax}%
\providecommand \BibitemShut  [1]{\csname bibitem#1\endcsname}%
\let\auto@bib@innerbib\@empty
%</preamble>
\bibitem [{\citenamefont {{Moudgalya}}\ and\ \citenamefont
  {{Motrunich}}(2022)}]{moudgalya2022exhaustive}%
  \BibitemOpen
  \bibfield  {author} {\bibinfo {author} {\bibfnamefont {S.}~\bibnamefont
  {{Moudgalya}}}\ and\ \bibinfo {author} {\bibfnamefont {O.~I.}\ \bibnamefont
  {{Motrunich}}},\ }\bibfield  {title} {\bibinfo {title} {{Exhaustive
  Characterization of Quantum Many-Body Scars using Commutant Algebras}},\
  }\href@noop {} {\bibfield  {journal} {\bibinfo  {journal} {arXiv e-prints}\ }
  (\bibinfo {year} {2022})},\ \Eprint {https://arxiv.org/abs/2209.03377}
  {arXiv:2209.03377 [cond-mat.str-el]} \BibitemShut {NoStop}%
\bibitem [{\citenamefont {Sachdev}(2011)}]{sachdev2011quantum}%
  \BibitemOpen
  \bibfield  {author} {\bibinfo {author} {\bibfnamefont {S.}~\bibnamefont
  {Sachdev}},\ }\href@noop {} {\emph {\bibinfo {title} {Quantum phase
  transitions}}}\ (\bibinfo  {publisher} {Cambridge university press},\
  \bibinfo {year} {2011})\BibitemShut {NoStop}%
\bibitem [{\citenamefont {Fradkin}(2013)}]{fradkin2013field}%
  \BibitemOpen
  \bibfield  {author} {\bibinfo {author} {\bibfnamefont {E.}~\bibnamefont
  {Fradkin}},\ }\href@noop {} {\emph {\bibinfo {title} {Field theories of
  condensed matter physics}}}\ (\bibinfo  {publisher} {Cambridge University
  Press},\ \bibinfo {year} {2013})\BibitemShut {NoStop}%
\bibitem [{\citenamefont {Zeng}\ \emph {et~al.}(2019)\citenamefont {Zeng},
  \citenamefont {Chen}, \citenamefont {Zhou}, \citenamefont {Wen} \emph
  {et~al.}}]{zeng2019quantum}%
  \BibitemOpen
  \bibfield  {author} {\bibinfo {author} {\bibfnamefont {B.}~\bibnamefont
  {Zeng}}, \bibinfo {author} {\bibfnamefont {X.}~\bibnamefont {Chen}}, \bibinfo
  {author} {\bibfnamefont {D.-L.}\ \bibnamefont {Zhou}}, \bibinfo {author}
  {\bibfnamefont {X.-G.}\ \bibnamefont {Wen}}, \emph {et~al.},\ }\href@noop {}
  {\emph {\bibinfo {title} {Quantum information meets quantum matter}}}\
  (\bibinfo  {publisher} {Springer},\ \bibinfo {year} {2019})\BibitemShut
  {NoStop}%
\bibitem [{\citenamefont {{D'Alessio}}\ \emph {et~al.}(2016)\citenamefont
  {{D'Alessio}}, \citenamefont {{Kafri}}, \citenamefont {{Polkovnikov}},\ and\
  \citenamefont {{Rigol}}}]{d2016quantum}%
  \BibitemOpen
  \bibfield  {author} {\bibinfo {author} {\bibfnamefont {L.}~\bibnamefont
  {{D'Alessio}}}, \bibinfo {author} {\bibfnamefont {Y.}~\bibnamefont
  {{Kafri}}}, \bibinfo {author} {\bibfnamefont {A.}~\bibnamefont
  {{Polkovnikov}}},\ and\ \bibinfo {author} {\bibfnamefont {M.}~\bibnamefont
  {{Rigol}}},\ }\bibfield  {title} {\bibinfo {title} {{From quantum chaos and
  eigenstate thermalization to statistical mechanics and thermodynamics}},\
  }\href {https://doi.org/10.1080/00018732.2016.1198134} {\bibfield  {journal}
  {\bibinfo  {journal} {Advances in Physics}\ }\textbf {\bibinfo {volume}
  {65}},\ \bibinfo {pages} {239} (\bibinfo {year} {2016})}\BibitemShut
  {NoStop}%
\bibitem [{\citenamefont {Mori}\ \emph {et~al.}(2018)\citenamefont {Mori},
  \citenamefont {Ikeda}, \citenamefont {Kaminishi},\ and\ \citenamefont
  {Ueda}}]{mori2018thermalization}%
  \BibitemOpen
  \bibfield  {author} {\bibinfo {author} {\bibfnamefont {T.}~\bibnamefont
  {Mori}}, \bibinfo {author} {\bibfnamefont {T.~N.}\ \bibnamefont {Ikeda}},
  \bibinfo {author} {\bibfnamefont {E.}~\bibnamefont {Kaminishi}},\ and\
  \bibinfo {author} {\bibfnamefont {M.}~\bibnamefont {Ueda}},\ }\bibfield
  {title} {\bibinfo {title} {Thermalization and prethermalization in isolated
  quantum systems: a theoretical overview},\ }\href
  {https://doi.org/10.1088/1361-6455/aabcdf} {\bibfield  {journal} {\bibinfo
  {journal} {Journal of Physics B: Atomic, Molecular and Optical Physics}\
  }\textbf {\bibinfo {volume} {51}},\ \bibinfo {pages} {112001} (\bibinfo
  {year} {2018})}\BibitemShut {NoStop}%
\bibitem [{\citenamefont {Quella}(2020)}]{quella2020symmetry}%
  \BibitemOpen
  \bibfield  {author} {\bibinfo {author} {\bibfnamefont {T.}~\bibnamefont
  {Quella}},\ }\bibfield  {title} {\bibinfo {title} {Symmetry-protected
  topological phases beyond groups: The $q$-deformed
  affleck-kennedy-lieb-tasaki model},\ }\href
  {https://doi.org/10.1103/PhysRevB.102.081120} {\bibfield  {journal} {\bibinfo
   {journal} {Phys. Rev. B}\ }\textbf {\bibinfo {volume} {102}},\ \bibinfo
  {pages} {081120} (\bibinfo {year} {2020})}\BibitemShut {NoStop}%
\bibitem [{\citenamefont {Lootens}\ \emph {et~al.}(2021)\citenamefont
  {Lootens}, \citenamefont {Fuchs}, \citenamefont {Haegeman}, \citenamefont
  {Schweigert},\ and\ \citenamefont {Verstraete}}]{lootens2021MPO}%
  \BibitemOpen
  \bibfield  {author} {\bibinfo {author} {\bibfnamefont {L.}~\bibnamefont
  {Lootens}}, \bibinfo {author} {\bibfnamefont {J.}~\bibnamefont {Fuchs}},
  \bibinfo {author} {\bibfnamefont {J.}~\bibnamefont {Haegeman}}, \bibinfo
  {author} {\bibfnamefont {C.}~\bibnamefont {Schweigert}},\ and\ \bibinfo
  {author} {\bibfnamefont {F.}~\bibnamefont {Verstraete}},\ }\bibfield  {title}
  {\bibinfo {title} {{Matrix product operator symmetries and intertwiners in
  string-nets with domain walls}},\ }\href
  {https://doi.org/10.21468/SciPostPhys.10.3.053} {\bibfield  {journal}
  {\bibinfo  {journal} {SciPost Phys.}\ }\textbf {\bibinfo {volume} {10}},\
  \bibinfo {pages} {53} (\bibinfo {year} {2021})}\BibitemShut {NoStop}%
\bibitem [{\citenamefont {Moudgalya}\ and\ \citenamefont
  {Motrunich}(2022)}]{moudgalya2021hilbert}%
  \BibitemOpen
  \bibfield  {author} {\bibinfo {author} {\bibfnamefont {S.}~\bibnamefont
  {Moudgalya}}\ and\ \bibinfo {author} {\bibfnamefont {O.~I.}\ \bibnamefont
  {Motrunich}},\ }\bibfield  {title} {\bibinfo {title} {Hilbert space
  fragmentation and commutant algebras},\ }\href
  {https://doi.org/10.1103/PhysRevX.12.011050} {\bibfield  {journal} {\bibinfo
  {journal} {Phys. Rev. X}\ }\textbf {\bibinfo {volume} {12}},\ \bibinfo
  {pages} {011050} (\bibinfo {year} {2022})}\BibitemShut {NoStop}%
\bibitem [{\citenamefont {McGreevy}(2023)}]{mcgreevy2022generalized}%
  \BibitemOpen
  \bibfield  {author} {\bibinfo {author} {\bibfnamefont {J.}~\bibnamefont
  {McGreevy}},\ }\bibfield  {title} {\bibinfo {title} {{Generalized Symmetries
  in Condensed Matter}},\ }\href
  {https://doi.org/10.1146/annurev-conmatphys-040721-021029} {\bibfield
  {journal} {\bibinfo  {journal} {Annual Review of Condensed Matter Physics}\
  }\textbf {\bibinfo {volume} {14}},\ \bibinfo {pages} {57} (\bibinfo {year}
  {2023})},\ \Eprint
  {https://arxiv.org/abs/https://doi.org/10.1146/annurev-conmatphys-040721-021029}
  {https://doi.org/10.1146/annurev-conmatphys-040721-021029} \BibitemShut
  {NoStop}%
\bibitem [{\citenamefont {Serbyn}\ \emph {et~al.}(2021)\citenamefont {Serbyn},
  \citenamefont {Abanin},\ and\ \citenamefont
  {Papi{\'{c}}}}]{serbyn2020review}%
  \BibitemOpen
  \bibfield  {author} {\bibinfo {author} {\bibfnamefont {M.}~\bibnamefont
  {Serbyn}}, \bibinfo {author} {\bibfnamefont {D.~A.}\ \bibnamefont {Abanin}},\
  and\ \bibinfo {author} {\bibfnamefont {Z.}~\bibnamefont {Papi{\'{c}}}},\
  }\bibfield  {title} {\bibinfo {title} {Quantum many-body scars and weak
  breaking of ergodicity},\ }\href {https://doi.org/10.1038/s41567-021-01230-2}
  {\bibfield  {journal} {\bibinfo  {journal} {Nature Physics}\ }\textbf
  {\bibinfo {volume} {17}},\ \bibinfo {pages} {675} (\bibinfo {year}
  {2021})}\BibitemShut {NoStop}%
\bibitem [{\citenamefont {Papi{\'{c}}}(2022)}]{papic2021review}%
  \BibitemOpen
  \bibfield  {author} {\bibinfo {author} {\bibfnamefont {Z.}~\bibnamefont
  {Papi{\'{c}}}},\ }\bibinfo {title} {{Weak Ergodicity Breaking Through the
  Lens of Quantum Entanglement}},\ in\ \href
  {https://doi.org/10.1007/978-3-031-03998-0_13} {\emph {\bibinfo {booktitle}
  {{Entanglement in Spin Chains: From Theory to Quantum Technology
  Applications}}}},\ \bibinfo {editor} {edited by\ \bibinfo {editor}
  {\bibfnamefont {A.}~\bibnamefont {Bayat}}, \bibinfo {editor} {\bibfnamefont
  {S.}~\bibnamefont {Bose}},\ and\ \bibinfo {editor} {\bibfnamefont
  {H.}~\bibnamefont {Johannesson}}}\ (\bibinfo  {publisher} {Springer
  International Publishing},\ \bibinfo {address} {Cham},\ \bibinfo {year}
  {2022})\ pp.\ \bibinfo {pages} {341--395}\BibitemShut {NoStop}%
\bibitem [{\citenamefont {Moudgalya}\ \emph {et~al.}(2022)\citenamefont
  {Moudgalya}, \citenamefont {Bernevig},\ and\ \citenamefont
  {Regnault}}]{moudgalya2021review}%
  \BibitemOpen
  \bibfield  {author} {\bibinfo {author} {\bibfnamefont {S.}~\bibnamefont
  {Moudgalya}}, \bibinfo {author} {\bibfnamefont {B.~A.}\ \bibnamefont
  {Bernevig}},\ and\ \bibinfo {author} {\bibfnamefont {N.}~\bibnamefont
  {Regnault}},\ }\bibfield  {title} {\bibinfo {title} {Quantum many-body scars
  and hilbert space fragmentation: a review of exact results},\ }\href
  {https://doi.org/10.1088/1361-6633/ac73a0} {\bibfield  {journal} {\bibinfo
  {journal} {Reports on Progress in Physics}\ }\textbf {\bibinfo {volume}
  {85}},\ \bibinfo {pages} {086501} (\bibinfo {year} {2022})}\BibitemShut
  {NoStop}%
\bibitem [{\citenamefont {Sala}\ \emph {et~al.}(2020)\citenamefont {Sala},
  \citenamefont {Rakovszky}, \citenamefont {Verresen}, \citenamefont {Knap},\
  and\ \citenamefont {Pollmann}}]{sala2020fragmentation}%
  \BibitemOpen
  \bibfield  {author} {\bibinfo {author} {\bibfnamefont {P.}~\bibnamefont
  {Sala}}, \bibinfo {author} {\bibfnamefont {T.}~\bibnamefont {Rakovszky}},
  \bibinfo {author} {\bibfnamefont {R.}~\bibnamefont {Verresen}}, \bibinfo
  {author} {\bibfnamefont {M.}~\bibnamefont {Knap}},\ and\ \bibinfo {author}
  {\bibfnamefont {F.}~\bibnamefont {Pollmann}},\ }\bibfield  {title} {\bibinfo
  {title} {Ergodicity breaking arising from {H}ilbert space fragmentation in
  dipole-conserving {H}amiltonians},\ }\href
  {https://doi.org/10.1103/PhysRevX.10.011047} {\bibfield  {journal} {\bibinfo
  {journal} {Phys. Rev. X}\ }\textbf {\bibinfo {volume} {10}},\ \bibinfo
  {pages} {011047} (\bibinfo {year} {2020})}\BibitemShut {NoStop}%
\bibitem [{\citenamefont {Khemani}\ \emph {et~al.}(2020)\citenamefont
  {Khemani}, \citenamefont {Hermele},\ and\ \citenamefont
  {Nandkishore}}]{khemani2020localization}%
  \BibitemOpen
  \bibfield  {author} {\bibinfo {author} {\bibfnamefont {V.}~\bibnamefont
  {Khemani}}, \bibinfo {author} {\bibfnamefont {M.}~\bibnamefont {Hermele}},\
  and\ \bibinfo {author} {\bibfnamefont {R.}~\bibnamefont {Nandkishore}},\
  }\bibfield  {title} {\bibinfo {title} {Localization from {H}ilbert space
  shattering: From theory to physical realizations},\ }\href
  {https://doi.org/10.1103/PhysRevB.101.174204} {\bibfield  {journal} {\bibinfo
   {journal} {Phys. Rev. B}\ }\textbf {\bibinfo {volume} {101}},\ \bibinfo
  {pages} {174204} (\bibinfo {year} {2020})}\BibitemShut {NoStop}%
\bibitem [{\citenamefont {Moudgalya}\ \emph {et~al.}()\citenamefont
  {Moudgalya}, \citenamefont {Prem}, \citenamefont {Nandkishore}, \citenamefont
  {Regnault},\ and\ \citenamefont {Bernevig}}]{moudgalya2019thermalization}%
  \BibitemOpen
  \bibfield  {author} {\bibinfo {author} {\bibfnamefont {S.}~\bibnamefont
  {Moudgalya}}, \bibinfo {author} {\bibfnamefont {A.}~\bibnamefont {Prem}},
  \bibinfo {author} {\bibfnamefont {R.}~\bibnamefont {Nandkishore}}, \bibinfo
  {author} {\bibfnamefont {N.}~\bibnamefont {Regnault}},\ and\ \bibinfo
  {author} {\bibfnamefont {B.~A.}\ \bibnamefont {Bernevig}},\ }\bibinfo {title}
  {{Thermalization and Its Absence within Krylov Subspaces of a Constrained
  Hamiltonian}},\ in\ \href {https://doi.org/10.1142/9789811231711_0009} {\emph
  {\bibinfo {booktitle} {Memorial Volume for Shoucheng Zhang}}},\
  Chap.~\bibinfo {chapter} {7}, pp.\ \bibinfo {pages} {147--209}\BibitemShut
  {NoStop}%
\bibitem [{\citenamefont {Yang}\ \emph {et~al.}(2020)\citenamefont {Yang},
  \citenamefont {Liu}, \citenamefont {Gorshkov},\ and\ \citenamefont
  {Iadecola}}]{yang2019hilbertspace}%
  \BibitemOpen
  \bibfield  {author} {\bibinfo {author} {\bibfnamefont {Z.-C.}\ \bibnamefont
  {Yang}}, \bibinfo {author} {\bibfnamefont {F.}~\bibnamefont {Liu}}, \bibinfo
  {author} {\bibfnamefont {A.~V.}\ \bibnamefont {Gorshkov}},\ and\ \bibinfo
  {author} {\bibfnamefont {T.}~\bibnamefont {Iadecola}},\ }\bibfield  {title}
  {\bibinfo {title} {Hilbert-space fragmentation from strict confinement},\
  }\href {https://doi.org/10.1103/PhysRevLett.124.207602} {\bibfield  {journal}
  {\bibinfo  {journal} {Phys. Rev. Lett.}\ }\textbf {\bibinfo {volume} {124}},\
  \bibinfo {pages} {207602} (\bibinfo {year} {2020})}\BibitemShut {NoStop}%
\bibitem [{\citenamefont {Deutsch}(1991)}]{deutsch1991quantum}%
  \BibitemOpen
  \bibfield  {author} {\bibinfo {author} {\bibfnamefont {J.~M.}\ \bibnamefont
  {Deutsch}},\ }\bibfield  {title} {\bibinfo {title} {Quantum statistical
  mechanics in a closed system},\ }\href
  {https://doi.org/10.1103/PhysRevA.43.2046} {\bibfield  {journal} {\bibinfo
  {journal} {Phys. Rev. A}\ }\textbf {\bibinfo {volume} {43}},\ \bibinfo
  {pages} {2046} (\bibinfo {year} {1991})}\BibitemShut {NoStop}%
\bibitem [{\citenamefont {Srednicki}(1994)}]{srednicki1994chaos}%
  \BibitemOpen
  \bibfield  {author} {\bibinfo {author} {\bibfnamefont {M.}~\bibnamefont
  {Srednicki}},\ }\bibfield  {title} {\bibinfo {title} {Chaos and quantum
  thermalization},\ }\href {https://doi.org/10.1103/PhysRevE.50.888} {\bibfield
   {journal} {\bibinfo  {journal} {Phys. Rev. E}\ }\textbf {\bibinfo {volume}
  {50}},\ \bibinfo {pages} {888} (\bibinfo {year} {1994})}\BibitemShut
  {NoStop}%
\bibitem [{\citenamefont {{Rigol}}\ \emph {et~al.}(2008)\citenamefont
  {{Rigol}}, \citenamefont {{Dunjko}},\ and\ \citenamefont
  {{Olshanii}}}]{rigol2008thermalization}%
  \BibitemOpen
  \bibfield  {author} {\bibinfo {author} {\bibfnamefont {M.}~\bibnamefont
  {{Rigol}}}, \bibinfo {author} {\bibfnamefont {V.}~\bibnamefont {{Dunjko}}},\
  and\ \bibinfo {author} {\bibfnamefont {M.}~\bibnamefont {{Olshanii}}},\
  }\bibfield  {title} {\bibinfo {title} {{Thermalization and its mechanism for
  generic isolated quantum systems}},\ }\href
  {https://doi.org/10.1038/nature06838} {\bibfield  {journal} {\bibinfo
  {journal} {Nature}\ }\textbf {\bibinfo {volume} {452}},\ \bibinfo {pages}
  {854} (\bibinfo {year} {2008})}\BibitemShut {NoStop}%
\bibitem [{\citenamefont {Polkovnikov}\ \emph {et~al.}(2011)\citenamefont
  {Polkovnikov}, \citenamefont {Sengupta}, \citenamefont {Silva},\ and\
  \citenamefont {Vengalattore}}]{polkovnikov2011colloquium}%
  \BibitemOpen
  \bibfield  {author} {\bibinfo {author} {\bibfnamefont {A.}~\bibnamefont
  {Polkovnikov}}, \bibinfo {author} {\bibfnamefont {K.}~\bibnamefont
  {Sengupta}}, \bibinfo {author} {\bibfnamefont {A.}~\bibnamefont {Silva}},\
  and\ \bibinfo {author} {\bibfnamefont {M.}~\bibnamefont {Vengalattore}},\
  }\bibfield  {title} {\bibinfo {title} {Colloquium: Nonequilibrium dynamics of
  closed interacting quantum systems},\ }\href
  {https://doi.org/10.1103/RevModPhys.83.863} {\bibfield  {journal} {\bibinfo
  {journal} {Reviews of Modern Physics}\ }\textbf {\bibinfo {volume} {83}},\
  \bibinfo {pages} {863} (\bibinfo {year} {2011})}\BibitemShut {NoStop}%
\bibitem [{\citenamefont {Bu{\v{c}}a}\ \emph {et~al.}(2019)\citenamefont
  {Bu{\v{c}}a}, \citenamefont {Tindall},\ and\ \citenamefont
  {Jaksch}}]{buca2019nonstationary}%
  \BibitemOpen
  \bibfield  {author} {\bibinfo {author} {\bibfnamefont {B.}~\bibnamefont
  {Bu{\v{c}}a}}, \bibinfo {author} {\bibfnamefont {J.}~\bibnamefont
  {Tindall}},\ and\ \bibinfo {author} {\bibfnamefont {D.}~\bibnamefont
  {Jaksch}},\ }\bibfield  {title} {\bibinfo {title} {Non-stationary coherent
  quantum many-body dynamics through dissipation},\ }\href
  {https://doi.org/10.1038/s41467-019-09757-y} {\bibfield  {journal} {\bibinfo
  {journal} {Nature Communications}\ }\textbf {\bibinfo {volume} {10}},\
  \bibinfo {pages} {1730} (\bibinfo {year} {2019})}\BibitemShut {NoStop}%
\bibitem [{\citenamefont {Medenjak}\ \emph {et~al.}(2020)\citenamefont
  {Medenjak}, \citenamefont {Bu\ifmmode~\check{c}\else \v{c}\fi{}a},\ and\
  \citenamefont {Jaksch}}]{medenjak2020isolated}%
  \BibitemOpen
  \bibfield  {author} {\bibinfo {author} {\bibfnamefont {M.}~\bibnamefont
  {Medenjak}}, \bibinfo {author} {\bibfnamefont {B.}~\bibnamefont
  {Bu\ifmmode~\check{c}\else \v{c}\fi{}a}},\ and\ \bibinfo {author}
  {\bibfnamefont {D.}~\bibnamefont {Jaksch}},\ }\bibfield  {title} {\bibinfo
  {title} {Isolated heisenberg magnet as a quantum time crystal},\ }\href
  {https://doi.org/10.1103/PhysRevB.102.041117} {\bibfield  {journal} {\bibinfo
   {journal} {Phys. Rev. B}\ }\textbf {\bibinfo {volume} {102}},\ \bibinfo
  {pages} {041117} (\bibinfo {year} {2020})}\BibitemShut {NoStop}%
\bibitem [{\citenamefont {Moudgalya}\ \emph {et~al.}(2020)\citenamefont
  {Moudgalya}, \citenamefont {Regnault},\ and\ \citenamefont
  {Bernevig}}]{moudgalya2020eta}%
  \BibitemOpen
  \bibfield  {author} {\bibinfo {author} {\bibfnamefont {S.}~\bibnamefont
  {Moudgalya}}, \bibinfo {author} {\bibfnamefont {N.}~\bibnamefont
  {Regnault}},\ and\ \bibinfo {author} {\bibfnamefont {B.~A.}\ \bibnamefont
  {Bernevig}},\ }\bibfield  {title} {\bibinfo {title}
  {$\ensuremath{\eta}$-pairing in {H}ubbard models: From spectrum generating
  algebras to quantum many-body scars},\ }\href
  {https://doi.org/10.1103/PhysRevB.102.085140} {\bibfield  {journal} {\bibinfo
   {journal} {Phys. Rev. B}\ }\textbf {\bibinfo {volume} {102}},\ \bibinfo
  {pages} {085140} (\bibinfo {year} {2020})}\BibitemShut {NoStop}%
\bibitem [{\citenamefont {Tang}\ \emph {et~al.}(2022)\citenamefont {Tang},
  \citenamefont {O'Dea},\ and\ \citenamefont {Chandran}}]{tang2020multi}%
  \BibitemOpen
  \bibfield  {author} {\bibinfo {author} {\bibfnamefont {L.-H.}\ \bibnamefont
  {Tang}}, \bibinfo {author} {\bibfnamefont {N.}~\bibnamefont {O'Dea}},\ and\
  \bibinfo {author} {\bibfnamefont {A.}~\bibnamefont {Chandran}},\ }\bibfield
  {title} {\bibinfo {title} {Multimagnon quantum many-body scars from tensor
  operators},\ }\href {https://doi.org/10.1103/PhysRevResearch.4.043006}
  {\bibfield  {journal} {\bibinfo  {journal} {Phys. Rev. Res.}\ }\textbf
  {\bibinfo {volume} {4}},\ \bibinfo {pages} {043006} (\bibinfo {year}
  {2022})}\BibitemShut {NoStop}%
\bibitem [{\citenamefont {Pakrouski}\ \emph {et~al.}(2020)\citenamefont
  {Pakrouski}, \citenamefont {Pallegar}, \citenamefont {Popov},\ and\
  \citenamefont {Klebanov}}]{pakrouski2020many}%
  \BibitemOpen
  \bibfield  {author} {\bibinfo {author} {\bibfnamefont {K.}~\bibnamefont
  {Pakrouski}}, \bibinfo {author} {\bibfnamefont {P.~N.}\ \bibnamefont
  {Pallegar}}, \bibinfo {author} {\bibfnamefont {F.~K.}\ \bibnamefont
  {Popov}},\ and\ \bibinfo {author} {\bibfnamefont {I.~R.}\ \bibnamefont
  {Klebanov}},\ }\bibfield  {title} {\bibinfo {title} {Many-body scars as a
  group invariant sector of {H}ilbert space},\ }\href
  {https://doi.org/10.1103/PhysRevLett.125.230602} {\bibfield  {journal}
  {\bibinfo  {journal} {Phys. Rev. Lett.}\ }\textbf {\bibinfo {volume} {125}},\
  \bibinfo {pages} {230602} (\bibinfo {year} {2020})}\BibitemShut {NoStop}%
\bibitem [{\citenamefont {Pakrouski}\ \emph {et~al.}(2021)\citenamefont
  {Pakrouski}, \citenamefont {Pallegar}, \citenamefont {Popov},\ and\
  \citenamefont {Klebanov}}]{pakrouski2021group}%
  \BibitemOpen
  \bibfield  {author} {\bibinfo {author} {\bibfnamefont {K.}~\bibnamefont
  {Pakrouski}}, \bibinfo {author} {\bibfnamefont {P.~N.}\ \bibnamefont
  {Pallegar}}, \bibinfo {author} {\bibfnamefont {F.~K.}\ \bibnamefont
  {Popov}},\ and\ \bibinfo {author} {\bibfnamefont {I.~R.}\ \bibnamefont
  {Klebanov}},\ }\bibfield  {title} {\bibinfo {title} {Group theoretic approach
  to many-body scar states in fermionic lattice models},\ }\href
  {https://doi.org/10.1103/PhysRevResearch.3.043156} {\bibfield  {journal}
  {\bibinfo  {journal} {Phys. Rev. Research}\ }\textbf {\bibinfo {volume}
  {3}},\ \bibinfo {pages} {043156} (\bibinfo {year} {2021})}\BibitemShut
  {NoStop}%
\bibitem [{\citenamefont {{Sun}}\ \emph {et~al.}(2022)\citenamefont {{Sun}},
  \citenamefont {{Popov}}, \citenamefont {{Klebanov}},\ and\ \citenamefont
  {{Pakrouski}}}]{sun2022majorana}%
  \BibitemOpen
  \bibfield  {author} {\bibinfo {author} {\bibfnamefont {Z.}~\bibnamefont
  {{Sun}}}, \bibinfo {author} {\bibfnamefont {F.~K.}\ \bibnamefont {{Popov}}},
  \bibinfo {author} {\bibfnamefont {I.~R.}\ \bibnamefont {{Klebanov}}},\ and\
  \bibinfo {author} {\bibfnamefont {K.}~\bibnamefont {{Pakrouski}}},\
  }\bibfield  {title} {\bibinfo {title} {{Majorana Scars as Group Singlets}},\
  }\bibfield  {journal} {\bibinfo  {journal} {arXiv e-prints}\ }\href
  {https://doi.org/10.48550/arXiv.2212.11914} {10.48550/arXiv.2212.11914}
  (\bibinfo {year} {2022}),\ \Eprint {https://arxiv.org/abs/2212.11914}
  {arXiv:2212.11914 [cond-mat.str-el]} \BibitemShut {NoStop}%
\bibitem [{\citenamefont {Klebanov}\ \emph {et~al.}(2018)\citenamefont
  {Klebanov}, \citenamefont {Milekhin}, \citenamefont {Popov},\ and\
  \citenamefont {Tarnopolsky}}]{klebanov2018spectra}%
  \BibitemOpen
  \bibfield  {author} {\bibinfo {author} {\bibfnamefont {I.~R.}\ \bibnamefont
  {Klebanov}}, \bibinfo {author} {\bibfnamefont {A.}~\bibnamefont {Milekhin}},
  \bibinfo {author} {\bibfnamefont {F.}~\bibnamefont {Popov}},\ and\ \bibinfo
  {author} {\bibfnamefont {G.}~\bibnamefont {Tarnopolsky}},\ }\bibfield
  {title} {\bibinfo {title} {Spectra of eigenstates in fermionic tensor quantum
  mechanics},\ }\href {https://doi.org/10.1103/PhysRevD.97.106023} {\bibfield
  {journal} {\bibinfo  {journal} {Phys. Rev. D}\ }\textbf {\bibinfo {volume}
  {97}},\ \bibinfo {pages} {106023} (\bibinfo {year} {2018})}\BibitemShut
  {NoStop}%
\bibitem [{\citenamefont {Gaitan}\ \emph {et~al.}(2020)\citenamefont {Gaitan},
  \citenamefont {Klebanov}, \citenamefont {Pakrouski}, \citenamefont
  {Pallegar},\ and\ \citenamefont {Popov}}]{gaitan2020hagedorn}%
  \BibitemOpen
  \bibfield  {author} {\bibinfo {author} {\bibfnamefont {G.}~\bibnamefont
  {Gaitan}}, \bibinfo {author} {\bibfnamefont {I.~R.}\ \bibnamefont
  {Klebanov}}, \bibinfo {author} {\bibfnamefont {K.}~\bibnamefont {Pakrouski}},
  \bibinfo {author} {\bibfnamefont {P.~N.}\ \bibnamefont {Pallegar}},\ and\
  \bibinfo {author} {\bibfnamefont {F.~K.}\ \bibnamefont {Popov}},\ }\bibfield
  {title} {\bibinfo {title} {Hagedorn temperature in large $n$ majorana quantum
  mechanics},\ }\href {https://doi.org/10.1103/PhysRevD.101.126002} {\bibfield
  {journal} {\bibinfo  {journal} {Phys. Rev. D}\ }\textbf {\bibinfo {volume}
  {101}},\ \bibinfo {pages} {126002} (\bibinfo {year} {2020})}\BibitemShut
  {NoStop}%
\bibitem [{\citenamefont {{Landsman}}(1998)}]{landsman1998lecture}%
  \BibitemOpen
  \bibfield  {author} {\bibinfo {author} {\bibfnamefont {N.~P.}\ \bibnamefont
  {{Landsman}}},\ }\bibfield  {title} {\bibinfo {title} {{Lecture notes on
  C*-algebras, Hilbert C*-modules, and quantum mechanics}},\ }\bibfield
  {journal} {\bibinfo  {journal} {arXiv e-prints}\ }\href
  {https://doi.org/10.48550/arXiv.math-ph/9807030}
  {10.48550/arXiv.math-ph/9807030} (\bibinfo {year} {1998}),\ \Eprint
  {https://arxiv.org/abs/math-ph/9807030} {arXiv:math-ph/9807030 [math-ph]}
  \BibitemShut {NoStop}%
\bibitem [{\citenamefont {Qi}\ and\ \citenamefont
  {Ranard}(2019)}]{qi2019determininglocal}%
  \BibitemOpen
  \bibfield  {author} {\bibinfo {author} {\bibfnamefont {X.-L.}\ \bibnamefont
  {Qi}}\ and\ \bibinfo {author} {\bibfnamefont {D.}~\bibnamefont {Ranard}},\
  }\bibfield  {title} {\bibinfo {title} {Determining a local {H}amiltonian from
  a single eigenstate},\ }\href {https://doi.org/10.22331/q-2019-07-08-159}
  {\bibfield  {journal} {\bibinfo  {journal} {{Quantum}}\ }\textbf {\bibinfo
  {volume} {3}},\ \bibinfo {pages} {159} (\bibinfo {year} {2019})}\BibitemShut
  {NoStop}%
\bibitem [{\citenamefont {Chertkov}\ \emph {et~al.}(2020)\citenamefont
  {Chertkov}, \citenamefont {Villalonga},\ and\ \citenamefont
  {Clark}}]{chertkov2020engineering}%
  \BibitemOpen
  \bibfield  {author} {\bibinfo {author} {\bibfnamefont {E.}~\bibnamefont
  {Chertkov}}, \bibinfo {author} {\bibfnamefont {B.}~\bibnamefont
  {Villalonga}},\ and\ \bibinfo {author} {\bibfnamefont {B.~K.}\ \bibnamefont
  {Clark}},\ }\bibfield  {title} {\bibinfo {title} {Engineering topological
  models with a general-purpose symmetry-to-hamiltonian approach},\ }\href
  {https://doi.org/10.1103/PhysRevResearch.2.023348} {\bibfield  {journal}
  {\bibinfo  {journal} {Phys. Rev. Research}\ }\textbf {\bibinfo {volume}
  {2}},\ \bibinfo {pages} {023348} (\bibinfo {year} {2020})}\BibitemShut
  {NoStop}%
\bibitem [{\citenamefont {Mark}\ and\ \citenamefont
  {Motrunich}(2020)}]{mark2020eta}%
  \BibitemOpen
  \bibfield  {author} {\bibinfo {author} {\bibfnamefont {D.~K.}\ \bibnamefont
  {Mark}}\ and\ \bibinfo {author} {\bibfnamefont {O.~I.}\ \bibnamefont
  {Motrunich}},\ }\bibfield  {title} {\bibinfo {title}
  {$\ensuremath{\eta}$-pairing states as true scars in an extended {Hubbard}
  model},\ }\href {https://doi.org/10.1103/PhysRevB.102.075132} {\bibfield
  {journal} {\bibinfo  {journal} {Phys. Rev. B}\ }\textbf {\bibinfo {volume}
  {102}},\ \bibinfo {pages} {075132} (\bibinfo {year} {2020})}\BibitemShut
  {NoStop}%
\bibitem [{\citenamefont {Nussinov}\ and\ \citenamefont
  {Ortiz}(2009)}]{nussinov2009bond}%
  \BibitemOpen
  \bibfield  {author} {\bibinfo {author} {\bibfnamefont {Z.}~\bibnamefont
  {Nussinov}}\ and\ \bibinfo {author} {\bibfnamefont {G.}~\bibnamefont
  {Ortiz}},\ }\bibfield  {title} {\bibinfo {title} {Bond algebras and exact
  solvability of {H}amiltonians: {S}pin ${S}=\frac{1}{2}$ multilayer systems},\
  }\href {https://doi.org/10.1103/PhysRevB.79.214440} {\bibfield  {journal}
  {\bibinfo  {journal} {Phys. Rev. B}\ }\textbf {\bibinfo {volume} {79}},\
  \bibinfo {pages} {214440} (\bibinfo {year} {2009})}\BibitemShut {NoStop}%
\bibitem [{\citenamefont {Cobanera}\ \emph {et~al.}(2010)\citenamefont
  {Cobanera}, \citenamefont {Ortiz},\ and\ \citenamefont
  {Nussinov}}]{cobanera2010unified}%
  \BibitemOpen
  \bibfield  {author} {\bibinfo {author} {\bibfnamefont {E.}~\bibnamefont
  {Cobanera}}, \bibinfo {author} {\bibfnamefont {G.}~\bibnamefont {Ortiz}},\
  and\ \bibinfo {author} {\bibfnamefont {Z.}~\bibnamefont {Nussinov}},\
  }\bibfield  {title} {\bibinfo {title} {Unified approach to quantum and
  classical dualities},\ }\href
  {https://doi.org/10.1103/PhysRevLett.104.020402} {\bibfield  {journal}
  {\bibinfo  {journal} {Phys. Rev. Lett.}\ }\textbf {\bibinfo {volume} {104}},\
  \bibinfo {pages} {020402} (\bibinfo {year} {2010})}\BibitemShut {NoStop}%
\bibitem [{\citenamefont {Cobanera}\ \emph {et~al.}(2011)\citenamefont
  {Cobanera}, \citenamefont {Ortiz},\ and\ \citenamefont
  {Nussinov}}]{cobenera2011bond}%
  \BibitemOpen
  \bibfield  {author} {\bibinfo {author} {\bibfnamefont {E.}~\bibnamefont
  {Cobanera}}, \bibinfo {author} {\bibfnamefont {G.}~\bibnamefont {Ortiz}},\
  and\ \bibinfo {author} {\bibfnamefont {Z.}~\bibnamefont {Nussinov}},\
  }\bibfield  {title} {\bibinfo {title} {The bond-algebraic approach to
  dualities},\ }\href {https://doi.org/10.1080/00018732.2011.619814} {\bibfield
   {journal} {\bibinfo  {journal} {Advances in Physics}\ }\textbf {\bibinfo
  {volume} {60}},\ \bibinfo {pages} {679} (\bibinfo {year} {2011})}\BibitemShut
  {NoStop}%
\bibitem [{\citenamefont {Harlow}(2017)}]{harlow2017}%
  \BibitemOpen
  \bibfield  {author} {\bibinfo {author} {\bibfnamefont {D.}~\bibnamefont
  {Harlow}},\ }\bibfield  {title} {\bibinfo {title} {The {R}yu--{T}akayanagi
  formula from quantum error correction},\ }\href
  {https://doi.org/10.1007/s00220-017-2904-z} {\bibfield  {journal} {\bibinfo
  {journal} {Communications in Mathematical Physics}\ }\textbf {\bibinfo
  {volume} {354}},\ \bibinfo {pages} {865} (\bibinfo {year}
  {2017})}\BibitemShut {NoStop}%
\bibitem [{\citenamefont {Kabernik}(2021)}]{kabernik2021reductions}%
  \BibitemOpen
  \bibfield  {author} {\bibinfo {author} {\bibfnamefont {O.}~\bibnamefont
  {Kabernik}},\ }\href@noop {} {\bibinfo {title} {Reductions in
  finite-dimensional quantum mechanics: from symmetries to operator algebras
  and beyond}} (\bibinfo {year} {2021}),\ \Eprint
  {https://arxiv.org/abs/2103.08226} {arXiv:2103.08226 [quant-ph]} \BibitemShut
  {NoStop}%
\bibitem [{\citenamefont {Zanardi}(2001)}]{zanardi2001virtual}%
  \BibitemOpen
  \bibfield  {author} {\bibinfo {author} {\bibfnamefont {P.}~\bibnamefont
  {Zanardi}},\ }\bibfield  {title} {\bibinfo {title} {Virtual quantum
  subsystems},\ }\href {https://doi.org/10.1103/PhysRevLett.87.077901}
  {\bibfield  {journal} {\bibinfo  {journal} {Phys. Rev. Lett.}\ }\textbf
  {\bibinfo {volume} {87}},\ \bibinfo {pages} {077901} (\bibinfo {year}
  {2001})}\BibitemShut {NoStop}%
\bibitem [{\citenamefont {Bartlett}\ \emph {et~al.}(2007)\citenamefont
  {Bartlett}, \citenamefont {Rudolph},\ and\ \citenamefont
  {Spekkens}}]{bartlett2007reference}%
  \BibitemOpen
  \bibfield  {author} {\bibinfo {author} {\bibfnamefont {S.~D.}\ \bibnamefont
  {Bartlett}}, \bibinfo {author} {\bibfnamefont {T.}~\bibnamefont {Rudolph}},\
  and\ \bibinfo {author} {\bibfnamefont {R.~W.}\ \bibnamefont {Spekkens}},\
  }\bibfield  {title} {\bibinfo {title} {Reference frames, superselection
  rules, and quantum information},\ }\href
  {https://doi.org/10.1103/RevModPhys.79.555} {\bibfield  {journal} {\bibinfo
  {journal} {Rev. Mod. Phys.}\ }\textbf {\bibinfo {volume} {79}},\ \bibinfo
  {pages} {555} (\bibinfo {year} {2007})}\BibitemShut {NoStop}%
\bibitem [{\citenamefont {Lidar}(2014)}]{lidar2014dfs}%
  \BibitemOpen
  \bibfield  {author} {\bibinfo {author} {\bibfnamefont {D.~A.}\ \bibnamefont
  {Lidar}},\ }\bibinfo {title} {Review of decoherence-free subspaces, noiseless
  subsystems, and dynamical decoupling},\ in\ \href
  {https://doi.org/https://doi.org/10.1002/9781118742631.ch11} {\emph {\bibinfo
  {booktitle} {Quantum Information and Computation for Chemistry}}}\ (\bibinfo
  {publisher} {John Wiley \& Sons, Ltd},\ \bibinfo {year} {2014})\ pp.\
  \bibinfo {pages} {295--354}\BibitemShut {NoStop}%
\bibitem [{\citenamefont {Fulton}\ and\ \citenamefont
  {Harris}(2013)}]{fulton2013representation}%
  \BibitemOpen
  \bibfield  {author} {\bibinfo {author} {\bibfnamefont {W.}~\bibnamefont
  {Fulton}}\ and\ \bibinfo {author} {\bibfnamefont {J.}~\bibnamefont
  {Harris}},\ }\href@noop {} {\emph {\bibinfo {title} {Representation theory: a
  first course}}},\ Vol.\ \bibinfo {volume} {129}\ (\bibinfo  {publisher}
  {Springer Science \& Business Media},\ \bibinfo {year} {2013})\BibitemShut
  {NoStop}%
\bibitem [{\citenamefont {Yang}(1989)}]{yang1989eta}%
  \BibitemOpen
  \bibfield  {author} {\bibinfo {author} {\bibfnamefont {C.~N.}\ \bibnamefont
  {Yang}},\ }\bibfield  {title} {\bibinfo {title} {\ensuremath{\eta} pairing
  and off-diagonal long-range order in a hubbard model},\ }\href
  {https://doi.org/10.1103/PhysRevLett.63.2144} {\bibfield  {journal} {\bibinfo
   {journal} {Phys. Rev. Lett.}\ }\textbf {\bibinfo {volume} {63}},\ \bibinfo
  {pages} {2144} (\bibinfo {year} {1989})}\BibitemShut {NoStop}%
\bibitem [{\citenamefont {Kitaev}(2003)}]{kitaev2003fault}%
  \BibitemOpen
  \bibfield  {author} {\bibinfo {author} {\bibfnamefont {A.}~\bibnamefont
  {Kitaev}},\ }\bibfield  {title} {\bibinfo {title} {Fault-tolerant quantum
  computation by anyons},\ }\href
  {https://doi.org/https://doi.org/10.1016/S0003-4916(02)00018-0} {\bibfield
  {journal} {\bibinfo  {journal} {Annals of Physics}\ }\textbf {\bibinfo
  {volume} {303}},\ \bibinfo {pages} {2 } (\bibinfo {year} {2003})}\BibitemShut
  {NoStop}%
\bibitem [{\citenamefont {Kitaev}\ and\ \citenamefont
  {Laumann}(2010)}]{kitaev2010topological}%
  \BibitemOpen
  \bibfield  {author} {\bibinfo {author} {\bibfnamefont {A.}~\bibnamefont
  {Kitaev}}\ and\ \bibinfo {author} {\bibfnamefont {C.}~\bibnamefont
  {Laumann}},\ }\bibfield  {title} {\bibinfo {title} {Topological phases and
  quantum computation},\ }\href@noop {} {\bibfield  {journal} {\bibinfo
  {journal} {Exact Methods in Low-dimensional Statistical Physics and Quantum
  Computing: Lecture Notes of the Les Houches Summer School: Volume 89, July
  2008}\ ,\ \bibinfo {pages} {101}} (\bibinfo {year} {2010})}\BibitemShut
  {NoStop}%
\bibitem [{\citenamefont {Nandkishore}\ and\ \citenamefont
  {Hermele}(2019)}]{fractonreview}%
  \BibitemOpen
  \bibfield  {author} {\bibinfo {author} {\bibfnamefont {R.~M.}\ \bibnamefont
  {Nandkishore}}\ and\ \bibinfo {author} {\bibfnamefont {M.}~\bibnamefont
  {Hermele}},\ }\bibfield  {title} {\bibinfo {title} {Fractons},\ }\href
  {https://doi.org/10.1146/annurev-conmatphys-031218-013604} {\bibfield
  {journal} {\bibinfo  {journal} {Annual Review of Condensed Matter Physics}\
  }\textbf {\bibinfo {volume} {10}},\ \bibinfo {pages} {295} (\bibinfo {year}
  {2019})}\BibitemShut {NoStop}%
\bibitem [{\citenamefont {Pretko}\ \emph {et~al.}(2020)\citenamefont {Pretko},
  \citenamefont {Chen},\ and\ \citenamefont {You}}]{pretko2020fracton}%
  \BibitemOpen
  \bibfield  {author} {\bibinfo {author} {\bibfnamefont {M.}~\bibnamefont
  {Pretko}}, \bibinfo {author} {\bibfnamefont {X.}~\bibnamefont {Chen}},\ and\
  \bibinfo {author} {\bibfnamefont {Y.}~\bibnamefont {You}},\ }\bibfield
  {title} {\bibinfo {title} {Fracton phases of matter},\ }\href
  {https://doi.org/10.1142/S0217751X20300033} {\bibfield  {journal} {\bibinfo
  {journal} {International Journal of Modern Physics A}\ }\textbf {\bibinfo
  {volume} {35}},\ \bibinfo {pages} {2030003} (\bibinfo {year}
  {2020})}\BibitemShut {NoStop}%
\bibitem [{\citenamefont {Poulin}(2005)}]{poulin2005stabilizer}%
  \BibitemOpen
  \bibfield  {author} {\bibinfo {author} {\bibfnamefont {D.}~\bibnamefont
  {Poulin}},\ }\bibfield  {title} {\bibinfo {title} {Stabilizer formalism for
  operator quantum error correction},\ }\href
  {https://doi.org/10.1103/PhysRevLett.95.230504} {\bibfield  {journal}
  {\bibinfo  {journal} {Phys. Rev. Lett.}\ }\textbf {\bibinfo {volume} {95}},\
  \bibinfo {pages} {230504} (\bibinfo {year} {2005})}\BibitemShut {NoStop}%
\bibitem [{\citenamefont {Kribs}\ \emph {et~al.}(2006)\citenamefont {Kribs},
  \citenamefont {Laflamme}, \citenamefont {Poulin},\ and\ \citenamefont
  {Lesosky}}]{kribs2006oqec}%
  \BibitemOpen
  \bibfield  {author} {\bibinfo {author} {\bibfnamefont {D.~W.}\ \bibnamefont
  {Kribs}}, \bibinfo {author} {\bibfnamefont {R.}~\bibnamefont {Laflamme}},
  \bibinfo {author} {\bibfnamefont {D.}~\bibnamefont {Poulin}},\ and\ \bibinfo
  {author} {\bibfnamefont {M.}~\bibnamefont {Lesosky}},\ }\bibfield  {title}
  {\bibinfo {title} {Operator quantum error correction},\ }\href@noop {}
  {\bibfield  {journal} {\bibinfo  {journal} {Quantum Info. Comput.}\ }\textbf
  {\bibinfo {volume} {6}},\ \bibinfo {pages} {382–399} (\bibinfo {year}
  {2006})}\BibitemShut {NoStop}%
\bibitem [{\citenamefont {Bacon}(2006)}]{bacon2006oqec}%
  \BibitemOpen
  \bibfield  {author} {\bibinfo {author} {\bibfnamefont {D.}~\bibnamefont
  {Bacon}},\ }\bibfield  {title} {\bibinfo {title} {Operator quantum
  error-correcting subsystems for self-correcting quantum memories},\ }\href
  {https://doi.org/10.1103/PhysRevA.73.012340} {\bibfield  {journal} {\bibinfo
  {journal} {Phys. Rev. A}\ }\textbf {\bibinfo {volume} {73}},\ \bibinfo
  {pages} {012340} (\bibinfo {year} {2006})}\BibitemShut {NoStop}%
\bibitem [{\citenamefont {Xu}\ and\ \citenamefont
  {Moore}(2004)}]{xu2004strong}%
  \BibitemOpen
  \bibfield  {author} {\bibinfo {author} {\bibfnamefont {C.}~\bibnamefont
  {Xu}}\ and\ \bibinfo {author} {\bibfnamefont {J.~E.}\ \bibnamefont {Moore}},\
  }\bibfield  {title} {\bibinfo {title} {Strong-weak coupling self-duality in
  the two-dimensional quantum phase transition of $p+ip$ superconducting
  arrays},\ }\href {https://doi.org/10.1103/PhysRevLett.93.047003} {\bibfield
  {journal} {\bibinfo  {journal} {Phys. Rev. Lett.}\ }\textbf {\bibinfo
  {volume} {93}},\ \bibinfo {pages} {047003} (\bibinfo {year}
  {2004})}\BibitemShut {NoStop}%
\bibitem [{\citenamefont {You}\ \emph {et~al.}(2018)\citenamefont {You},
  \citenamefont {Devakul}, \citenamefont {Burnell},\ and\ \citenamefont
  {Sondhi}}]{you2018subsystem}%
  \BibitemOpen
  \bibfield  {author} {\bibinfo {author} {\bibfnamefont {Y.}~\bibnamefont
  {You}}, \bibinfo {author} {\bibfnamefont {T.}~\bibnamefont {Devakul}},
  \bibinfo {author} {\bibfnamefont {F.~J.}\ \bibnamefont {Burnell}},\ and\
  \bibinfo {author} {\bibfnamefont {S.~L.}\ \bibnamefont {Sondhi}},\ }\bibfield
   {title} {\bibinfo {title} {Subsystem symmetry protected topological order},\
  }\href {https://doi.org/10.1103/PhysRevB.98.035112} {\bibfield  {journal}
  {\bibinfo  {journal} {Phys. Rev. B}\ }\textbf {\bibinfo {volume} {98}},\
  \bibinfo {pages} {035112} (\bibinfo {year} {2018})}\BibitemShut {NoStop}%
\bibitem [{\citenamefont {Zhou}\ \emph {et~al.}(2021)\citenamefont {Zhou},
  \citenamefont {Zhang}, \citenamefont {Pollmann},\ and\ \citenamefont
  {You}}]{zhou2021fractal}%
  \BibitemOpen
  \bibfield  {author} {\bibinfo {author} {\bibfnamefont {Z.}~\bibnamefont
  {Zhou}}, \bibinfo {author} {\bibfnamefont {X.-F.}\ \bibnamefont {Zhang}},
  \bibinfo {author} {\bibfnamefont {F.}~\bibnamefont {Pollmann}},\ and\
  \bibinfo {author} {\bibfnamefont {Y.}~\bibnamefont {You}},\ }\href@noop {}
  {\bibinfo {title} {Fractal quantum phase transitions: Critical phenomena
  beyond renormalization}} (\bibinfo {year} {2021}),\ \Eprint
  {https://arxiv.org/abs/2105.05851} {arXiv:2105.05851 [cond-mat.str-el]}
  \BibitemShut {NoStop}%
\bibitem [{\citenamefont {Wildeboer}\ \emph {et~al.}(2022)\citenamefont
  {Wildeboer}, \citenamefont {Iadecola},\ and\ \citenamefont
  {Williamson}}]{wildeboer2021symmetryprotected}%
  \BibitemOpen
  \bibfield  {author} {\bibinfo {author} {\bibfnamefont {J.}~\bibnamefont
  {Wildeboer}}, \bibinfo {author} {\bibfnamefont {T.}~\bibnamefont
  {Iadecola}},\ and\ \bibinfo {author} {\bibfnamefont {D.~J.}\ \bibnamefont
  {Williamson}},\ }\bibfield  {title} {\bibinfo {title} {Symmetry-protected
  infinite-temperature quantum memory from subsystem codes},\ }\href
  {https://doi.org/10.1103/PRXQuantum.3.020330} {\bibfield  {journal} {\bibinfo
   {journal} {PRX Quantum}\ }\textbf {\bibinfo {volume} {3}},\ \bibinfo {pages}
  {020330} (\bibinfo {year} {2022})}\BibitemShut {NoStop}%
\bibitem [{\citenamefont {Zadnik}\ and\ \citenamefont
  {Fagotti}(2021)}]{zadnik2021folded}%
  \BibitemOpen
  \bibfield  {author} {\bibinfo {author} {\bibfnamefont {L.}~\bibnamefont
  {Zadnik}}\ and\ \bibinfo {author} {\bibfnamefont {M.}~\bibnamefont
  {Fagotti}},\ }\bibfield  {title} {\bibinfo {title} {{The Folded Spin-1/2 XXZ
  Model: I. Diagonalisation, Jamming, and Ground State Properties}},\ }\href
  {https://doi.org/10.21468/SciPostPhysCore.4.2.010} {\bibfield  {journal}
  {\bibinfo  {journal} {SciPost Phys. Core}\ }\textbf {\bibinfo {volume} {4}},\
  \bibinfo {pages} {10} (\bibinfo {year} {2021})}\BibitemShut {NoStop}%
\bibitem [{\citenamefont {Essler}\ \emph {et~al.}(2005)\citenamefont {Essler},
  \citenamefont {Frahm}, \citenamefont {G{\"o}hmann}, \citenamefont
  {Kl{\"u}mper},\ and\ \citenamefont {Korepin}}]{essler2005one}%
  \BibitemOpen
  \bibfield  {author} {\bibinfo {author} {\bibfnamefont {F.~H.}\ \bibnamefont
  {Essler}}, \bibinfo {author} {\bibfnamefont {H.}~\bibnamefont {Frahm}},
  \bibinfo {author} {\bibfnamefont {F.}~\bibnamefont {G{\"o}hmann}}, \bibinfo
  {author} {\bibfnamefont {A.}~\bibnamefont {Kl{\"u}mper}},\ and\ \bibinfo
  {author} {\bibfnamefont {V.~E.}\ \bibnamefont {Korepin}},\ }\href@noop {}
  {\emph {\bibinfo {title} {The one-dimensional {Hubbard} model}}}\ (\bibinfo
  {publisher} {Cambridge University Press},\ \bibinfo {year}
  {2005})\BibitemShut {NoStop}%
\bibitem [{\citenamefont {Georgi}(2000)}]{georgi2000lie}%
  \BibitemOpen
  \bibfield  {author} {\bibinfo {author} {\bibfnamefont {H.}~\bibnamefont
  {Georgi}},\ }\href@noop {} {\emph {\bibinfo {title} {Lie algebras in particle
  physics}}}\ (\bibinfo  {publisher} {Taylor \& Francis},\ \bibinfo {year}
  {2000})\BibitemShut {NoStop}%
\bibitem [{\citenamefont {Bulycheva}\ \emph {et~al.}(2018)\citenamefont
  {Bulycheva}, \citenamefont {Klebanov}, \citenamefont {Milekhin},\ and\
  \citenamefont {Tarnopolsky}}]{bulycheva2018spectra}%
  \BibitemOpen
  \bibfield  {author} {\bibinfo {author} {\bibfnamefont {K.}~\bibnamefont
  {Bulycheva}}, \bibinfo {author} {\bibfnamefont {I.~R.}\ \bibnamefont
  {Klebanov}}, \bibinfo {author} {\bibfnamefont {A.}~\bibnamefont {Milekhin}},\
  and\ \bibinfo {author} {\bibfnamefont {G.}~\bibnamefont {Tarnopolsky}},\
  }\bibfield  {title} {\bibinfo {title} {Spectra of operators in large $n$
  tensor models},\ }\href {https://doi.org/10.1103/PhysRevD.97.026016}
  {\bibfield  {journal} {\bibinfo  {journal} {Phys. Rev. D}\ }\textbf {\bibinfo
  {volume} {97}},\ \bibinfo {pages} {026016} (\bibinfo {year}
  {2018})}\BibitemShut {NoStop}%
\bibitem [{\citenamefont {Pakrouski}\ \emph {et~al.}(2019)\citenamefont
  {Pakrouski}, \citenamefont {Klebanov}, \citenamefont {Popov},\ and\
  \citenamefont {Tarnopolsky}}]{pakrouski2019spectrum}%
  \BibitemOpen
  \bibfield  {author} {\bibinfo {author} {\bibfnamefont {K.}~\bibnamefont
  {Pakrouski}}, \bibinfo {author} {\bibfnamefont {I.~R.}\ \bibnamefont
  {Klebanov}}, \bibinfo {author} {\bibfnamefont {F.}~\bibnamefont {Popov}},\
  and\ \bibinfo {author} {\bibfnamefont {G.}~\bibnamefont {Tarnopolsky}},\
  }\bibfield  {title} {\bibinfo {title} {Spectrum of majorana quantum mechanics
  with $\mathrm{O}(4{)}^{3}$ symmetry},\ }\href
  {https://doi.org/10.1103/PhysRevLett.122.011601} {\bibfield  {journal}
  {\bibinfo  {journal} {Phys. Rev. Lett.}\ }\textbf {\bibinfo {volume} {122}},\
  \bibinfo {pages} {011601} (\bibinfo {year} {2019})}\BibitemShut {NoStop}%
\bibitem [{\citenamefont {{Gurau}}\ and\ \citenamefont
  {{Ryan}}(2012)}]{gurau2012colored}%
  \BibitemOpen
  \bibfield  {author} {\bibinfo {author} {\bibfnamefont {R.}~\bibnamefont
  {{Gurau}}}\ and\ \bibinfo {author} {\bibfnamefont {J.~P.}\ \bibnamefont
  {{Ryan}}},\ }\bibfield  {title} {\bibinfo {title} {{Colored Tensor Models - a
  Review}},\ }\href {https://doi.org/10.3842/SIGMA.2012.020} {\bibfield
  {journal} {\bibinfo  {journal} {SIGMA}\ }\textbf {\bibinfo {volume} {8}},\
  \bibinfo {eid} {020} (\bibinfo {year} {2012})},\ \Eprint
  {https://arxiv.org/abs/1109.4812} {arXiv:1109.4812 [hep-th]} \BibitemShut
  {NoStop}%
\bibitem [{\citenamefont {{Klebanov}}\ \emph {et~al.}(2018)\citenamefont
  {{Klebanov}}, \citenamefont {{Popov}},\ and\ \citenamefont
  {{Tarnopolsky}}}]{klebanov2018TASI}%
  \BibitemOpen
  \bibfield  {author} {\bibinfo {author} {\bibfnamefont {I.~R.}\ \bibnamefont
  {{Klebanov}}}, \bibinfo {author} {\bibfnamefont {F.}~\bibnamefont
  {{Popov}}},\ and\ \bibinfo {author} {\bibfnamefont {G.}~\bibnamefont
  {{Tarnopolsky}}},\ }\bibfield  {title} {\bibinfo {title} {{TASI Lectures on
  Large $N$ Tensor Models}},\ }\bibfield  {journal} {\bibinfo  {journal} {arXiv
  e-prints}\ }\href {https://doi.org/10.48550/arXiv.1808.09434}
  {10.48550/arXiv.1808.09434} (\bibinfo {year} {2018}),\ \Eprint
  {https://arxiv.org/abs/1808.09434} {arXiv:1808.09434 [hep-th]} \BibitemShut
  {NoStop}%
\bibitem [{\citenamefont {Witten}(2019)}]{witten2019SYK}%
  \BibitemOpen
  \bibfield  {author} {\bibinfo {author} {\bibfnamefont {E.}~\bibnamefont
  {Witten}},\ }\bibfield  {title} {\bibinfo {title} {An syk-like model without
  disorder},\ }\href {https://doi.org/10.1088/1751-8121/ab3752} {\bibfield
  {journal} {\bibinfo  {journal} {Journal of Physics A: Mathematical and
  Theoretical}\ }\textbf {\bibinfo {volume} {52}},\ \bibinfo {pages} {474002}
  (\bibinfo {year} {2019})}\BibitemShut {NoStop}%
\bibitem [{\citenamefont {Rosenhaus}(2019)}]{rosenhaus2019SYK}%
  \BibitemOpen
  \bibfield  {author} {\bibinfo {author} {\bibfnamefont {V.}~\bibnamefont
  {Rosenhaus}},\ }\bibfield  {title} {\bibinfo {title} {An introduction to the
  syk model},\ }\href {https://doi.org/10.1088/1751-8121/ab2ce1} {\bibfield
  {journal} {\bibinfo  {journal} {Journal of Physics A: Mathematical and
  Theoretical}\ }\textbf {\bibinfo {volume} {52}},\ \bibinfo {pages} {323001}
  (\bibinfo {year} {2019})}\BibitemShut {NoStop}%
\bibitem [{\citenamefont {Mark}\ \emph {et~al.}(2020)\citenamefont {Mark},
  \citenamefont {Lin},\ and\ \citenamefont {Motrunich}}]{mark2020unified}%
  \BibitemOpen
  \bibfield  {author} {\bibinfo {author} {\bibfnamefont {D.~K.}\ \bibnamefont
  {Mark}}, \bibinfo {author} {\bibfnamefont {C.-J.}\ \bibnamefont {Lin}},\ and\
  \bibinfo {author} {\bibfnamefont {O.~I.}\ \bibnamefont {Motrunich}},\
  }\bibfield  {title} {\bibinfo {title} {Unified structure for exact towers of
  scar states in the {A}ffleck-{K}ennedy-{L}ieb-{T}asaki and other models},\
  }\href {https://doi.org/10.1103/PhysRevB.101.195131} {\bibfield  {journal}
  {\bibinfo  {journal} {Phys. Rev. B}\ }\textbf {\bibinfo {volume} {101}},\
  \bibinfo {pages} {195131} (\bibinfo {year} {2020})}\BibitemShut {NoStop}%
\bibitem [{\citenamefont {Yang}(1991)}]{yang1992remarks}%
  \BibitemOpen
  \bibfield  {author} {\bibinfo {author} {\bibfnamefont {C.~N.}\ \bibnamefont
  {Yang}},\ }\bibfield  {title} {\bibinfo {title} {Remarks and generalizations
  about su2×su2 symmetry of {Hubbard} models},\ }\href
  {https://doi.org/https://doi.org/10.1016/0375-9601(91)90020-9} {\bibfield
  {journal} {\bibinfo  {journal} {Physics Letters A}\ }\textbf {\bibinfo
  {volume} {161}},\ \bibinfo {pages} {292 } (\bibinfo {year}
  {1991})}\BibitemShut {NoStop}%
\bibitem [{\citenamefont {Kraus}\ \emph {et~al.}(2007)\citenamefont {Kraus},
  \citenamefont {Wolf},\ and\ \citenamefont {Cirac}}]{kraus2007quantum}%
  \BibitemOpen
  \bibfield  {author} {\bibinfo {author} {\bibfnamefont {C.~V.}\ \bibnamefont
  {Kraus}}, \bibinfo {author} {\bibfnamefont {M.~M.}\ \bibnamefont {Wolf}},\
  and\ \bibinfo {author} {\bibfnamefont {J.~I.}\ \bibnamefont {Cirac}},\
  }\bibfield  {title} {\bibinfo {title} {Quantum simulations under
  translational symmetry},\ }\href {https://doi.org/10.1103/PhysRevA.75.022303}
  {\bibfield  {journal} {\bibinfo  {journal} {Phys. Rev. A}\ }\textbf {\bibinfo
  {volume} {75}},\ \bibinfo {pages} {022303} (\bibinfo {year}
  {2007})}\BibitemShut {NoStop}%
\bibitem [{\citenamefont {Zeier}\ and\ \citenamefont
  {Schulte-Herbrüggen}(2011)}]{zeier2011symmetry}%
  \BibitemOpen
  \bibfield  {author} {\bibinfo {author} {\bibfnamefont {R.}~\bibnamefont
  {Zeier}}\ and\ \bibinfo {author} {\bibfnamefont {T.}~\bibnamefont
  {Schulte-Herbrüggen}},\ }\bibfield  {title} {\bibinfo {title} {Symmetry
  principles in quantum systems theory},\ }\href
  {https://doi.org/10.1063/1.3657939} {\bibfield  {journal} {\bibinfo
  {journal} {Journal of Mathematical Physics}\ }\textbf {\bibinfo {volume}
  {52}},\ \bibinfo {pages} {113510} (\bibinfo {year} {2011})}\BibitemShut
  {NoStop}%
\bibitem [{\citenamefont {Zimbor\'as}\ \emph {et~al.}(2015)\citenamefont
  {Zimbor\'as}, \citenamefont {Zeier}, \citenamefont {Schulte-Herbr\"uggen},\
  and\ \citenamefont {Burgarth}}]{zimboras2015symmetry}%
  \BibitemOpen
  \bibfield  {author} {\bibinfo {author} {\bibfnamefont {Z.}~\bibnamefont
  {Zimbor\'as}}, \bibinfo {author} {\bibfnamefont {R.}~\bibnamefont {Zeier}},
  \bibinfo {author} {\bibfnamefont {T.}~\bibnamefont {Schulte-Herbr\"uggen}},\
  and\ \bibinfo {author} {\bibfnamefont {D.}~\bibnamefont {Burgarth}},\
  }\bibfield  {title} {\bibinfo {title} {Symmetry criteria for quantum
  simulability of effective interactions},\ }\href
  {https://doi.org/10.1103/PhysRevA.92.042309} {\bibfield  {journal} {\bibinfo
  {journal} {Phys. Rev. A}\ }\textbf {\bibinfo {volume} {92}},\ \bibinfo
  {pages} {042309} (\bibinfo {year} {2015})}\BibitemShut {NoStop}%
\bibitem [{\citenamefont {Moudgalya}\ and\ \citenamefont
  {Motrunich}(tion)}]{szminprep}%
  \BibitemOpen
  \bibfield  {author} {\bibinfo {author} {\bibfnamefont {S.}~\bibnamefont
  {Moudgalya}}\ and\ \bibinfo {author} {\bibfnamefont {O.~I.}\ \bibnamefont
  {Motrunich}},\ }\href@noop {} {\  (\bibinfo {year} {in
  preparation})}\BibitemShut {NoStop}%
\bibitem [{\citenamefont {Alicea}\ and\ \citenamefont
  {Fendley}(2016)}]{alicea2016topological}%
  \BibitemOpen
  \bibfield  {author} {\bibinfo {author} {\bibfnamefont {J.}~\bibnamefont
  {Alicea}}\ and\ \bibinfo {author} {\bibfnamefont {P.}~\bibnamefont
  {Fendley}},\ }\bibfield  {title} {\bibinfo {title} {Topological phases with
  parafermions: Theory and blueprints},\ }\href
  {https://doi.org/10.1146/annurev-conmatphys-031115-011336} {\bibfield
  {journal} {\bibinfo  {journal} {Annual Review of Condensed Matter Physics}\
  }\textbf {\bibinfo {volume} {7}},\ \bibinfo {pages} {119} (\bibinfo {year}
  {2016})}\BibitemShut {NoStop}%
\bibitem [{\citenamefont {Fendley}(2016)}]{fendley2016strong}%
  \BibitemOpen
  \bibfield  {author} {\bibinfo {author} {\bibfnamefont {P.}~\bibnamefont
  {Fendley}},\ }\bibfield  {title} {\bibinfo {title} {Strong zero modes and
  eigenstate phase transitions in the {XYZ}/interacting majorana chain},\
  }\href {https://doi.org/10.1088/1751-8113/49/30/30lt01} {\bibfield  {journal}
  {\bibinfo  {journal} {Journal of Physics A: Mathematical and Theoretical}\
  }\textbf {\bibinfo {volume} {49}},\ \bibinfo {pages} {30LT01} (\bibinfo
  {year} {2016})}\BibitemShut {NoStop}%
\bibitem [{\citenamefont {{Moudgalya}}\ and\ \citenamefont
  {{Motrunich}}(2023)}]{moudgalya2022numerical}%
  \BibitemOpen
  \bibfield  {author} {\bibinfo {author} {\bibfnamefont {S.}~\bibnamefont
  {{Moudgalya}}}\ and\ \bibinfo {author} {\bibfnamefont {O.~I.}\ \bibnamefont
  {{Motrunich}}},\ }\bibfield  {title} {\bibinfo {title} {{Numerical Methods
  for Detecting Symmetries and Commutant Algebras}},\ }\href@noop {} {\bibfield
   {journal} {\bibinfo  {journal} {arXiv e-prints}\ } (\bibinfo {year}
  {2023})},\ \Eprint {https://arxiv.org/abs/2302.03028} {arXiv:2302.03028
  [cond-mat.str-el]} \BibitemShut {NoStop}%
\bibitem [{\citenamefont {Lidar}\ and\ \citenamefont
  {Birgitta~Whaley}(2003)}]{lidar2003decoherencereview}%
  \BibitemOpen
  \bibfield  {author} {\bibinfo {author} {\bibfnamefont {D.~A.}\ \bibnamefont
  {Lidar}}\ and\ \bibinfo {author} {\bibfnamefont {K.}~\bibnamefont
  {Birgitta~Whaley}},\ }\bibinfo {title} {Decoherence-free subspaces and
  subsystems},\ in\ \href {https://doi.org/10.1007/3-540-44874-8_5} {\emph
  {\bibinfo {booktitle} {Irreversible Quantum Dynamics}}},\ \bibinfo {editor}
  {edited by\ \bibinfo {editor} {\bibfnamefont {F.}~\bibnamefont {Benatti}}\
  and\ \bibinfo {editor} {\bibfnamefont {R.}~\bibnamefont {Floreanini}}}\
  (\bibinfo  {publisher} {Springer Berlin Heidelberg},\ \bibinfo {address}
  {Berlin, Heidelberg},\ \bibinfo {year} {2003})\ pp.\ \bibinfo {pages}
  {83--120}\BibitemShut {NoStop}%
\bibitem [{\citenamefont {Holbrook}\ \emph {et~al.}(2003)\citenamefont
  {Holbrook}, \citenamefont {Kribs},\ and\ \citenamefont
  {Laflamme}}]{holbrook2003noiseless}%
  \BibitemOpen
  \bibfield  {author} {\bibinfo {author} {\bibfnamefont {J.~A.}\ \bibnamefont
  {Holbrook}}, \bibinfo {author} {\bibfnamefont {D.~W.}\ \bibnamefont
  {Kribs}},\ and\ \bibinfo {author} {\bibfnamefont {R.}~\bibnamefont
  {Laflamme}},\ }\bibfield  {title} {\bibinfo {title} {Noiseless subsystems and
  the structure of the commutant in quantum error correction},\ }\href
  {https://doi.org/10.1023/B:QINP.0000022737.53723.b4} {\bibfield  {journal}
  {\bibinfo  {journal} {Quantum Information Processing}\ }\textbf {\bibinfo
  {volume} {2}},\ \bibinfo {pages} {381} (\bibinfo {year} {2003})}\BibitemShut
  {NoStop}%
\bibitem [{\citenamefont {Haber}(2021)}]{haber2021useful}%
  \BibitemOpen
  \bibfield  {author} {\bibinfo {author} {\bibfnamefont {H.~E.}\ \bibnamefont
  {Haber}},\ }\bibfield  {title} {\bibinfo {title} {{Useful relations among the
  generators in the defining and adjoint representations of SU(N)}},\ }\href
  {https://doi.org/10.21468/SciPostPhysLectNotes.21} {\bibfield  {journal}
  {\bibinfo  {journal} {SciPost Phys. Lect. Notes}\ ,\ \bibinfo {pages} {21}}
  (\bibinfo {year} {2021})}\BibitemShut {NoStop}%
\bibitem [{\citenamefont {Yang}\ and\ \citenamefont {Zhang}()}]{yang1990so}%
  \BibitemOpen
  \bibfield  {author} {\bibinfo {author} {\bibfnamefont {C.~N.}\ \bibnamefont
  {Yang}}\ and\ \bibinfo {author} {\bibfnamefont {S.~C.}\ \bibnamefont
  {Zhang}},\ }\bibinfo {title} {So(4) symmetry in a hubbard model},\ in\ \href
  {https://doi.org/10.1142/9789814449021_0020} {\emph {\bibinfo {booktitle}
  {Selected Papers of Chen Ning Yang II}}},\ pp.\ \bibinfo {pages}
  {154--161}\BibitemShut {NoStop}%
\bibitem [{\citenamefont {Fujii}\ \emph {et~al.}(2007)\citenamefont {Fujii},
  \citenamefont {Oike},\ and\ \citenamefont {Suzuki}}]{fujii2007more}%
  \BibitemOpen
  \bibfield  {author} {\bibinfo {author} {\bibfnamefont {K.}~\bibnamefont
  {Fujii}}, \bibinfo {author} {\bibfnamefont {H.}~\bibnamefont {Oike}},\ and\
  \bibinfo {author} {\bibfnamefont {T.}~\bibnamefont {Suzuki}},\ }\bibfield
  {title} {\bibinfo {title} {More on the isomorphism $su(2) \otimes su(2) \cong
  so(4)$},\ }\href {https://doi.org/10.1142/S0219887807002120} {\bibfield
  {journal} {\bibinfo  {journal} {International Journal of Geometric Methods in
  Modern Physics}\ }\textbf {\bibinfo {volume} {04}},\ \bibinfo {pages} {471}
  (\bibinfo {year} {2007})}\BibitemShut {NoStop}%
\end{thebibliography}%

\appendix 
\onecolumngrid
\section{The Double Commutant Theorem (DCT)}\label{app:doublecommutant}
In this appendix, we sketch a proof of the DCT of Thm.~\ref{thm:dct}, focusing on finite-dimensional Hilbert spaces, closely following the discussion in \cite{landsman1998lecture}.
\dct*
\begin{proof}
Let us denote the centralizer of $\mC$ as $\mD$.
Since we know that $\mA \subseteq \mD$, to show that $\mD = \mA$, it is sufficient to show $\mD \subseteq \mA$.
Consider an orthonormal basis $\{\ket{v_1}, \ket{v_2}, \cdots, \ket{v_D}\}$ for the Hilbert space $\mH$. 
We then need to show that for any operator $T \in \mD$, there exists an operator $S \in \mA$ such that $T\ket{v_i} = S\ket{v_i}\;\;\forall i,\;\; 1 \leq i \leq D$. 
For some particular $\ket{v_i}$, we can consider the subspace $\mM_i \subseteq \mH$ defined as $\mM_i \defn \{A \ket{v_i} : A \in \mA\}$, and the orthogonal projector $\Pi_i$ onto $\mM_i$. 
It straightforwardly follows that the subspace $\mM_i$ is invariant under the action of any operator in $\mA$, and we obtain $A \Pi_i \mH \subseteq \Pi_i \mH$ if $A \in \mA$.
Hence we obtain   
\begin{equation}
    \Pi_i^\perp A \Pi_i = 0 \;\;\implies \;\; A \Pi_i = \Pi_i A \Pi_i,
\label{eq:perpcondition}
\end{equation}
where $\Pi_i^\perp \defn \mathds{1} - \Pi_i$ is the projector onto the orthogonal subspace.
Further, since $\mA$ is a $\dagger$-algebra, we can w.l.o.g. choose $A$ in Eq.~(\ref{eq:perpcondition}) to be a Hermitian (i.e., $A = A^\dagger$) basis element of $\mA$, and take the Hermitian conjugate of both sides of Eq.~(\ref{eq:perpcondition}) to obtain
\begin{equation}
    \Pi_i A = \Pi_i A \Pi_i\;\;\implies\;\;[\Pi_i, A] = 0\;\;\forall\;A \in \mA\;\;\implies \Pi_i \in \mC. 
\label{eq:projcommute}
\end{equation}
Since any $T \in \mD$ satisfies $[T, \Pi_i] = 0$ by definition, we obtain
\begin{equation}
 T\Pi_i \ket{v_i} = \Pi_i (T\ket{v_i}) \in \mM_i.   
\label{eq:TPicondition}
\end{equation}
Since $\mA$ is a unital algebra, $\mathds{1} \ket{v_i} = \ket{v_i} \in \mM_i$, hence using the definition of $\Pi_i$, we have $\Pi_i \ket{v_i} = \ket{v_i}$.
Then, using Eq.~(\ref{eq:TPicondition}) we obtain $T\ket{v_i} \in \mM_i$. 
Finally, using the definition of $\mM_i$, we have $T\ket{v_i} = S\ket{v_i}$ for some $S \in \mA$. 
While this completes the proof for a particular $\ket{v_i}$, we also need to show that there exists such an operator $S \in \mA$ that is $i$-independent.
Hence we work with a different space $\mH^{(D)} \defn \bigoplus_{i=1}^D \mH$ that consists of $D$ copies of the original Hilbert space.
We then define the vector $\ket{v} \defn \bigoplus_{i=1}^D{\ket{v_i}}$, such that $\ket{v_i}$ is in the $i$-th copy of $\mH$.
The vector $\ket{v}$ can be considered as a specific $1 \times D$ (block-)column vector whose entries are $\ket{v_1}, \ket{v_2}, \dots, \ket{v_D}$.
Operators on this space, denoted by $\mL(\mH^{(D)})$ can be considered to be $D \times D$ (block-)matrices with entries from the space of operators $\mL(\mH)$. 
We can further define the algebra $\mA^{(D)} \defn \{\bigoplus_{i=1}^D A |\ A \in \mA\} \subseteq \mL(\mH^{(D)})$; this consists of diagonal (block-)matrices with the same $A \in \mA$ on the diagonal.
$\mA^{(D)}$ inherits several properties of $\mA$, and in particular it is a unital $\dagger$-algebra.
Further, it is easy to show that any element of the centralizer of $\mA^{(D)}$ in $\mL(\mH^{(D)})$ is a $D \times D$ matrix with elements in the commutant algebra $\mC$.
As a consequence, any operator of the form $T^{(D)} \defn \bigoplus_{i=1}^D{T}$, where $T \in \mD$, belongs to the centralizer of the centralizer (or, the double-centralizer) of $\mA^{(D)}$, which we denote by $\mD^{(D)}$.
We can then follow the logic in the previous paragraph, with the appropriate substitutions $\ket{v_i} \rightarrow \ket{v}$, $\mA \rightarrow \mA^{(D)}$, $\mH \rightarrow \mH^{(D)}$, and deduce that for any $T^{(D)} \in \mD^{(D)}$, there exists an operator $S^{(D)} \in \mA^{(D)}$ such that $T^{(D)}\ket{v} = S^{(D)}\ket{v}$.
Since $S^{(D)} = \bigoplus_{i =1}^D{S}$ for some $S \in \mA$, this shows that for any $T \in \mD$, we have $T\ket{v_1} \oplus \cdots \oplus T\ket{v_D} = S\ket{v_1} \oplus \cdots \oplus S\ket{v_D}$, showing that $T = S \in \mA$.
This concludes the proof that $\mD = \mA$. 
\end{proof}
\section{Locality and the DCT} \label{app:localityDCT}
In this appendix, we discuss some aspects of the restrictions on local terms with a given set of conserved quantities due to the DCT of Thm.~\ref{thm:dct}, and also prove Lems.~\ref{lem:strloc} and \ref{lem:typeI}.
Specifically, suppose we want to find all local Hamiltonians that commute with some desired set of conserved quantities.
We will assume that these conserved quantities generate a commutant algebra that is the centralizer of a local algebra, where the latter is generated from some strictly local bond terms and perhaps also few extensive local operators as in the dynamical $SU(2)$ case of Sec.~\ref{subsec:dynamicalSU2}.
Using the DCT, we know that any such symmetric Hamiltonian can be produced by the generators of the local algebra.
However, the DCT does not tell us how many generators are needed and does not exclude possibility that this involves products of an extensive number of generators spread over the entire system.
Some variants of concern are that perhaps some linear combination of products of extensive numbers of bond generator terms can turn out to be a local or an extensive local operator due to some cancellations, or that something like this happens for combinations involving products of an extensive local generator (or its powers) and an extensive number of bond generator terms.
(Of course, an extensive local generator multiplied by a finite number of bond generators is a non-local operator and would not be allowed as a Hamiltonian, but the DCT does not provide any restrictions on the number of factors and summands.)
As we will see, in some cases we can impose restrictions on the range of the local terms we are interested in and still use the DCT to restrict their structure.
\subsection{Arbitrary on-site symmetries}\label{subsec:DCTonsite}
We start with the discussion of on-site symmetries such as $Z_2$, $U(1)$, or regular $SU(2)$, shown in cases of \#1 - \#5 of Tab.~\ref{tab:conventionalexamples}.
As we discuss in Secs.~\ref{subsec:spinlessfermions} and \ref{subsec:spinfulfermions}, these results are also applicable to cases \#1 and \#3a-b of Tab.~\ref{tab:spinlessbondalgebra} and also to cases \#1a-c, \#3a-b, and \#4 of Tab.~\ref{tab:spinfulbondalgebra}.
In the case of on-site symmetries, the commutant algebra $\mC$ of a bond algebra $\mA$ can be considered to be generated by the (full family of) on-site unitaries themselves, i.e., we have
\begin{equation}
    \hU(g) = \prod_j{\hu_j(g)},\;\;\;\;
    \mC = \lgen \{\hU(g)\} \rgen.
\label{eq:onsiteunitary}
\end{equation}
where $\hu_j(g)$ is a unitary operator on site $j$ with a label $g$, usually chosen from a discrete or Lie group. 
For example, in the case of $Z_2$ symmetry, we have non-trivial $\hu_j(g) = \sigma^z_j$.
In the case of a continuous on-site symmetry, there is a continuous family of unitary operators $\{\hU(g)\}$. 
For example, in spin-1/2 systems, $U(1)$ symmetry is represented by the unitary $\hU$ of Eq.~(\ref{eq:onsiteunitary}) with $\hu_j(\theta) = \exp(i \theta \sigma^z_j)$ for arbitrary $\theta$; whereas $\hu_j(\vec{\alpha}) = \exp(i \vec{\alpha}\cdot\vec{\sigma}_j)$ in the case of $SU(2)$ symmetry for arbitrary values of $\vec{\alpha}$, where $\vec{\sigma}_j$ is the vector of Pauli matrices $(\sigma^x_j, \sigma^y_j, \sigma^z_j)$.
Since the unitary of Eq.~(\ref{eq:onsiteunitary}) is on-site, we can also define a natural restriction of the unitary and commutant to a region $R$ as
\begin{equation}
    \hU_R(g) \defn \prod_{j \in R}{\hu_j(g)},\;\;\;
    \mC_R = \lgen \{\hU_R(g)\} \rgen, 
\label{eq:restrictedunitary}
\end{equation}
where a region is simply defined to be a collection of lattice sites. 
We now move onto the proofs of the important lemmas, restated here for easy reference.
\strloc*
\begin{proof}
Consider any strictly local operator $h_R = \hO_{\text{loc-}R}$ in a contiguous region $R$ that is symmetric under an on-site symmetry.
By definition it commutes with the algebra generated by the (family of) unitary operators $\hU(g)$ of the form of Eq.~(\ref{eq:onsiteunitary}).
Hence it is clear that any such operator should commute with the unitary restricted to the region $R$, i.e.,
\begin{equation}
    [\hO_{\text{loc-}R}, \hU(g)] = 0 \;\;\implies [\hO_{\text{loc-}R}, \hU_R(g)] = 0,\;\;\forall g,
\label{eq:hRcondition}
\end{equation}
where $\hU_R(g)$ is the unitary restricted to the region $R$, defined in Eq.~(\ref{eq:restrictedunitary}).
Hence, applying the DCT to the region $R$, $\hO_{\text{loc-}R}$ is part of the centralizer of $\mC_R$ w.r.t.\ the many-body Hilbert space on the region $R$, which is the algebra of all operators with support in the region $R$ that commute with the family of on-site unitaries $\hU_R(g)$, which we refer to as $\mA_R$.
However, depending on the shape of the region $R$, this might not be the same as the algebra $\mA_{\text{gen} \in R}$, which is generated by the generators of the original bond algebra $\mA$ that have support completely in the region $R$. 
For a general proof, we need to introduce a ``regular generation" assumption on the bond algebras $\mA$ corresponding to the commutant $\mC$ generated by the on-site unitary symmetry.
In particular we assume that the bond and commutant algebras are centralizers of each other for sufficiently regular finite lattices of varying sizes, which is equivalent to saying that for any such regular region $R$, the commutant of the bond algebra $\mA_{\text{gen} \in R}$ is precisely $\mC_R$.
This usually holds for bond algebras generated by some homogeneous set of generators of bounded range on the lattice and is checked as part of specific commutant exhaustion proofs such as those in App.~\ref{app:commutantexhaustion}. 
In particular for many examples of bond algebras generated by on-site and nearest-neighbor terms, it is easy to check in each case that the algebra $\mA_R$ is generated by the subset of generators of $\mA$ that lie completely within the region $R$, and also that $\mA_{\text{gen} \in R} = \mA_R$ in the notation used above.
For example, in the $Z_2$ case we can directly check that the bond algebra generated by $\{S^x_{j_k} S^x_{j_k+1}, k=1,2,\dots,m-1\}$ and $\{S^z_{j_k},k=1,2,\dots, m\}$ has commutant precisely generated by $\prod_{k=1}^m Z_{j_k}$, where the sites $\{j_k\}$ can be any contiguous region of $m$ sites arranged in any geometry in any number of dimensions; for such regions $\mA_{\text{gen} \in R} = \mA_R$ in the notation used above.
Another example is the $SU(2)$ case, where from a proof similar to App.~\ref{subsec:regularsu2proof} we know that any $SU(2)$-symmetric operator on sites $j_1, j_2, \dots, j_m$ can be generated by $\{\vec{S}_{j_k} \cdot \vec{S}_{j_{k+1}}, k = 1, 2, \dots, m-1 \}$, where again $\{j_k\}$ can label any contiguous region of $m$ sites.
With this ``regular generation" assumption, it is clear that even if the region $R$ is itself not regular, i.e., if $\mA_R \neq \mA_{\text{gen} \in R}$, one can always embed it in a bounded regular region $R' \supseteq R$ such that the algebra $\mA_{\text{gen} \in R'} = \mA_R' \supseteq \mA_R$.
Hence any operator $\hO_{\text{loc-}R}$ that satisfies Eq.~(\ref{eq:hRcondition}) can be generated by the generators of the $\mA$ restricted to a bounded region $R'$.
\end{proof}

\typeI*
\begin{proof}
Moving on to extensive-local operators, consider a Hamiltonian $H$ of the form
\begin{equation}
    H = \hO_\text{ext-loc} = \sum_R{\hA_R},
\label{eq:extlocreg}
\end{equation}
where $\hA_R$ is any operator with support \textit{everywhere}\footnote{A precise definition using expansions in terms of Pauli strings is given below.} in a local region $R$ (which can be any group of sites with separation bounded by some fixed number $r_{\max}$), and the sum over $R$ runs over distinct regions distributed all over the entire lattice.
We typically have in mind cases where $\hO_\text{ext-loc}$ is translation invariant, i.e., where $\hA_{R_1}$ and $\hA_{R_2}$ have the same ``form" for regions $R_1$ and $R_2$ that are related by a simple translation.
However, the following argument is completely general and holds for all operators $\hO_\text{ext-loc}$ that can be decomposed as Eq.~(\ref{eq:extlocreg}) where the sum runs over some set of distinct regions (not necessarily extensively many of them) with no relations among the $\hA_R$'s.
Symmetric operators of the form of Eq.~(\ref{eq:extlocreg}) should satisfy
\begin{equation}
    [\sum_R{\hA_R}, \hU] = 0,\;\;\implies\;\;\sum_R{\hU_{\bar{R}}  [\hA_R, \hU_R]} = 0, 
\label{eq:extloccomm}
\end{equation}
where $\bar{R}$ is the complement of the region $R$. 
We now show that terms that appear in Eq.~(\ref{eq:extloccomm}) corresponding to distinct regions $R_1$ and $R_2$ are orthogonal to each other. 
In particular, we wish to show that the Frobenius inner product between the terms in Eq.~(\ref{eq:extloccomm}) vanishes, i.e.,
\begin{equation}
    \Tr([\hA_{R_1}, \hU_{R_1}]^\dagger \hU_{\bar{R}_1}^\dagger \hU_{\bar{R}_2} [\hA_{R_2}, \hU_{R_2}]) = 0.
\label{eq:innerprod}
\end{equation}
Without loss of generality, we assume $\hA_R$ is traceless and we can expand $\hA_R$ in terms of ``Pauli strings" (or its generalizations) $\{\hP^\mu_R\}$ with support precisely in the region $R$ (i.e., with non-trivial Pauli matrices everywhere in $R$ and identity matrices everywhere in the region $\bar{R}$) as $\hA_R = \sum_{\mu}{c_\mu \hP^\mu_R}$.
The terms that appear in the expansion of the LHS of Eq.~(\ref{eq:innerprod}) after using this Pauli string expansion are of one of the following forms:
\begin{equation}
    \Tr(\hP^\mu_{R_1} \hU^\dagger_{R_1} \hU^\dagger_{\bar{R}_1} \hU_{\bar{R}_2} \hP^\nu_{R_2} \hU_{R_2}),\;\;\Tr(\hU^\dagger_{R_1} \hP^\mu_{R_1} \hU^\dagger_{\bar{R}_1} \hU_{\bar{R}_2} \hP^\nu_{R_2} \hU_{R_2}),\;\;\Tr(\hP^\mu_{R_1} \hU^\dagger_{R_1} \hU^\dagger_{\bar{R}_1} \hU_{\bar{R}_2} \hU_{R_2} \hP^\nu_{R_2}),\;\;\Tr(\hU^\dagger_{R_1} \hP^\mu_{R_1} \hU^\dagger_{\bar{R}_1}  \hU_{\bar{R}_2} \hU_{R_2} \hP^\nu_{R_2}).
\label{eq:paulipossibleterms}
\end{equation}
Since all the operators that appear within the parentheses in these terms are on-site operators, i.e., they can be expressed as $\prodal{j}{}{\ho_j}$ for single-site operators $\{\ho_j\}$, the trace decomposes as follows: $\text{Tr}(\prodal{j}{}{\ho_j}) = \prodal{j}{}{\text{Tr}(\ho_j)}$.
Further, using the expressions in Eq.~(\ref{eq:paulipossibleterms}), it is easy to show that as long as $R_1 \neq R_2$, for each of those terms there is at least one site $j$ (can be any site that belongs to only one of the two regions) for which one of the following holds:
\begin{equation}
\ho_j = \hu^\dagger_j \sigma^{\mu_j}_j \hu_j,\;\;  \ho_j = \hu^\dagger_j \hu_j \sigma^{\mu_j}_j,\;\;\ho_j = \sigma^{\mu_j}_j \hu^\dagger_j \hu_j,
\label{eq:ojpossibilities}
\end{equation}
where $\sigma^{\mu_j}_j$ is some Pauli matrix that is not the identity.
In all the cases in Eq.~(\ref{eq:ojpossibilities}), it is easy to see that $\text{Tr}(\ho_j) = 0$.
Hence all the terms in Eq.~(\ref{eq:paulipossibleterms}) vanish when $R_1 \neq R_2$, proving Eq.~(\ref{eq:innerprod}).
Since the sum in Eq.~(\ref{eq:extloccomm}) runs over distinct regions $R$, using Eq.~(\ref{eq:innerprod}) we obtain that $[\hA_R, \hU_R] = 0$ for any extensive local symmetric operator of the form of Eq.~(\ref{eq:extlocreg}) (and the specified precise localization conditions on $A_R$, which can be always achieved for any extensive-local operator). 
This completes the proof of the Lemma that any extensive local operator or Hamiltonian $H$ that is symmetric under an on-site symmetry can be expressed as a sum of strictly local terms $\{h_{R}\}$ that are also symmetric, i.e., $H = \sum_R{h_R}$.
Combined with Lem.~\ref{lem:strloc} under the regular generation assumption (which was not needed until this point), we can conclude that for any $R$ the corresponding $\hA_R$ can be generated by bond generators contained completely within a sufficiently regular region $R'$ covering $R$ as described earlier.
In the case of several examples of bond algebras $\mA$ generated by homogeneous nearest-neighbor and on-site terms, this can also be generated by bond generators contained strictly within the region $R$, as also discussed earlier.
\end{proof}

Note that Lems.~\ref{lem:typeI} and \ref{lem:strloc} are very general and apply to \textit{all} on-site symmetries, including $U(1)$, $SU(2)$, and $Z_2$.
Further, it also generalizes straightforwardly to systems with multiple on-site symmetries.
\subsection{Dynamical \texorpdfstring{$SU(2)$}{} symmetry}\label{subsec:DCTdynamical}
The situation is a bit more complicated but also more interesting in the dynamical $SU(2)$ case where the commutant is generated by $\vec{S}_\text{tot}^2$ and $S^z_\text{tot}$.
In contrast to other examples discussed in Tab.~\ref{tab:conventionalexamples}, the dynamical $SU(2)$ is not an on-site symmetry, hence the arguments of the previous section do not apply.
Nevertheless, we will show that any extensive local Hamiltonian that commutes with $\vec{S}_\text{tot}^2$ must be a sum of an SU(2)-invariant extensive local operator and a Zeeman term proportional to $S^z_{\tot}$.
We again start by considering a strictly local operator $\hO_{\text{loc-}R}$ that commutes with $\vec{S}_{\tot}^2$.
Here the localization region $R$ again means a collection of sites that are separated by at most $r_{\max}$ and are not necessarily contiguous, i.e., a local set of points; this is the meaning for $R$'s everywhere below.
We can write
\begin{equation}
\vec{S}_\text{tot}^2 = (\vec{S}_{R,\text{tot}} + \vec{S}_{\bar{R},\text{tot}})^2 = \vec{S}_{R,\text{tot}}^2 + \vec{S}_{\bar{R},\text{tot}}^2 + 2 \vec{S}_{R,\text{tot}} \cdot \vec{S}_{\bar{R},\text{tot}} ~,
\end{equation}
where $\bar{R}$ is a complement to the region $R$ in the whole system and $\vec{S}_{\bar{R},\text{tot}} \equiv \sum_{j \in \bar{R}} \vec{S}_j = \sum_{j \not\in R} \vec{S}_j$.
Since $\hO_{\text{loc-}R}$ commutes with $\vec{S}_{\bar{R}, \tot}$, we can write
\begin{equation}
[\hO_{\text{loc-}R}, \vec{S}_\tot^2] = [\hO_{\text{loc-}R}, \vec{S}_{R,\tot}^2] + 2 [\hO_{\text{loc-}R}, \vec{S}_{R,\tot}] \cdot \vec{S}_{\bar{R},\tot}. \label{eq:comm_OlocR_Stot2}
\end{equation}
For this commutator to be zero (assuming non-empty regions $R$ and $\bar{R}$) we clearly must have $[\hO_{\text{loc-}R}, \vec{S}_{R,\tot}] = 0$.
This also implies that $[\hO_{\text{loc-}R}, \vec{S}_{\tot}] = 0$; hence $\hO_{\text{loc-}R}$ must have the regular $SU(2)$ symmetry, thus reducing to the case considered in the previous subsection.
Hence all strictly local operators with the dynamical $SU(2)$ symmetry that are localized within a region $R$ can be generated using regular $SU(2)$-symmetric bond generators $\{\vec{S}_j \cdot \vec{S}_{j+1}\}$ restricted to a contiguous region containing $R$.
Note that this also shows that it is not possible to write down a bond algebra (i.e., an algebra generated by strictly local operators) that has the full commutant as $\mC_{\dyn SU(2)} = \lgen \vec{S}^2_{\tot}, S^z_{\tot}\rgen$ -- any such bond algebra will have the larger algebra $\mC_{SU(2)} = \lgen S^x_{\tot}, S^y_{\tot}, S^z_{\tot} \rgen$ as its commutant.
We now use this result to prove Lem.~\ref{lem:dynsym}.

\dynsym*
\begin{proof}
Consider now an extensive local operator as in Eq.~(\ref{eq:extlocreg}).
Using Eq.~(\ref{eq:comm_OlocR_Stot2}) we see that $[\hO_\text{ext-loc}, \vec{S}_\text{tot}^2]$ can be expressed as
\begin{equation}
    [\hO_\text{ext-loc}, \vec{S}_\text{tot}^2] = \sumal{R}{}{[\hA_R, \vec{S}^2_{R, \tot}]} + 2\sumal{R}{}{[\hA_R, \vec{S}_{R, \tot}]\cdot\vec{S}_{\bar{R},\tot}} = \sumal{R}{}{[\hA_R, \vec{S}^2_{R, \tot}]} + 2\sumal{R}{}{\sumal{\alpha}{}{\sumal{\ell \in \bar{R}}{}{[\hA_R, S_{R,\tot}^\alpha
    ] S_\ell^\alpha
    }}}.
\label{eq:extlocregexpand}
\end{equation}
We then expand the contributions in terms of Pauli strings (or its generalizations).
We again assume w.l.o.g.\ that each $\hA_R$ contains only Pauli strings with support precisely in the region $R$ (i.e., with non-identity matrices on all sites in the region); we denote such Pauli strings as $\{\hP^\mu_{R}\}$.
It is clear that $[\hA_R, S_{R,\tot}^\alpha]$ also contains only precisely such Pauli strings, while $[\hA_R, \vec{S}_{R,\tot}^2]$ can contain in addition strings with identity on one of the sites (to be precise, we will denote this larger set of Pauli strings as $\{\hat{\tilde{P}}^{\tilde{\mu}}_R \}$ but the only thing that is important below is that they are all supported in a subset of the region $R$).
Writing these commutators explicitly, the expression in Eq.~(\ref{eq:extlocregexpand}) has the following form
\begin{equation}
    [\hA_R, \vec{S}^2_{R,\tot}] = \sumal{\tilde{\mu}}{}{a^R_{\tilde{\mu}}
    \hat{\tilde{P}}^{\tilde{\mu}}_R},\;\;
    [\hA_R, S^\alpha_{R, \tot}] = \sumal{\mu}{}{b^R_{\mu\alpha} \hP^\mu_R} \;\; \implies \;\;
    [\hO_{\text{ext-loc}}, \vec{S}^2_{\tot}] = \sumal{R,\tilde{\mu}}{}{a^R_{\tilde{\mu}} \hat{\tilde{P}}^{\tilde{\mu}}_R} + 4 \sumal{R,\mu,\alpha,\ell\in \bar{R}}{}{b^R_{\mu\alpha} \hP^{\mu}_{R}\sigma^\alpha_\ell},
\label{eq:extlocpauliexpand}
\end{equation}
$\{a^R_\mu\}, \{b^R_{\mu\alpha}\}$ are some sets of coefficients and $\sigma^\alpha_{\ell}$ is the Pauli matrix on site $\ell$.
Focusing on the second sum in Eq.~(\ref{eq:extlocpauliexpand}), consider the summand for a particular $\ell$ that is sufficiently far away from the region $R$, e.g., separated by more than $(r_\text{max}+1)$ from every site in $R$ (we implicitly assume sufficiently large system size $L$ that such $\ell$ exists). 
It is easy to see that for every region $R$ with number of sites $|R| \geq 2$, there is an appropriate choice of $\ell$ such that the Pauli string $\hP^\mu_R\sigma^\nu_\ell$, which has a support in the region $R \cup \{\ell\}$, is orthogonal to \textit{any other} Pauli string that appears in Eq.~(\ref{eq:extlocpauliexpand}). 
Hence by imposing $[\hO_{\text{ext-loc}}, \vec{S}^2_{\tot}] = 0$, we necessarily obtain that $b^R_{\mu\alpha} = 0$ or $[\hA_R, \vec{S}_{R, \tot}] = 0$ for every region $R$ such that $|R| \geq 2$. 
Hence any such term in $\hO_{\text{ext-loc}}$ with $|R| \geq 2$ should be $SU(2)$-symmetric, and the question of their generatability is already answered in the previous subsection.
We still need to consider the case where the regions $R$ consist of a single site, i.e., $|R| = 1$.
In that case, the most general extensive local operator is of the form $\hO_{\text{ext-loc}} = \sumal{j, \mu}{}{h_{j,\mu}\sigma^\mu_j}$.
Using Eq.~(\ref{eq:extlocpauliexpand}), we obtain 
\begin{equation}
    [\hO_{\text{ext-loc}}, \vec{S}^2_{\tot}] = \frac{1}{2} \sumal{j, \mu, \alpha}{}{\sumal{\ell \neq j}{}{h_{j,\mu} [\sigma^\mu_j, \sigma^\alpha_j] \sigma^\alpha_\ell}} = i \sumal{j, \mu, \alpha, \beta}{}{\sumal{\ell \neq j}{}{\epsilon_{\mu\alpha\beta} h_{j,\mu} \sigma^\beta_j \sigma^\alpha_\ell}} = i \sumal{(j, \ell)}{}{\sumal{\mu, \alpha, \beta}{}{(\epsilon_{\mu\alpha\beta} h_{j,\mu} + \epsilon_{\mu\beta\alpha} h_{\ell,\mu})\sigma^\beta_j \sigma^\alpha_\ell}},
\label{eq:onesiteexpr}
\end{equation} 
where we have used $S^\alpha_j = \sigma^\alpha_j/2$, the sum over $(j,\ell)$ denotes unordered pairs of distinct sites on the lattice, and $\epsilon_{\alpha\beta\gamma}$ is the totally antisymmetric Levi-Civita symbol. 
It is easy to see that imposing $[\hO_{\text{ext-loc}}, \vec{S}^2_{\tot}] = 0$ in Eq.~(\ref{eq:onesiteexpr}) requires that $h_{j,\mu} = h_{\ell,\mu}$ for all pairs of sites $(j, \ell)$; hence $\hO_{\text{ext-loc}}$ that satisfies that condition should be a Zeeman term of the form $\hO_{\text{ext-loc}} = \sumal{\mu}{}{c_\mu S^\mu_{\tot}}$. 
Further requiring that $[\hO_{\text{ext-loc}}, S^z_{\tot}] = 0$, we obtain $\hO_{\text{ext-loc}} = S^z_{\tot}$ as the unique extensive local operator of the form of Eq.~(\ref{eq:extlocreg}) with the regions $R$ that satisfy $|R| = 1$.
This leads to the conclusion that any extensive local Hamiltonian with the dynamical $SU(2)$ symmetry is a linear combination of strictly local $SU(2)$-symmetric terms along with a Zeeman field term, e.g., $S^z_{\tot}$. 
Hence for any such extensive local Hamiltonian, all multiplets that would be degenerate in the regular $SU(2)$-invariant case (e.g., the ferromagnetic multiplet with the largest $S_\text{tot} = L/2$) will be split into towers of equally-spaced states with the same splitting given by the strength of the Zeeman field term $S^z_{\tot}$.
\end{proof}

\section{Details on Spinless Free-Fermion Bond Algebras}
\label{app:spinlessfermion}
We discuss examples of free-fermion bond algebras generated starting from a natural subset of elementary terms of Eq.~(\ref{eq:spinlessgens}), the results discussed here are summarized in Tab.~\ref{tab:spinlessbondalgebra}.
We start by listing some useful set of commutators of elementary terms, which can be derived using Eq.~(\ref{eq:TAcommutation}):
\begin{gather}
    i [T^{(r)}_{j,k}, T^{(r)}_{k,l}] = T^{(i)}_{j,l},\;\;\;
    i [T^{(r)}_{j,k}, T^{(i)}_{k,l}] = -T^{(r)}_{j,l},\;\;\;i\, [T^{(i)}_{j,k}, T^{(i)}_{k,l}] = -T^{(i)}_{j,l},\label{eq:generatelongrange}\\
    i\,[n_j, T^{(r)}_{j,k}] = T^{(i)}_{j,k},\;\;\; i\,[n_j, T^{(i)}_{j,k}] = -T^{(r)}_{j,k},\;\;i\, [T^{(r)}_{j,k}, T^{(i)}_{j,k}] = 2(n_j - n_k).\label{eq:othercomms}
\end{gather}
As we will see below, natural subsets of these generators provide representations of standard Lie algebras.
We will often use the familiar names of these algebras to also refer to the specific representations in the full fermionic Hilbert spaces.
\subsection{Hoppings with chemical potentials}\label{subsec:spinlesscomplexchemical}
We start with the most general case where the bond algebra $\mA_{c,\mu}$ is generated by all the nearest-neighbor hopping terms and on-site chemical potential terms, i.e., 
\begin{equation}
    \mA_{c, \mu} \defn \lgen \{T^{(\Upsilon)}_{j,k}\}_{\tnn}, \{n_j\} \rgen,\;\;\; \Upsilon \in \{r, i, c\}.
\label{eq:spinlesscmu}
\end{equation}
The equivalence between the three cases -- only real hoppings, only imaginary hoppings, or both real and imaginary hoppings -- follows from Eq.~(\ref{eq:othercomms}), i.e., they can be generated from one another in the presence of chemical potential terms.
$\mA_{c,\mu}$ is the bond algebra that corresponds to families of free-fermion systems with nearest-neighbor hopping terms and on-site chemical potential terms.
Although only nearest-neighbor hopping terms are included in the bond algebra generators, it is easy to see that repetitive applications of Eqs.~(\ref{eq:generatelongrange}) and (\ref{eq:othercomms}) on the generators can be used to construct hopping terms between any pair of sites on any lattice, hence they are all part of the algebra $\mA_{c,\mu}$.
In fact, we could reduce the number of generators even further, e.g., to 
$T^{(i)}_{j,k}$ along one path visiting all sites of the lattice and a single $n_j$; however, it is natural to include all relevant local terms from the start.
The Lie algebra generated by these terms in this case is (a representation of) $\fu(N)$, the full algebra of $N \times N$ Hermitian matrices.
The corresponding bond algebra is the enveloping algebra $\mU(\fu(N))$, and the terms in the Lie algebra can also be interpreted as generators of the group $U(N)$.
The bond algebra $\mA_{c,\mu}$ has two singlets -- the vacuum state $\ket{\Omega}$ with no fermions, and the fully occupied state $\sket{\bar{\Omega}}$ with $N$ fermions. 
Indeed, all hopping terms $\{T^{(c)}_{j,k}\}_{\tnn}$ annihilate these states and they are eigenstates of the chemical potential terms $\{n_j\}$, and it is easy to see that there are no other simultaneous eigenstates of all $\{n_j\}$ and $\{T^{(c)}_{j,k}\}_{\tnn}$. 
These singlets are non-degenerate, since they differ in their eigenvalues under the operators $\{n_j\}$. 
Hence, apart from the algebra generated by $N_{\tot}$ (denoted by $\lgen N_{\tot} \rgen$), the singlet projectors $\ketbra{\Omega}{\Omega}$ and $\ketbra{\bar{\Omega}}{\bar{\Omega}}$ should be a part of the commutant $\mC_{c,\mu}$ of $\mA_{c,\mu}$.
However, these projectors are both included in the algebra $\lgen N_{\tot} \rgen$, since they can be expressed in terms of $N_{\tot}$ [e.g., $\ketbra{\bar{\Omega}}{\bar{\Omega}} = \prodal{m = 0}{N-1}{(N_{\tot} - m)/(N - m)}$].
Hence the full commutant algebra is just given by
\begin{equation}
\mC_{c,\mu} = \lgen N_{\tot}\rgen.
\end{equation}
Moreover, the commutant can also be viewed as being generated by the full family of on-site unitaries of the form $\exp(i \alpha N_{\tot}) = \prod_j{\exp(i\alpha n_j)}$.
As discussed in Sec.~\ref{subsec:spinlessfermions}, this property is useful for the application of the DCT since Lems.~\ref{lem:strloc} and \ref{lem:typeI} apply.
Note that since these unitaries can be viewed as elements of a $U(1)$ group, the symmetry is typically referred to as $U(1)$.
Since the commutant $\mC_{c, \mu}$ is Abelian, the common center of $\mA_{c, \mu}$ and $\mC_{c, \mu}$ is simply the commutant itself (see Sec.~\ref{subsec:Hilbertdecomp}), i.e.,  $\mZ_{c, \mu} = \mC_{c, \mu} = \lgen N_{\tot} \rgen$.
\subsection{Complex hoppings without chemical potentials}\label{subsec:spinlesscomplex}
\subsubsection{Bond and Commutant Algebra}
We now include complex hopping terms and exclude the chemical potentials $\{n_j\}$, and study the algebra 
\begin{equation}
    \mA_{c} = \lgen \{T^{(c)}_{j,k}\}_{\tnn}\rgen \defn \lgen \{T^{(r)}_{j,k}\}_{\tnn}, \{T^{(i)}_{j,k}\}_{\tnn}\rgen. 
\label{eq:spinlessc}
\end{equation}
$\mA_c$ is the bond algebra that corresponds to families of free-fermion systems with complex hopping terms and no on-site potentials.
As discussed in App.~\ref{subsec:spinlesscomplexchemical}, using Eq.~(\ref{eq:generatelongrange}), it is easy to show that complex hopping terms between any pair of sites on any lattice can be generated from the set of nearest-neighbor complex hopping terms, and they are all part of $\mA_c$.
Furthermore, using the last of Eq.~(\ref{eq:othercomms}) we can also generate $n_j - n_k$.
The Lie algebra of all the $N(N-1)/2$ complex hopping terms and $N-1$ independent $n_j - n_k$ terms is the Lie algebra of \textit{traceless} $N \times N$ Hermitian matrices $\{A\}$ in Eq.~(\ref{eq:TAdefnspinless}), which is $\mf{su}(N)$.
These hopping terms form the generators of the group $SU(N)$, and the associative bond algebra $\mA_c$ is the UEA $\mU(\mf{su}(N))$~\cite{pakrouski2020many}.  
The bond algebra $\mA_{c}$ has two singlets -- the vacuum state $\ket{\Omega}$, and the fully occupied state $\sket{\bar{\Omega}}$ -- same as those of the algebra $\mA_{c,\mu}$. 
However, in this case, the singlets are \textit{degenerate}, since all generators $\{T^{(c)}_{j,k}\}_{\tnn}$ annihilate these states. 
Apart from the algebra generated by $N_{\tot}$ (denoted by $\lgen N_{\tot} \rgen$), the operators $\ketbra{\Omega}{\Omega}$, $\ketbra{\Omega}{\bar{\Omega}}$, $\ketbra{\bar{\Omega}}{\Omega}$, and $\ketbra{\bar{\Omega}}{\bar{\Omega}}$ are all part of the commutant $\mC_c$ of $\mA_c$.
As discussed in Sec.~\ref{subsec:spinlesscomplexchemical}, the projectors onto the singlets $\ketbra{\Omega}{\Omega}$ and $\ketbra{\bar{\Omega}}{\bar{\Omega}}$ are included in the algebra $\lgen N_{\tot} \rgen$, hence the full commutant in this case is 
\begin{equation}
\mC_c = \lgen N_{\tot}, \ketbra{\bar{\Omega}}{\Omega}\rgen,
\end{equation}
where we have abused notation for $\lgen \cdots \rgen$ to implicitly include closure under the Hermitian conjugate $\dagger$ operation.
Another way to obtain this expression for the commutant is to observe that the operators $\prod_j{\cd_j}$ ($=\ketbra{\bar{\Omega}}{\Omega}$) and $\prod_j{c_j}$ ($=\ketbra{\Omega}{\bar{\Omega}}$ up to a sign due to fermion ordering) straightforwardly commute with all of the generators $\{T^{(c)}_{j,k}\}_{\tnn}$ (hence are part of $\mC_c$) but do not commute with $N_{\tot}$ (hence are not part of $\lgen N_{\tot}\rgen$).
Note that we do not know of any useful ``group" representation of $\mC_c$ that is analogous to the $U(1)$ interpretation of $\mC_{c,\mu}$, and it is more convenient to think of it as an algebra.
\subsubsection{Center}
Note that here the commutant $\mC_{c}$ is non-Abelian, since $N_{\tot}$ and $\ketbra{\Omega}{\bar{\Omega}}$ do not commute. 
We expect the center of $\mC_{c}$ to be generated by some polynomials of $N_{\tot}$ that commute with $\ketbra{\Omega}{\bar{\Omega}}$.
Consider such a polynomial $P(N_{\tot})$ where $P(x) = \sum_{m > 0}{a_m x^m}$ and we have excluded the identity $x^0$ which is trivially in the center.
Since $\ketbra{\Omega}$ and $\sket{\bar{\Omega}}$ are eigenstates of $N_{\tot}$ with eigenvalues $0$ and $N$, commutation with $\ketbra{\Omega}{\bar{\Omega}}$ is equivalent to  requiring $P(N) = P(0)$ where we know $P(0) = 0$ .
The set of all such polynomials is given by $f(N_{\tot}) N_{\tot} (N - N_{\tot})$, where $f$ can be an arbitrary polynomial, and the center can then be written as 
\begin{equation}
    \mZ_c = \lgen \{N_{\tot}^\alpha (N - N_{\tot}):\;\; \alpha \geq 1\}\rgen.
\label{eq:Zc}
\end{equation}
(The shown operators with $\alpha = 1, 2, \dots N-1$, together with the identity operator, in fact span $\mZ_c$.) 
Note that we are not able to identify a simpler set of generators for $\mZ_c$.
\subsubsection{Double Commutant Theorem}\label{subsubsec:DCTcomplexhopping}
We now discuss the application of the DCT (Thm.~\ref{thm:dct}) to this case.
In particular, we wish to understand the structure of local operators within the algebra $\mA_c$ by studying local operators that commute with (the generators of) $\mC_c$.
Since we know that $\mA_c \subset \mA_{c,\mu}$, local operators in $\mA_c$ are also local operators in $\mA_{c,\mu}$, where we understand their structure due to the on-site symmetry structure of $\mC_{c,\mu}$ (see Sec.~\ref{subsec:spinlessfermions}), and Lems.~\ref{lem:strloc} and \ref{lem:typeI} apply.
Hence, local operators within $\mA_c$ are those in $\mA_{c,\mu}$ that commute with $\ketbra{\Omega}{\bar{\Omega}}$ and $\ketbra{\bar{\Omega}}{\Omega}$, or equivalently and w.l.o.g., operators that annihilate the states $\ket{\Omega}$ and $\sket{\bar{\Omega}}$ (recalling that these are singlets of $\mA_{c,\mu}$ and $\mathds{1} \in \mA_{c,\mu}$). 
Focusing on strictly local operators $\hO_R \in \mA_{c,\mu}$ and keeping in mind that $\ket{\Omega}$ and $\sket{\bar{\Omega}}$ are product states (vacuum state and the fully filled state respectively), the condition $\hO_R\ket{\Omega} = 0 = \hO_R\sket{\bar{\Omega}}$ is equivalent to $\hO_R \ket{\Omega}_R = \hO_R \sket{\bar{\Omega}}_R = 0$, where $\ket{\Omega}_R$ and $\sket{\bar{\Omega}}_R$ are the vacuum and fully filled states on the sites in the region $R$.
Hence, applying the DCT to operators in the region $R$, $\hO_R$ should belong to the centralizer of the algebra $\lgen N_{R, \tot} , (\ketbra{\Omega}{\bar{\Omega}})_R\rgen$, where $N_{R,\tot} \defn \sum_{j \in R}{n_j}$.
For contiguous regions $R$, it is easy to show that this is the algebra generated by nearest-neighbor complex hopping terms restricted to the region $R$.
We then turn to operators that are sums of strictly local operators (which include extensive local operators) that are of the form $\hO_{\text{ext-loc}} = \sum_R{\hA_R}$, where $\hA_R$ has support everywhere in the region $R$, and the sum over $R$ is over distinct contiguous regions (not necessarily extensively many of them).
As we have shown in the previous section, for any $\hO_{\text{ext-loc}} \in \mA_{c,\mu}$, the strictly local operators $\{\hA_R\}$ are also a part of the bond algebra $\mA_{c,\mu}$.
Hence $\hA_R$ can be expressed as polynomials in the generators of $\mA_{c,\mu}$, i.e., $\hA_R = f_R(\{n_j\}_R, \{T^{(\ast)}_{j,k}\}_R)$, where the $\{\cdot\}_R$ denotes the restriction of the set to operators with support completely within the region $R$, and $f_R$ is some polynomial that is generically $R$-dependent.
Requiring $\hO_{\text{ext-loc}}$ to vanish on the states $\ket{\Omega}$ and $\sket{\bar{\Omega}}$ and noting that these states are singlets of the algebra $\mA_{c,\mu}$, we obtain the conditions $\sum_{R}{f_R(\{n_j = 0\}_R, \{T^{(\ast)}_{j,k} = 0\}_R)} = 0$ and $\sum_{R}{f_R(\{n_j = 1\}_R, \{T^{(\ast)}_{j,k} = 0\}_R)} = 0$.
For completely arbitrary polynomials $\{f_R\}$, we are not able to additionally constrain the structure of the local terms. 
Indeed, there are operators of the form of $\hO_{\text{ext-loc}}$ that are in $\mA_c$, where it is clear that the individual parts $\hA_R$ cannot be expressed in terms of generators of $\mA_c$ restricted to the region $R$. 
One such example is $\hO_{\text{ext-loc}} = n_j - n_k$, where $j$ and $k$ are two sites that are far from each other. 
Here, it is clear that $n_j$ and $n_k$ are not individually part of the bond algebra $\mA_c$ (since they do not commute with $\ketbra{\bar{\Omega}}{\Omega}$), but their difference is.
However, we can obtain additional constraints if we are interested in a restricted class of extensive local operators, in particular ones where the individual terms $\hA_{R_1}$ and $\hA_{R_2}$ corresponding to distinct regions $R_1$ and $R_2$ have the same ``forms", i.e., they are  related by translation.
This class of operators includes translation-invariant operators that frequently appear in physics applications.
For such operators, we can write $\hA_R = f(\{n_j\}_R, \{T^{(\ast)}_{j,k}\}_R)$, where $f$ is a polynomial independent of the region $R$.
The aforementioned conditions on $\hO_{\text{ext-loc}}$ then imply that $f(\{n_j = 0\}_R, \{T^{(\ast)}_{j,k} = 0\}_R) = f(\{n_j = 1\}_R, \{T^{(\ast)}_{j,k} = 0\}_R) = 0$, and hence $\hA_R\ket{\Omega} = \hA_R\sket{\bar{\Omega}} = 0$.
$\hA_R$ is hence a strictly local operator within the algebra $\mA_c$, and arguments from  earlier in this section apply.
Hence all such extensive local operators in $\mA_c$, including translation-invariant ones, are linear combinations of strictly local operators in $\mA_c$.
We close this section with a few more words about the example of $\hO_\text{ext-loc} = n_j - n_k \in \mA_c$.
As already mentioned, its strictly 1-local parts $n_j$ and $n_k$ do not belong to $\mA_c$ itself, which is one specific distinction from the cases with only on-site unitary symmetries.
On one hand, the full term can be written as a sum of 2-local terms from $\mA_c$, e.g., on a one-dimensional chain $n_j - n_k = \sum_{\ell=j}^{k-1} (n_\ell - n_{\ell+1}) = \sum_{\ell=j}^{k-1}\frac{i}{2} [T_{\ell,\ell+1}^{(r)}, T_{\ell,\ell+1}^{(i)}]$ [using Eq.~(\ref{eq:othercomms})].
However, this expression contains nearest-neighbor generators of $\mA_c$ on a path between sites $j$ and $k$, and the need for such generators ``in-between" $j$ and $k$ seems unavoidable.
Indeed, suppose we could produce this operator by generators localized in non-intersecting regions $R_1$ and $R_2$ containing $j$ and $k$ respectively, $n_j - n_k = \sum_{\gamma} A_{R_1,\gamma} A_{R_2,\gamma}$, where the R.H.S. is the most general such form with $A_{R_1,\gamma}, A_{R_2,\gamma} \in \mA_c$ (note that $A_{R_1,\gamma}$ and $A_{R_2,\gamma}$ can also be identities).
We can see that the R.H.S. has the same eigenvalue on the states $\ket{\Omega}$ and $\sket{\bar{\Omega}}$ containing ``parts'' $\ket{\Omega}_{R_1} \otimes \ket{\Omega}_{R_2}$ and $\sket{\bar{\Omega}}_{R_1} \otimes \sket{\bar{\Omega}}_{R_2}$ respectively {\it and} also on product states containing $\ket{\Omega}_{R_1} \otimes \sket{\bar{\Omega}}_{R_2}$, which is not the case for the L.H.S., i.e., we have a contradiction.
At present, the significance of the need for ``in-between'' generators is not clear to us.
This example with $j$ and $k$ arbitrary far from each other hints at some additional ``non-local'' structure (even though this $\hO_{\text{ext-loc}}$ can be written as a sum of 2-local terms generated from $\mA_c$), but we are not able to quantify it more precisely. 
Moreover, we can also write down several other types of one-dimensional examples that appear to have varying degrees of such ``non-locality".
For example, we have $\hO'_{\text{ext-loc}} = \sum_i (n_{2i} - n_{2i+1}) \in \mA_c$ where the individual 1-local parts also do  not belong to $\mA_c$, but its generation from the nearest-neighbor hopping generators feels more genuinely local.
On the other extreme, we also have $\hO^{''}_{\text{ext-loc}} = \sum_{i=1}^{L/2} n_i - \sum_{i=L/2+1}^{L} n_i \in \mA_c$, which can still be written as a sum of 2-local terms generated from $\mA_c$, but any such writing necessarily involves extensively large coefficients in front of the 2-local terms.
We leave finding precise formulations and locality distinctions of possible operators in $\mA_c$ for future work.
\subsection{Pure imaginary hoppings}\label{subsec:spinlessimaginary}
We now remove real hoppings from the generators of $\mA_c$, and consider the bond algebra $\mA_i$ generated by only nearest-neighbor imaginary hoppings and no chemical potentials:
\begin{equation}
    \mA_{i} = \lgen \{T^{(i)}_{j,k}\}_{\tnn} \rgen.
\label{eq:spinlessimag}
\end{equation}
Starting from nearest-neighbor imaginary hopping terms, the repeated application of Eq.~(\ref{eq:generatelongrange}) generates imaginary hopping terms of all ranges. 
The Lie algebra of all these imaginary hopping terms is the algebra of $N \times N$ purely imaginary Hermitian (hence antisymmetric) matrices $\{A\}$ in Eq.~(\ref{eq:TAdefnspinless}), which is $\mf{so}(N)$.  
These terms $\{T^{(i)}_{j,k}\}$ can then be interpreted as generators of $SO(N)$, and the bond algebra is $\mA_i = \mU(\mf{so}(N))$.
Similar to $\mA_c$ in Sec.~\ref{subsec:spinlesscomplex}, $\mA_i$ possesses two degenerate singlets $\ket{\Omega}$ and $\sket{\bar{\Omega}}$.
A simple proof is as follows.
Any bond algebra singlet $\ket{\Psi}$ is by definition a simultaneous eigenstate of all 
$T^{(i)}_{j,k}$. 
Then using Eq.~(\ref{eq:generatelongrange}) we immediately conclude that $T^{(i)}_{j,k} \ket{\Psi} = 0$ for any pair $j,k$.
Finally, using $(T^{(i)}_{j,k})^2 = n_j + n_k - 2 n_j n_k = (n_j - n_k)^2$, we conclude that such a $\ket{\Psi}$ cannot contain configurations with unequal occupations for any pair of sites, hence it must be in the span of $\ket{\Omega}$ and $\sket{\bar{\Omega}}$.
Alternatively, one can reach the same conclusion using the identity $\sum_{j < k} (T^{(i)}_{j,k})^2 = N_\text{tot} (N - N_\text{tot})$.
In addition to $N_{\tot}$, the commutant $\mC_i$ of $\mA_{i}$ contains a ``particle-hole" $Z_2$ symmetry, generated by the unitary operator $Q_X$ defined in Eq.~(\ref{eq:QXQXtdefns}).
It is easy to verify that $[Q_X, T^{(i)}_{j,k}] = 0$ for all $j, k$. 
The full commutant is then given by 
\begin{equation}
    \mC_i = \lgen N_{\tot}, Q_X \rgen.
\end{equation}
(That this exhausts the commutant is proved in App.~\ref{app:commutantexhaustion}).
Note that $\mC_i$ as defined also contains the operators $\ketbra{\Omega}{\bar{\Omega}}$ and $\ketbra{\bar{\Omega}}{\Omega}$, since $Q_X\ket{\Omega} = \sket{\bar{\Omega}}$ and $\ketbra{\Omega}{\Omega}$ is a part of $\lgen N_{\tot}\rgen$ (see discussion in Sec.~\ref{subsec:spinlesscomplexchemical}).
Some obvious subgroups of $\mC_i$ are $Z_2$, generated by $Q_X$, and $U(1)$, generated by $N_{\tot}$.
Moreover, the full commutant can also be viewed as being generated by the full family of on-site unitaries corresponding to these groups, i.e., $\exp(i \alpha N_{\tot}) = \prod_j{\exp(i\alpha n_j)}$ for $U(1)$ and $Q_X$ for $Z_2$.
This property is useful for the application of the DCT, as Lems.~\ref{lem:strloc} and \ref{lem:typeI} apply.
Since $Q_X$ and $N_{\tot}$ do not commute (instead, they  satisfy $Q_X N_\tot + N_\tot Q_X = Q_X N$ or equivalently $Q_X N_\tot Q_X^\dagger = N - N_\tot$), the full group generated by these is not a simple product of the two symmetry groups, and is usually referred to as $U(1) \rtimes Z_2$.
While the commutant $\mC_i$ is non-Abelian since $Q_X$ does not commute with $N_{\tot}$, it is easy to check that $Q_X$ does commute with $(N_{\tot} - \frac{N}{2})^2$.
Hence we conjecture that the center here is given by 
\begin{equation}
    \mZ_i = \lgen (N_{\tot} - \frac{N}{2})^2 \rgen.
\end{equation}
One might wonder if a distinct bond algebra can be constructed by adding on-site chemical potential terms to the list of generators of $\mA_i$. 
However, using Eq.~(\ref{eq:othercomms}), we obtain that this combination can generate real hoppings; hence eventually pure imaginary hoppings and on-site chemical potentials generate the algebra $\mA_{c,\mu}$ discussed in Sec.~\ref{subsec:spinlesscomplexchemical}.  
\subsection{Real hoppings}\label{subsec:spinlessreal}
Finally, we briefly comment on the bond algebra generated from only nearest-neighbor real hoppings, defined as,
\begin{equation}
    \mA_r \defn \lgen \{T^{(r)}_{j,k}\}_{\tnn}\rgen,
\label{eq:Ardefn}
\end{equation}
analogous to $\mA_i$ in Sec.~\ref{subsec:spinlessimaginary}.
Using Eq.~(\ref{eq:generatelongrange}), we note that imaginary hoppings can be generated starting from real hoppings, hence the set of all real hopping terms between any pair of sites does not form a closed algebra. 
On a non-bipartite lattice, this eventually generates all complex hoppings, and the resulting bond algebra $\mA_r$ is equal to $\mA_c$ discussed in Sec.~\ref{subsec:spinlesscomplex}.
Nevertheless, on a \textit{bipartite} lattice, nearest-neighbor real hoppings do generate a different closed Lie algebra that consists of real hoppings between different sublattices and imaginary hoppings within the same sublattice.
Using Eq.~(\ref{eq:generatelongrange}), it is easy to show that such an algebra is closed, and no other hopping terms are generated.  
Indeed, in terms of the matrices $\{A\}$ of Eq.~(\ref{eq:TAdefnspinless}), this is the Lie algebra of $N \times N$ Hermitian matrices of the form
\begin{equation}
    \left(
    \begin{array}{c|c}
    \textrm{Im} & \textrm{Re} \\
    \hline
    \textrm{Re} & \textrm{Im}
    \end{array}
    \right),
\label{eq:realentrymatrix}
\end{equation}
in a grouping of basis by sublattice, where $\textrm{Im}$ and $\textrm{Re}$ denote purely imaginary and real entries respectively, and the size of each diagonal block in Eq.~(\ref{eq:realentrymatrix}) corresponds to the number of sites in each sublattice.
This Lie algebra is mathematically equivalent to the previous case of pure imaginary hopping (i.e., the Lie algebra is $\mf{so}(N)$ and the bond algebra $\mA_r$ is isomorphic to $\mA_i$), as they can be mapped onto each other by transformation $c_j \to i c_j$ (implemented using $e^{-i\frac{\pi}{2} n_j} c_j e^{i\frac{\pi}{2} n_j} = i c_j$) on one of the sublattices.
This also results in an isomorphism of the commutant $\mC_r$ of $\mA_r$, and $\mC_i = \lgen N_{\tot}, Q_X \rgen$ of $\mA_i$.
While $N_{\tot}$ is invariant under the sublattice transformation, $Q_X$ transforms to $\tQ_X$, where both are shown in Eq.~(\ref{eq:QXQXtdefns}).
Hence we obtain 
\begin{equation}
    \mC_r = \lgen N_{\tot}, \tQ_X\rgen.
\end{equation} 
Since $N_{\tot}$ remains invariant under the sublattice transformation, the center $\mZ_r$ is the same as $\mZ_i$ and is given by $\mZ_r = \lgen \left(N_{\tot} - \frac{N}{2} \right)^2 \rgen$.
Further, due to the isomorphism between the algebras $(\mA_r, \mC_r)$ and $(\mA_i, \mC_i)$ all the results on the group structure of the commutant and the application of the DCT discussed in Sec.~\ref{subsec:spinlessimaginary} apply here.
\section{Details on Spinful Free-Fermion Bond Algebras}\label{app:spinfulfermion}
We discuss examples of free-fermion bond algebras generated starting from a natural subset of spinful elementary terms of Eq.~(\ref{eq:spinfulgens}) and their commutants.
The results presented here are summarized in Tab.~\ref{tab:spinfulbondalgebra}.
Note that in many of the cases below, we are not able to \textit{prove} that the commutants are exhausted by the algebras we mention.
However, in all such cases, we are able to verify our conjectures using numerical techniques~\cite{moudgalya2022numerical} up to fairly large system sizes.
On the other hand, the exhaustive determination of all singlets of the bond algebra is easier, and in all cases we can prove our claims analytically (even if we do not always show the proofs).
We start by listing some useful set of commutators of elementary terms, which can be derived using Eq.~(\ref{eq:TAcommutation}):
\begin{gather}
    i [T^{(r)}_{j,k}, T^{(r)}_{k,l}] = T^{(i)}_{j,l},\;\;\;
    i [T^{(r)}_{j,k}, T^{(i)}_{k,l}] = -T^{(r)}_{j,l},\;\;\;i\, [T^{(i)}_{j,k}, T^{(i)}_{k,l}] = -T^{(i)}_{j,l},\label{eq:generatelongrangespin}\\
    i\,[K_j, T^{(r)}_{j,k}] = T^{(i)}_{j,k},\;\;\; i\,[K_j, T^{(i)}_{j,k}] = -T^{(r)}_{j,k},\;\;i\, [T^{(r)}_{j,k}, T^{(i)}_{j,k}] = 2(K_j - K_k),\label{eq:othercommsspin1}\\
    i\,[M_j, T^{(r)}_{j,k}] = \wT^{(i)}_{j,k},\;\;\; i\,[M_j, T^{(i)}_{j,k}] = -\wT^{(r)}_{j,k},\;\;\;
    i\,[M_j, \wT^{(r)}_{j,k}] = T^{(i)}_{j,k},\;\;\; i\,[M_j, \wT^{(i)}_{j,k}] = -T^{(r)}_{j,k}, \label{eq:othercommsspin2}\\
    i [T^{(r)}_{j,k}, \wT^{(r)}_{k,l}] = \wT^{(i)}_{j,l},\;\;\;
    i [T^{(r)}_{j,k}, \wT^{(i)}_{k,l}] = i [\wT^{(r)}_{j,k}, T^{(i)}_{k,l}] =  -\wT^{(r)}_{j,l},\;\;\;i\, [T^{(i)}_{j,k}, \wT^{(i)}_{k,l}] = -\wT^{(i)}_{j,l},\;\; \nn \\
    i [\wT^{(r)}_{j,k}, \wT^{(r)}_{k,l}] = T^{(i)}_{j,l},\;\;\;
    i [\wT^{(r)}_{j,k}, \wT^{(i)}_{k,l}] = -T^{(r)}_{j,l},\;\;\;i\, [\wT^{(i)}_{j,k}, \wT^{(i)}_{k,l}] = -T^{(i)}_{j,l},\;\label{eq:generatelongrangetild}
\end{gather}
where we have defined
\begin{equation}
    \wT^{(r)}_{j,k} \defn \sumal{\sigma \in \{\uparrow, \downarrow\}}{}{s_\sigma (\cd_{j,\sigma} c_{k,\sigma} + \cd_{k,\sigma} c_{j,\sigma})},\quad
    \wT^{(i)}_{j,k} \defn \sumal{\sigma \in \{\uparrow, \downarrow\}}{}{i s_\sigma (\cd_{j,\sigma} c_{k,\sigma} - \cd_{k,\sigma} c_{j,\sigma})},\quad s_\uparrow = +1,\;\;s_\downarrow = -1.
\label{eq:tildeTdefn}
\end{equation} 
Further, in this appendix, we also list some useful relations between Casimir elements in some of the spinful cases discussed in Tab.~\ref{tab:spinfulbondalgebra}.
For concreteness, we use the following definition of the quadratic Casimir of a Lie group $G$ with generators $\{R^a\}$ of the corresponding Lie algebra (representations of $\{R^a_\text{def}\}$)~\cite{haber2021useful}:
\begin{equation}
    C^{G}_2 \defn \sum_a{(R^a)^2},\;\;\;\Tr(R^a_\text{def} R^b_\text{def}) = \frac{1}{2}\delta_{a,b}\ \text{in the defining representation.}
\label{eq:casimirdefn}
\end{equation}
For example, the defining representation $\{R^a_\text{def}\}$ is that of $n \times n$ matrices from the appropriate Lie algebras for $U(n)$, $SU(n)$, and $SO(n)$.
\subsection{Hoppings with chemical potentials}\label{subsec:spincomplexmu}
\subsubsection{Bond and Commutant Algebra}
We start with the general case that is symmetric between the two spins, where the bond algebra $\mA_{c, \mu}$ is generated by nearest-neighbor hopping elementary terms and on-site chemical potential terms, i.e.,
\begin{equation}
    \mA_{c, \mu} \defn \lgen \{T^{(\Upsilon)}_{j,k}\}_{\tnn}, \{K_j\}\rgen,\;\;
    \Upsilon
    \in \{r, i, c\}. 
\label{eq:spinfulcmu}
\end{equation}
Similar to the spinless fermion cases discussed in Sec.~\ref{subsec:spinlessfermions}, nearest-neighbor hopping terms along with chemical potential terms are sufficient to generate complex hopping terms of all ranges as a consequence of Eq.~(\ref{eq:generatelongrangespin}).
The Lie algebra generated by the terms in Eq.~(\ref{eq:spinfulcmu}) is then $\mathfrak{u}(N)$, the algebra of all matrices $A$ in Eq.~(\ref{eq:TAdefinitionspin}) with $A^\uparrow = A^\downarrow$.
The $N^2$ terms in the Lie algebra are then generators of the group $U(N)$, and the bond algebra $\mA_{c,\mu}$ is the enveloping algebra $\mU(\mathfrak{u}(N))$.
In addition to conserving particle number and spin, all the elementary terms in the generators of $\mA_{c,\mu}$ in Eq.~(\ref{eq:spinfulcmu}) are spin-$SU(2)$ symmetric, i.e., they all commute with the operators $\{S^\alpha_{\tot}\}$ defined in Eq.~(\ref{eq:Salphadefn}).
The full commutant of $\mA_{c,\mu}$ is given by
\begin{equation}
    \mC_{c,\mu} = \lgen \{S^\alpha_{\tot}\}, N_{\tot} \rgen.
\end{equation}
Note that $\mC_{c,\mu}$ also contains the $SU(2)$ raising and lowering operators $S^+_{\tot}$ and $S^-_{\tot}$ as well as the $SU(2)$ Casimir operator $\vec{S}_{\tot}^2 \defn \sum_\alpha{(S^\alpha_{\tot})^2}$. 
Since the $S^\alpha_{\tot}$'s are the generators of an $SU(2)$ Lie group and $N_{\tot}$ is the generator of a $U(1)$ group that commutes with all of the $S^\alpha_{\tot}$'s, the commutant is associated with the group $SU(2) \times U(1)$. 
Moreover, the full commutant can also be viewed as being generated by the full family of on-site unitaries corresponding to these groups, i.e., $\exp(i \sum_{\alpha}{\theta_\alpha S^\alpha_{\tot}}) = \prod_j{\exp(i\sum_\alpha{\theta_\alpha S^\alpha_j})}$ for $SU(2)$ and $\exp(i\alpha N_{\tot}) = \prod_j{\exp(i\alpha K_j)}$ for $U(1)$.
This property is useful for the application of the DCT since Lems.~\ref{lem:strloc} and \ref{lem:typeI} apply.
The bond algebra $\mA_{c,\mu}$ has several singlets.
The vacuum $\ket{\Omega} \defn \ket{0 \cdots 0}$ and the anti-vacuum $\sket{\bar{\Omega}} \defn \ket{\updownarrow \cdots \updownarrow}$ (where $0$ and $\updownarrow$ denote empty and doubly-occupied sites) are annihilated by all the hopping terms $\{T^{(c)}_{j,k}\}_{\tnn}$ and are eigenstates (with eigenvalues $0$ and $2$ respectively) of the on-site chemical potential terms $\{K_j\}$ -- hence they are non-degenerate singlets. 
Another set of singlets are the spin polarized ``ferromagnetic states" $\ket{F} \defn \ket{\uparrow \cdots \uparrow}$ and $\sket{\bar{F}} \defn \ket{\downarrow \cdots \downarrow}$, which are also annihilated by $\{T^{(c)}_{j,k}\}_{\tnn}$ and are eigenstates of the $\{K_j \}$ (with eigenvalue $1$).
In fact, since $S^\pm_{\tot}$ are a part of the commutant $\mC_{c, \mu}$, all the states of the ferromagnetic multiplet, i.e., $\ket{F_n} \defn (S^-_{\tot})^n \ket{F}$,  are singlets of $\mA_{c,\mu}$ that are degenerate with $\ket{F}$ and $\sket{\bar{F}}$ (of course, $\ket{F_N} \sim \sket{\bar{F}}$).  
It is easy to show that all the singlet projectors $\ketbra{\Omega}{\Omega}$, $\ketbra{\bar{\Omega}}{\bar{\Omega}}$, $\ketbra{F}{F}$, $\ketbra{\bar{F}}{\bar{F}}$ can be expressed in terms of $S^z_{\tot}$ and $N_{\tot}$ [e.g.,  $\ketbra{F}{F} = \prod_{m < N/2} (S^z_{\tot} - m)/(N/2 - m)$, where $m$ runs over all allowed integer and half-integer eigenvalues of $S^z_{\tot}$ other than the maximal value $N/2$] and are part of $\mC_{c,\mu}$ as defined previously, and so are $\ketbra{F_n}{F_m} = (S^-_{\tot})^n \ketbra{F}{F} (S^+_{\tot})^m$ (since $S^\pm_{\tot} \in \mC_{c,\mu}$).
\subsubsection{Center and Casimir Relations}
Since the commutant $\mC_{c, \mu}$ consists of the spin $SU(2)$ symmetry and the particle number $U(1)$ symmetry that are ``independent" of each other, its center $\mZ_{c, \mu}$ is generated by $\vec{S}^2_{\tot}$ (the Casimir element of the spin $SU(2)$ symmetry) and $N_{\tot}$ (the generator of the particle number  $U(1)$ symmetry). Hence, we obtain
\begin{equation}
    \mZ_{c, \mu} = \lgen \vec{S}^2_{\tot}, N_{\tot} \rgen.
\label{eq:Zcmu}
\end{equation} 
From the perspective of the bond algebra $\mA_{c,\mu}$, since its generators are those of $U(N)$, its center should contain all Casimir operators of $U(N)$, including its quadratic Casimir $C^{U(N)}_2$. [The case of the linear Casimir $C^{U(N)}_1$ is trivial since it is simply proportional to $N_\tot$.]
Since the centers of $\mC_{c,\mu}$ and $\mA_{c,\mu}$ coincide to $\mZ_{c,\mu}$ of Eq.~(\ref{eq:Zcmu}), we obtain $C^{U(N)}_2 \in \lgen \vec{S}^2_{\tot}, N_{\tot}\rgen$; hence $C^{U(N)}_2$, $C^{SU(2)}_2 \defn \vec{S}^2_{\tot}$, and $N_{\tot}$ must be related.
Indeed, we can find the relation
\begin{equation}
C^{U(N)}_2 \defn \sum_{j < k} \left[ \left(\frac{T_{j,k}^{(r)}}{2} \right)^2 + \left(\frac{T_{j,k}^{(i)}}{2}\right)^2 \right] + \sum_j \left(\frac{K_j}{\sqrt{2}}\right)^2 = \frac{1}{4}N_\tot (2N + 4 - N_\tot) - \vec{S}_\tot^2. 
\label{eq:quadraticUN}
\end{equation}
Defining a $U(1)$ charge $Q \defn N_{\tot} - N$, we can rewrite Eq.~(\ref{eq:quadraticUN}) as
\begin{equation}
C^{U(N)}_2 + C^{SU(2)}_2 = \frac{(N + Q)(N - Q + 4)}{4},
\end{equation}
which resembles Casimir relations discussed in \cite{klebanov2018spectra, gaitan2020hagedorn}.
In this case, since the center $\mZ_{c,\mu}$ of Eq.~(\ref{eq:Zcmu}) is equal to the Casimir algebra of $SU(2) \times U(1)$, given by $\mZ_{SU(2) \times U(1)} = \lgen \vec{S}^2_{\tot}, N_{\tot}\rgen$, the decomposition of the Hilbert space in terms of the bond and commutant algebra is equivalent to the decomposition into irreducible representations of $U(N) \times (SU(2) \times U(1))$.
Equivalently, the Hilbert space within a given $N_{\tot}$ sector can be partitioned according to irreducible representations of $U(N) \times SU(2)$, as pointed out in \cite{pakrouski2020many}.
Note that the same construction also applies to the cases \#1b and \#1c in Tab.~\ref{tab:spinfulbondalgebra}, which are isomorphic to this case, as we will discuss in App.~\ref{subsec:spinfulimagh} and App.~\ref{subsec:spinfulreal}.
\subsection{Complex hoppings}\label{subsec:spincomplex}
\subsubsection{Bond and Commutant Algebra}
We now consider complex hopping terms without the chemical potential terms $\{K_j\}$ and study the algebra
\begin{equation}
    \mA_{c} = \lgen \{T^{(c)}_{j,k}\}_{\tnn} \rgen \defn \lgen \{T^{(r)}_{j,k}\}_{\tnn}, \{T^{(i)}_{j,k}\}_{\tnn} \rgen.
\label{eq:spinfulc}
\end{equation}
Similar to the case discussed in App.~\ref{subsec:spincomplexmu}, hopping terms of all ranges as well as terms such as $K_j - K_k$ are generated from the nearest-neighbor terms as a consequence of Eq.~(\ref{eq:generatelongrangespin}) and the last of Eq.~(\ref{eq:othercommsspin1}).
The Lie algebra generated by the terms in Eq.~(\ref{eq:spinfulc}) is then $\mathfrak{su}(N)$, the algebra of all \textit{traceless} matrices $A$ in Eq.~(\ref{eq:TAdefinitionspin}) with $A^\uparrow = A^\downarrow$.
The $N^2-1$ terms in the Lie algebra are then generators of the group $SU(N)$, and the bond algebra $\mA_{c}$ is the enveloping algebra $\mU(\mathfrak{su}(N))$.
Since the bond algebra $\mA_c$ is a subalgebra of $\mA_{c,\mu}$, the commutant $\mC_c$ is at least $\mC_{c,\mu}$, i.e., it contains the algebra generated by the spin-$SU(2)$ generators $\{S^\alpha_{\tot}\}$ and the particle number operator $N_{\tot}$.
However, the full commutant is larger, which is evident by studying the singlets of $\mA_{c}$.
Although the singlets of $\mA_c$ are the same as those of $\mA_{c,\mu}$ in App.~\ref{subsec:spincomplexmu}, i.e., they are given by the ferromagnetic multiplet $\{(S^-_{\tot})^n\ket{F}\}$, the vacuum $\ket{\Omega}$, and the antivacuum $\sket{\bar{\Omega}}$, they are \textit{degenerate} here, since all of the generators of $\mA_c$ ($\{T^{(c)}_{j,k}\}_{\tnn}$) vanish on these states.
This implies that operators such as $\ketbra{\bar{\Omega}}{\Omega} = \prod_j{\cd_{j,\uparrow} \cd_{j,\downarrow}}$, $\ketbra{F}{\Omega} = \prod_j{\cd_{j,\uparrow}} (1 - n_{j,\downarrow})$, and their Hermitian conjugates are also part of the commutant $\mC_c$.
These are not part of the algebra $\mC_{c,\mu}$ since they do not commute with the operators $N_\text{tot}, \vec{S}_{\tot}^2$, which are in the center of the algebra $\mC_{c,\mu}$. 
The full commutant is thus given by
\begin{equation}
    \mC_c = \lgen \{S^\alpha_{\tot}\}, N_{\tot}, \ketbra{\bar{\Omega}}{\Omega}, \ketbra{F}{\Omega} \rgen.
\end{equation}
It is easy to see that the addition of $\ketbra{\bar{\Omega}}{\Omega}$ and $\ketbra{F}{\Omega}$ (and their Hermitian conjugates implicit in our $\lgen \dots \rgen$) to the list of generators is sufficient to ensure that all other ``ket-bra" operators of singlets are part of the algebra. 
Unlike the case with chemical potential terms, we do not find any useful ``group" interpretation of the commutant $\mC_{c}$, although $SU(2) \times U(1)$ is an obvious subgroup generated by $\{S^\alpha_{\tot}\}$ and $N_{\tot}$.
\subsubsection{Center and Casimir Relations}
Note the commutant $\mC_c$ contains additional generators $\ketbra{\bar{\Omega}}{\Omega}$ and $\ketbra{F}{\Omega}$ compared to the previous case $\mC_{c,\mu}$.
Hence we expect that the center $\mZ_{c}$ will be smaller than $\mZ_{c,\mu}$ in the previous case, and we expect it to be generated by some polynomials of $\vec{S}^2_{\tot}$ and $N_{\tot}$ (in order to ensure that operators in $\mZ_c$ commute with $\{S^\alpha_{\tot}\}$ and $N_{\tot}$).
Consider such a polynomial $P(\vec{S}_\tot^2, N_\tot) \in \mZ_c$, $P(x,y) = \sum_{m,n; m+n>0} a_{m,n} x^m y^n$, where we have excluded the identity $x^0 y^0$ which is trivially in the center.
Since $\ket{\Omega}$, $\sket{\bar{\Omega}}$, $\ket{F}$ are all eigenstates of both $\vec{S}_\tot^2$ and $N_\tot$ with eigenvalues $(\vec{S}_\tot^2, N_\tot) = (0,0), (0,2N), (\frac{N}{2}(\frac{N}{2}+1),N)$, commutation with $\ketbra{\bar{\Omega}}{\Omega}$ and $\ketbra{F}{\Omega}$ is equivalent to requiring
\begin{equation}
    P(0,2N) = P\left(\frac{N}{2} \left(\frac{N}{2}+1 \right), N \right) = P(0,0) = 0 ~.
\label{eq:polyroots}
\end{equation}
Hence the set of operators in the center can be determined from the set of polynomials with the above three conditions.
Three lowest-degree polynomials satisfying these conditions are
\begin{equation}
P_1 \defn \vec{S}^2_{\tot}  + \frac{N+2}{4N} N_{\tot} (N_{\tot} - 2N) ~, \quad
P_2 \defn \vec{S}^2_\tot \left[\vec{S}^2_\tot - \frac{N}{2} \left(\frac{N}{2}+1 \right) \right] ~, \quad
P_3 \defn N_\tot (N_\tot - N) (N_\tot - 2N) ~.
\label{eq:poly4Zc}
\end{equation}
We conjecture that the center $\mZ_c$ is spanned by (a linearly independent subset of) operators of the form $f(\vec{S}_\tot^2, N_\tot) P_{1,2,3}(\vec{S}_\tot^2, N_\tot)$ where $f$ can be an arbitrary polynomial.
From the perspective of the bond algebra $\mA_c$, since its generators are those of $SU(N)$, its center should contain all Casimir operators of $SU(N)$, including its quadratic Casimir $C^{SU(N)}_2$. 
However, since the centers of $\mA_c$ and $\mC_c$ coincide to $\mZ_c$ of Eq.~(\ref{eq:Zc}), $C^{SU(N)}_2$ should be related to some specific combination of $C^{SU(2)}_2 \defn \vec{S}^2_{\tot}$ and $N_{\tot}$ that belongs to $\mZ_c$.
Indeed we find the relation\footnote{In Eq.~(\ref{eq:quadraticSUN}), we have obtained $C^{SU(N)}_2$ using the fact that $N_{\tot}$ corresponds to the identity $\mathds{1}_{N \times N}$ in the defining representation.
Since $SU(N)$ and $U(N)$ only differ by $\mathds{1}_{N \times N}$ in the defining representation, their Casimirs are related by $C^{U(N)}_2 = C^{SU(N)}_2 + \frac{1}{2N}\mathds{1}^2_{N \times N}$ in that representation, where the specific factor is to ensure normalization of Eq.~(\ref{eq:casimirdefn}).
}
\begin{equation}
C^{SU(N)}_2 \defn \sum_{j < k} \left[ \left(\frac{T_{j,k}^{(r)}}{2} \right)^2 + \left(\frac{T_{j,k}^{(i)}}{2}\right)^2 \right] + \sum_j \left(\frac{K_j}{\sqrt{2}}\right)^2 - \left(\frac{N_{\tot}}{\sqrt{2N}}\right)^2 = \frac{N+2}{4N} N_\tot (2N - N_\tot) - \vec{S}_\tot^2 = -P_1,
%\implies C^{SU(N)}_2 + C^{SU(2)}_2 = \frac{N+2}{4N}(N^2 - Q^2),
\label{eq:quadraticSUN}
\end{equation}
where $P_1$ is defined in Eq.~(\ref{eq:poly4Zc}).
Defining a $U(1)$ charge $Q := N_\tot - N$, and the $SU(2)$ Casimir operator $C^{SU(2)}_2 \defn \vec{S}^2_{\tot}$, we can rewrite Eq.~(\ref{eq:quadraticSUN}) as
\begin{equation}
C^{SU(N)}_2 + C^{SU(2)}_2 = \frac{N+2}{4N}(N^2 - Q^2),
\end{equation}
which is the Casimir relation derived in \cite{klebanov2018spectra, gaitan2020hagedorn}.
In this case, the center $\mZ_c$ is clearly smaller than the Casimir algebra of $SU(2) \times U(1)$, given by $\mZ_{SU(2) \times U(1)} = \lgen \vec{S}^2_{\tot}, N_{\tot} \rgen$.
Hence, as discussed in Sec.~\ref{subsec:groupdecomp}, the Hilbert space partitioning in terms of the irreducible representations of the group $SU(N) \times (SU(2) \times U(1))$ is different from the decomposition in terms of bond and commutant algebras. 
This group partitioning misses some properties of Hamiltonians from the bond algebra, e.g., it is oblivious to the fact that the vacuum $\ket{\Omega}$ and the antivacuum $\sket{\bar{\Omega}}$ are degenerate with each other and the ferromagnet $\ket{F}$.
\subsubsection{Double Commutant Theorem}\label{subsubsec:DCTcomplexspin}
We now discuss the application of the DCT of Thm.~\ref{thm:dct} to this case, which closely follows the discussion in App.~\ref{subsubsec:DCTcomplexhopping}.
Following similar arguments, it is easy to show that local operators within $\mA_c$ are those in $\mA_{c,\mu}$ that w.l.o.g. annihilate the states $\ket{\Omega}$, $\sket{\bar{\Omega}}$, and $\ket{F}$. 
Since these are product states, strictly local operators $\hO_R$ with support in a region $R$ that annihilate these states also annihilate the states $\ket{\Omega}_R$, $\sket{\bar{\Omega}}_R$, and $\ket{F}_R$, which are the restrictions of the three states to the region $R$.
Applying the DCT to operators in region $R$, $\hO_R$ should belong to the centralizer of the algebra $\lgen \{S^\alpha_{R,\tot}\}, N_{R, \tot} , (\ketbra{\bar{\Omega}}{\Omega})_R, (\ketbra{F}{\Omega})_R\rgen$, where $S^\alpha_{R,\tot} \defn \sum_{j \in R}{S^\alpha_j}$ and $N_{R,\tot} \defn \sum_{j \in R}{n_j}$. 
For contiguous regions $R$, this is the algebra generated by nearest-neighbor complex hopping terms in $\mA_c$ restricted to the region $R$.
Turning to sums of local (or extensive local) operators that are in $\mA_{c,\mu}$ and are of the form $\hO_{\text{ext-loc}} = \sum_R{\hA_R}$ for contiguous regions $R$, we can express $\hA_R$ as polynomials in the generators of $\mA_{c,\mu}$, i.e., $\hA_R = f_R(\{K_j\}_R, \{T^{(\ast)}_{j,k}\}_R)$.
While here too we are not able to make any general statements on arbitrary $\hO_{\text{ext-loc}}$, we can use the fact that states $\ket{\Omega}$, $\sket{\bar{\Omega}}$, and $\ket{F}$ are singlets of $\mA_{c,\mu}$ and follow arguments similar to those in App.~\ref{subsubsec:DCTcomplexspin} to show that when the polynomial $f$ is $R$-independent (e.g., when $\hO_{\text{ext-loc}}$ is translation-invariant), $\hO_{\text{ext-loc}}$ vanishes on these states only if each $\hA_R$ annihilates these. 
Hence all translation invariant extensive local operators in $\mA_c$ are linear combinations of strictly local operators in $\mA_c$ considered above.
\subsection{Pure imaginary hoppings}\label{subsec:spinimaginary}
\subsubsection{Bond and Commutant Algebra}
We further remove the real hoppings from the list of generators of $\mA_c$ in Eq.~(\ref{eq:spinfulc}), and study the algebra
\begin{equation}
    \mA_i \defn \lgen \{T^{(i)}_{j,k}\}_{\tnn} \rgen.
\label{eq:spinfulimag}
\end{equation}
According to Eq.~(\ref{eq:generatelongrangespin}), purely imaginary hoppings of all ranges are generated starting from nearest-neighbor terms.
The corresponding Lie algebra is then $\mathfrak{so}(N)$, the algebra of all purely imaginary antisymmetric matrices $A$ in Eq.~(\ref{eq:TAdefinitionspin}) with $A^\uparrow = A^\downarrow$.
The $\binom{N}{2}$ terms in the Lie algebra are then generators of the group $SO(N)$, and the bond algebra $\mA_{i}$ is the enveloping algebra $\mU(\mathfrak{so}(N))$.
Since $\mA_i$ is contained in $\mA_c$ of Eq.~(\ref{eq:spinfulc}), its commutant $\mC_i$ contains the algebra $\mC_c$.
In addition to the spin-$SU(2)$ and particle number symmetries generated by $\{S^\alpha_{\tot}\}$ and $N_{\tot}$ (which are a part of $\mC_c$), $\mC_i$ also contains a ``pseudospin" $SU(2)$ symmetry~\cite{yang1990so,mark2020eta} generated by operators $\eta^x_0$, $\eta^y_0$, and $\eta^z_0$, defined in Eq.~(\ref{eq:etaalphadefns}).
Further, similar to the spinless case, $\mC_i$ contains two obvious $Z_2$ symmetries $Q^\uparrow_X$ and $Q^\downarrow_X$ defined in Eq.~(\ref{eq:QXQXtdefnsspin}).
However, these symmetries are not independent in the presence of $\vec{S}_{\tot}$ conservation, since they can be related by a spin rotation around the $\hat{y}$-axis, i.e., $e^{-i \pi S^y_\tot} Q^\uparrow_X e^{i \pi S^y_\tot} = Q_X^\downarrow$.\footnote{This can be verified using the on-site spin rotation identity
$e^{i \phi \vec{n} \cdot \vec{S}_j}
\begin{pmatrix} 
c_{j,\uparrow} \\ c_{j,\downarrow} \end{pmatrix} 
e^{-i \phi \vec{n} \cdot \vec{S}_j}= \exp\left( \frac{-i \phi \vec{n} \cdot \vec{\sigma}}{2} \right)
\begin{pmatrix} 
c_{j,\uparrow} \\ c_{j,\downarrow} 
\end{pmatrix}$
where $\vec{n}$ is any unit vector, $\vec{\sigma}$ is the vector of Pauli matrices, and $\vec{S}_j$ are the spin-1/2 operators on site $j$.
Substituting $\phi = -\pi$ and $\vec{n} = \hat{y}$ results in the transformations $c_{j,\uparrow} \rightarrow c_{j,\downarrow}$ and $c_{j,\downarrow} \rightarrow -c_{j,\uparrow}$.}
In a similar way we can obtain a particle-hole tranformation acting only on fermions in a definite polarization along any other spin quantization direction.
The full commutant is given by 
\begin{equation}
    \mC_i = \lgen \{\eta^\alpha_0\}, \{S^\alpha_{\tot}\}, \{Q^\sigma_X\} \rgen.
\label{eq:Ciapp}
\end{equation}
Since $\{\eta^\alpha_0\}$ and $\{S^\alpha_{\tot}\}$ generate $SU(2)$ groups and commute with each other, $\mC_i$ has an $SU(2) \times SU(2)$ subgroup.
Further, $\mC_i$ has other $Z_2$ subgroups generated by $Q^\uparrow_X$ and $Q^\downarrow_X$ (or variants for other quantization directions). 
Moreover, the full commutant can also be viewed as being generated by the full family of on-site unitaries corresponding to these groups, i.e., $\exp(i \sum_{\alpha}{\theta_\alpha S^\alpha_{\tot}})$ and $\exp(i \sum_{\alpha}{\theta_\alpha \eta^\alpha_{0}})$ for the two $SU(2)$ symmetries and $Q^\sigma_X$ of Eq.~(\ref{eq:QXQXtdefnsspin}) for the two $Z_2$ symmetries.
This property is useful for the application of the DCT, as discussed in Sec.~\ref{subsec:spinfulfermions}.
However, $\{Q^\sigma_X\}$ do not commute with either of the $SU(2)$ generators (nor among themselves for odd $N$), hence the full group structure of $\mC_i$ is more involved, and here we do not attempt to define it precisely (though one can certainly write out the groups generated by the identified on-site unitary symmetries explicitly\footnote{On a single site, $N=1$, $\{S^\alpha\}$ generates $SU(2)$ acting within the states $\ket{\uparrow}$ and $\ket{\downarrow}$; $\{\eta^\alpha\}$ generates $SU(2)$ acting within $\sket{\bar{\Omega}}$ and $\ket{\Omega}$; while the unitary $Q^\downarrow$ swaps the two blocks up to overall signs, $\ket{\uparrow} \leftrightarrow \sket{\bar{\Omega}}, \ket{\downarrow} \leftrightarrow \ket{\Omega}$.
The generated group is then isomorphic to that of $4 \times 4$ matrices of the form $\begin{pmatrix} U_1 & 0 \\ 0 & U_2 \end{pmatrix}$ and $\begin{pmatrix} 0 & U_3 \\ U_4 & 0 \end{pmatrix}$ with $U_{1,2,3,4} \in SU(2)$.}
).
Note also that the listed generators of $\mC_i$ are not a minimal set:
Besides dropping one of the $Q^\sigma_X$ in the presence of $\{S^\alpha_{\tot} \}$, we could in fact also drop $\{\eta^\alpha_0\}$ since they can be generated as 
\begin{equation}
    Q^\downarrow_X S^{x,y}_{\tot} (Q^\downarrow_X)^\dagger = -\eta_0^{x,y}, ~ Q^\downarrow_X S^z_{\tot} (Q^\downarrow_X)^\dagger = \eta_0^z\quad \text{or}\quad Q^\uparrow_X S^{x,z}_{\tot} (Q^\uparrow_X)^\dagger = -\eta_0^{x,z}, ~ Q^\uparrow_X S^y_{\tot} (Q^\uparrow_X)^\dagger = \eta_0^y. 
\label{eq:etaSrelation}
\end{equation}
$\mC_i$ as defined also contains ``ket-bra" operators formed from the singlets of $\mA_i$, which are the states of the ferromagnetic tower $\{(S^-_{\tot})^n\ket{F}\}$ and those of the ``vacuum tower" $\{(\ed_0)^n\ket{\Omega}\}$ built upon the vacuum state (we will also refer to the latter as ``$\eta$-pairing'' states, as they are zero-momentum versions of C.~N.~Yang's $\eta$-pairing states~\cite{yang1989eta}).
It is easy to verify that all the singlets are annihilated by $\{T^{(i)}_{j,k}\}_{\tnn}$, hence they are degenerate.  
However, operators such as $\ketbra{F}{\Omega}$ and $\ketbra{\bar{\Omega}}{\Omega}$ that needed to be explicitly included in the list of generators of $\mC_{c}$ are implicitly part of $\mC_i$ as defined. 
To see this, note that $\ketbra{\Omega}{\Omega} \in \lgen N_{\tot}\rgen \subseteq \mC_i$.
Hence $\ketbra{F}{\Omega} = Q^\uparrow_X \ketbra{\Omega}{\Omega} \in \mC_i$ and $\ketbra{\bar{\Omega}}{\Omega} = (\eta^\dagger_0)^N\ketbra{\Omega}{\Omega} \in \mC_i$, since $Q^\uparrow_X$ and $\eta^\dagger_0$ are in $\mC_i$.
Note that if one adds on-site chemical potential terms $\{K_j\}$ to the list of generators of $\mA_i$, all real hopping terms are generated according to Eq.~(\ref{eq:othercommsspin1}), hence the resulting bond algebra is identical to $\mA_{c,\mu}$ of Eq.~(\ref{eq:spinfulcmu}).
\subsubsection{Center and Casimir Relations}
Note that the commutant $\mC_i$ contains both the spin and pseudospin $SU(2)$ symmetries, hence we might expect its center to consist of the corresponding Casimir operators $\vec{S}^2_{\tot}$ and $\vec{\eta}^2_0$. 
However, these operators do not commute with the operators $\{Q^\sigma_X\}$, as evident from Eq.~(\ref{eq:etaSrelation}).
Nevertheless, it is easy to verify that a linear combination of these Casimir operators $(\vec{S}^2_{\tot} + \vec{\eta}^2_0)$ does commute with both $Q^\uparrow_X$ and $Q^\downarrow_X$.
Hence we conjecture that 
\begin{equation}
    \mZ_i = \lgen (\vec{S}^2_{\tot}  + \vec{\eta}^2_0)\rgen.
\label{eq:Zi}
\end{equation}
From the perspective of the bond algebra $\mA_i$, since it consists of the generators of $SO(N)$, its center should contain all Casimir operators of $SO(N)$, including its quadratic Casimir $C^{SO(N)}_2$. 
Similar to the previous cases, since the centers of $\mA_i$ and $\mC_i$ coincide to $\mZ_i$, we expect that $C^{SO(N)}_2$ should be expressible in terms of a polynomial of $(\vec{S}^2_{\tot} + \vec{\eta}^2_0)$ that generates $\mZ_i$. 
Indeed, we find the relation
\begin{equation}
    C^{SO(N)}_2 \defn \sum_{j<k} \left(\frac{T_{jk}^{(i)}}{2} \right)^2 =  \frac{N}{4}\left(\frac{N}{2}+1 \right) - \frac{1}{2}\left(\vec{S}_\tot^2 + \vec{\eta}^2_0\right),
\label{eq:quadraticSON}
\end{equation}
Defining the quadratic Casimir operator for the group $SU(2) \times SU(2)$ (which is identical to that for the group $SO(4)$) as $C^{SO(4)}_2 \defn \frac{1}{2} (\vec{S}_\tot^2 + \vec{\eta}^2_0)$,\footnote{This definition can be verified in its defining representation, where the generators are of the form $\{R\, \sigma^\alpha\otimes \mathds{1}_{2 \times 2} \, R^\dagger, R\, \mathds{1}_{2 \times 2} \otimes \sigma^\alpha \, R^\dagger\}$, where $\sigma^\alpha$ are the Pauli matrices and $R$ is a fixed $4 \times 4$ unitary matrix~\cite{fujii2007more}.} we can rewrite Eq.~(\ref{eq:quadraticSON}) as
\begin{equation}
    C^{SO(N)}_2 + C^{SO(4)}_2 = \frac{N}{4}\left(\frac{N}{2} + 1\right),
\end{equation}
which is identical to the Casimir relations derived in \cite{klebanov2018spectra, gaitan2020hagedorn}.
In this case, the center $\mZ_i$ is clearly smaller than the Casimir algebra of $SU(2) \times SU(2)$, which is same as the Casimir algebra of $SO(4)$, given by $\mZ_{SO(4)} = \lgen \vec{S}^2_{\tot}, \vec{\eta}^2_{0} \rgen$.
Hence, as discussed in App.~\ref{app:partitioning}, the Hilbert space partitioning in terms of the irreducible representations of the group $SO(N) \times (SU(2) \times SU(2))$ [equivalent to $SO(N) \times SO(4)$ when $N$ is even] is different from the decomposition in terms of the bond and commutant algebras. 
This group partitioning misses some properties of Hamiltonians from the bond algebra, e.g., it is oblivious to the fact that the $\eta$-pairing tower of states $\{(\ed)^n\ket{\Omega}\}$ and the ferromagnetic tower $\{(S^-_{\tot})^n\ket{F}\}$ are degenerate with each other.
\subsection{Hoppings with chemical potentials and magnetic fields}
We now consider the bond algebra with on-site magnetic field terms $\{M_j\}$ added to the list of generators of $\mA_{c,\mu}$ in Eq.~(\ref{eq:spinfulcmu}), i.e.,
\begin{equation}
    \mA_{c,\mu,h} \defn \lgen \{T^{(\Upsilon})_{j,k}\}_{\tnn}, \{K_j\}, \{M_j\}\rgen,\;\; \Upsilon \in \{r, i, c\}.
\label{eq:Acmuh}
\end{equation}
Using Eqs.~(\ref{eq:generatelongrangespin})-(\ref{eq:othercommsspin2}), it is easy to see that all possible spin and particle number conserving quadratic terms are generated, and the Lie algebra generated is $\fu(N) \oplus \fu(N)$, that of the terms of Eq.~(\ref{eq:TAdefinitionspin}) with independent $A^\uparrow$ and $A^\downarrow$.
These terms can hence be viewed as generators of $U(N) \times U(N)$, and the bond algebra is the corresponding enveloping algebra $\mU(\fu(N) \oplus \fu(N))$.
Moreover, since the addition of $\{M_j\}$ generates terms such as $\{\wT^{(c)}_{j,k}\}$ (as a consequence of Eq.~(\ref{eq:othercommsspin2})), we can alternatively use generators with decoupled hopping terms of the two spins $\uparrow$ and $\downarrow$, which can be expressed as linear combinations of $\{T^{(c)}_{j,k}\}$ and $\{\wT^{(c)}_{j,k}\}$.
Hence the resulting bond algebra can also be viewed as (a product of) two copies  of the bond algebra for spinless fermions discussed in App.~\ref{subsec:spinlesscomplexchemical}.
The presence of on-site magnetic fields $\{M_j\}$ breaks the spin-$SU(2)$ symmetry in $\mC_{c,\mu}$ discussed in App.~\ref{subsec:spincomplexmu} to a $U(1)$ subgroup.
Hence the full commutant here is given by
\begin{equation}
    \mC_{c,\mu,h} = \lgen S^z_{\tot}, N_{\tot}\rgen.
\end{equation}
Since $S^z_{\tot}$ and $N_{\tot}$ generate $U(1)$ groups and commute with each other, this commutant is associated with the group $U(1) \times U(1)$. 
Moreover, the full commutant can also be viewed as being generated by the full family of on-site unitaries corresponding to the two $U(1)$ groups, i.e., $\exp(i \alpha S^z_{\tot})$ and $\exp(i \beta N_{\tot})$.
This property is useful for the application of the DCT, since Lems.~\ref{lem:strloc} and \ref{lem:typeI} apply.
Note that the commutant can also be written as $\mC_{c,\mu,h} = \lgen N^\uparrow_{\tot}, N^\downarrow_{\tot}\rgen$, where $N^\sigma_{\tot}$ is the total number operator for particles of spin $\sigma$, satisfying $N^\uparrow_{\tot} = (N_{\tot} + 2 S^z_{\tot})/2$ and $N^\downarrow_{\tot} = (N_{\tot} - 2 S^z_{\tot})/2$.
This shows that the commutant can also be viewed as (a product of) two copies of the commutant for spinless fermions discussed in App.~\ref{subsec:spinlesscomplexchemical}.
$\mA_{c,\mu, h}$ admits four singlets, which are the states $\ket{F}$, $\sket{\bar{F}}$, $\ket{\Omega}$, $\sket{\bar{\Omega}}$, and it is easy to see that projectors onto these singlets are a part of $\mC_{c,\mu,h}$ as defined.
They are all non-degenerate since they differ in the eigenvalues of the $\{M_j\}$ or $\{K_j\}$ operators. 
Since the commutant $\mC_{c, \mu,h}$ is evidently Abelian, the center in this case is equal to the commutant itself, i.e., $\mZ_{c, \mu, h} = \lgen S^z_{\tot}, N_{\tot} \rgen$.
\subsection{Complex hoppings with magnetic fields and without chemical potentials}\label{subsec:spincomplexmag}
\subsubsection{Bond and Commutant Algebras}
We now consider complex hopping terms in the presence of on-site magnetic fields $\{M_j \}$, exclude the chemical potential terms $\{K_j\}$, and study the algebra
\begin{equation}
    \mA_{c,h} \defn \lgen \{T^{(c)}_{j,k}\}_{\tnn}, \{M_j\} \rgen = \lgen \{T^{(r)}_{j,k}\}_{\tnn}, \{T^{(i)}_{j,k}\}_{\tnn}, \{M_j\} \rgen.
\label{eq:Ach}
\end{equation}
Using Eqs.~(\ref{eq:generatelongrangespin}) and (\ref{eq:othercommsspin2}), it is easy to see that all complex hoppings of the form $\{T^{(c)}_{j,k}\}$ and $\{\wT^{(c)}_{j,k}\}$ are generated starting from the elementary terms $\{T^{(c)}_{j,k}\}_{\tnn}$ and $\{M_j\}$, and also independent $K_j - K_k$ terms.
The Lie algebra generated then consists of $2N^2 - 1$ terms and is that of \textit{traceless} matrices of the form of Eq.~(\ref{eq:TAdefinitionspin}) (i.e., with the constraint $\Tr(A^\uparrow) + \Tr(A^\downarrow) = 0$), given by $(\fu(N) \oplus \fu(N))/\fu(1)$.
These terms can also be viewed as generators of $(U(N) \times U(N))/U(1)$, and the bond algebra is the corresponding enveloping algebra $\mU((\fu(N) \oplus \fu(N))/\fu(1))$.
Since $\mA_{c,h}$ is contained in $\mA_{c,\mu,h}$, its commutant $\mC_{c,h}$ contains $\mC_{c,\mu,h} = \lgen S^z_{\tot}, N_{\tot} \rgen$ but is larger, and this is evident from the properties of the singlets of $\mA_{c,h}$.
The bond algebra $\mA_{c, h}$ has the same four singlets as $\mA_{c,\mu,h}$, i.e., $\ket{F}$, $\sket{\bar{F}}$, $\ket{\Omega}$, $\sket{\bar{\Omega}}$.
However, in this case, $\ket{\Omega}$ and $\sket{\bar{\Omega}}$ are degenerate, since they do not differ under the eigenvalues of $\{M_j\}$. 
This degeneracy implies that operators $\ketbra{\bar{\Omega}}{\Omega} = \prod_j{\cd_{j,\uparrow} \cd_{j,\downarrow}}$ 
and its Hermitian conjugate are part of $\mC_{c,h}$.
Hence the full commutant is given by 
\begin{equation}
    \mC_{c,h} = \lgen S^z_{\tot}, N_{\tot}, \ketbra{\bar{\Omega}}{\Omega}\rgen,
\end{equation} 
where the closure under Hermitian conjugation is implicit in the notation $\lgen \cdots \rgen$.  
Similar to many other commutants discussed in this Appendix, we do not find any useful group interpretation of this commutant $\mC_{c,h}$.
\subsubsection{Center}
Note that the commutant $\mC_{c,h}$ contains the additional generator $\ketbra{\bar{\Omega}}{\Omega}$ compared to the previous case $\mC_{c,\mu,h}$.
Hence we expect that the center $\mZ_{c,h}$ will be smaller than $\mZ_{c,\mu,h}$ in the previous case.
Since $S^z_{\tot}$ commutes with $\ketbra{\bar{\Omega}}{\Omega}$ and $N_{\tot}$, it is one of the generators of $\mZ_{c,h}$, and the other elements of the center can be determined by demanding that they commute with $\ketbra{\bar{\Omega}}{\Omega}$.
Following the discussion for the center $\mZ_c$ in App.~\ref{subsec:spinlesscomplex}, we can show that
\begin{equation}
    \mZ_{c,h} = \lgen S^z_{\tot}, \{N_{\tot}^\alpha (2N - N_{\tot})\;:\;\alpha \geq 1\}\rgen,
\end{equation}
and we are not able to determine a more compact expression for the minimal set of generators.
\subsubsection{Double Commutant Theorem}\label{subsubsec:DCTspincomplexmag}
We now discuss the application of the DCT of Thm.~\ref{thm:dct} to this case.
Following similar arguments as in Apps.~\ref{subsubsec:DCTcomplexhopping} and \ref{subsubsec:DCTcomplexspin}, it is easy to show that local operators within $\mA_{c,h}$ are those in $\mA_{c,\mu,h}$ that w.l.o.g. annihilate the states $\ket{\Omega}$ and $\sket{\bar{\Omega}}$. 
Since these are product states, strictly local operators $\hO_R$ with support in a region $R$ that annihilate these states also annihilate the states $\ket{\Omega}_R$, $\sket{\bar{\Omega}}_R$, which are the restrictions of these states to the region $R$.
Applying the DCT to operators in a contiguous region $R$, we can show that any strictly local operator $\hO_R$ should be within the algebra generated by nearest-neighbor generators in $\mA_{c,h}$ restricted to the region $R$.
Turning to sums of local (or extensive local) operators $\hO_{\text{ext-loc}} \in \mA_{c,h}$ and noting that the states $\ket{\Omega}$ and $\sket{\bar{\Omega}}$ are singlets of $\mA_{c,\mu, h}$, we can follow arguments similar to those in Apps.~\ref{subsubsec:DCTcomplexspin} and \ref{subsubsec:DCTcomplexhopping} to show that any translation-invariant $\hO_{\text{ext-loc}}$ is a linear combination of strictly local terms that vanish on these singlet states considered above.
\subsection{Pure imaginary hoppings with magnetic fields}\label{subsec:spinfulimagh}
We further remove the real hopping terms $\{T^{(r)}_{j,k}\}_{\tnn}$ from the bond algebra of Eq.~(\ref{eq:Ach}), and we study the algebra
\begin{equation}
    \mA_{i,h} \defn \lgen \{T^{(i)}_{j,k}\}_{\tnn}, \{M_j\} \rgen.
\label{eq:Aih}
\end{equation}
Using Eqs.~(\ref{eq:generatelongrangespin}), (\ref{eq:othercommsspin2}), and (\ref{eq:generatelongrangetild}), all hoppings of the form $\{T^{(i)}_{j,k}\}$ and $\{\wT^{(r)}_{j,k}\}$ are generated starting from the elementary terms $\{T^{(i)}_{j,k}\}_{\tnn}$ and $\{M_j\}$, and it is easy to verify that they form a closed Lie algebra.
The Lie algebra generated consists of $N^2$ terms and is that of matrices of the form of Eq.~(\ref{eq:TAdefinitionspin}) with the additional constraint that $A^\downarrow = - (A^\uparrow)^\ast$.
Since the $A^\downarrow$ is uniquely determined by $A^\uparrow$, the Lie algebra of such matrices is $\fu(N)$, the Lie algebra of $N \times N$ Hermitian matrices.
These terms can also be viewed as generators of $U(N)$, and the bond algebra is the corresponding enveloping algebra $\mU(\fu(N))$.
Note that this is isomorphic to the case discussed in App.~\ref{subsec:spincomplexmu}, and this can be precisely established with the transformation $c_{j,\downarrow} \leftrightarrow \cd_{j,\downarrow}$.
This transformation leaves $T^{(i)}_{j,k}$ invariant, maps $\wT^{(r)}_{j,k} \leftrightarrow T^{(r)}_{j,k}$ and $M_j \leftrightarrow (K_j - 1)$. 
Hence the bond algebra of Eq.~(\ref{eq:Aih}) maps to $\lgen \{T^{(i)}_{j,k}\}_{\tnn}, \{K_j\} \rgen$, which is the same as $\mA_{c,\mu}$ of Eq.~(\ref{eq:spinfulcmu}).
Continuing in the representation of this subsection, since $\mA_{i,h}$ is contained in $\mA_{c,h}$, its commutant $\mC_{i,h}$ contains $\mC_{c,h}$, but is larger.
The absence of real hopping terms in the generators of $\mA_{i,h}$ restores the pseudospin $SU(2)$ symmetry generated by the operators $\{\eta^\alpha_0\}$ defined in Eq.~(\ref{eq:etaalphadefns}).
Hence the complete commutant is given by 
\begin{equation}
    \mC_{i,h} = \lgen S^z_{\tot}, \{\eta^\alpha_0\} \rgen,
\end{equation}
and this also contains the ``ket-bra" operators formed using degenerate singlets. 
$\mA_{i, h}$ has two singlets $\ket{F}$ and $\sket{\bar{F}}$, as well as the eta-pairing tower of degenerate states $\{(\ed_0)^n \ket{\Omega}\}$, and the former two and the tower are all non-degenerate with each other. 
As discussed in App.~\ref{subsec:spinimaginary}, the ``ket-bra" operators within the eta-pairing states are a part of $\lgen \{\eta^\alpha_0\}\rgen$, and the projectors onto the other singlets are a part of $\lgen S^z_{\tot}\rgen$.
Since $S^z_{\tot}$ and $\{\eta^\alpha_0\}$ in $\mC_{i,h}$ are the generators of $U(1)$ and $SU(2)$ groups, the commutant $\mC_{i,h}$ is associated with the group $U(1) \times SU(2)$.
This is also evident from the isomorphism between the bond algebras $\mA_{c,\mu}$ and $\mA_{i,h}$, which implies that the commutants $\mC_{c,\mu}$ and $\mC_{i,h}$ are also isomorphic. 
Indeed, the transformation $c_{j,\downarrow} \leftrightarrow \cd_{j,\downarrow}$ interchanges the spin and pseudospin operators, i.e., $\{S^\alpha_{\tot}\} \leftrightarrow \{\eta^\alpha_0\}$ (up to unimportant signs when this transformation is implemented via $Q_X^\downarrow$, see Eq.~(\ref{eq:etaSrelation})), hence mapping the commutants $\mC_{c,\mu} = \lgen \{S^\alpha_{\tot}\}, \eta^z_0\} \rgen \leftrightarrow \lgen S^z_{\tot}, \{\eta^\alpha_0\} \rgen = \mC_{i,h}$.
Hence, similar to $\mC_{c,\mu}$, the full commutant $\mC_{i,h}$ can also be viewed as being generated by the full family of on-site unitaries corresponding to the $U(1)$ and $SU(2)$ groups, a property useful for the application of the DCT, as discussed in Sec.~\ref{subsec:spinfulfermions}.
Similar to the case of $\mC_{c,\mu}$, since the commutant $\mC_{i, h}$ consists of ``independent" pseudospin $SU(2)$ and spin $U(1)$ symmetries, the center $\mZ_{i, h}$ is generated by $\vec{\eta}^2_0$ (the Casimir element of the psueodspin $SU(2)$ symmetry) and $S^z_{\tot}$ (the generator of the spin $U(1)$ symmetry). 
Hence, we obtain 
\begin{equation}
    \mZ_{i, h} = \lgen S^z_{\tot}, \vec{\eta}^2_0 \rgen.
\end{equation} 
Further, we note that the due to the isomorphism of the bond and commutant algebras in case with the ones in App.~\ref{subsec:spincomplexmu}, the Casimir relations discussed there (particularly Eq.~(\ref{eq:quadraticUN})) is also valid here after the appropriate transformations.
\subsection{Real hoppings}\label{subsec:spinfulreal}
Finally we comment on bond algebras similar to $\mA_i$ and $\mA_{i,h}$ of Eqs.~(\ref{eq:spinfulimag}) and (\ref{eq:Aih}), with the imaginary hopping terms $\{T^{(i)}_{j,k}\}_{\tnn}$ substituted by real hoppings $\{T^{(r)}_{j,k}\}_{\tnn}$, given by
\begin{equation}
    \mA_r \defn \lgen \{T^{(r)}_{j,k}\}_{\tnn} \rgen,\;\;\;\mA_{r,h} \defn \lgen \{T^{(r)}_{j,k}\}_{\tnn}, \{M_j\} \rgen.
\label{eq:Ars}
\end{equation}
Similar to the case of spinless fermions with real hoppings discussed in App.~\ref{subsec:spinlessreal}, on a non-bipartite lattice, it is easy to see that all complex hoppings $\{T^{(c)}_{j,k}\}$ are eventually generated, and the algebras $\mA_r$  and $\mA_{r,h}$ are equal to $\mA_c$ and $\mA_{c,h}$ of Eqs.~(\ref{eq:spinfulc}) and (\ref{eq:Ach}) respectively.
However, on a bipartite lattice, distinct closed algebras are generated, which can be derived using Eqs.~(\ref{eq:generatelongrangespin})-(\ref{eq:generatelongrangetild}).
The Lie algebra generated by $\{T^{(r)}_{j,k}\}_{\tnn}$ consists of real hoppings $\{T^{(r)}_{j,k}\}$ between different sublattices and imaginary hoppings $\{T^{(i)}_{j,k}\}$ within the same sublattices.
Further, the addition of on-site magnetic field terms $\{M_j\}$ also results in the generation of hopping terms such as $\{\wT^{(i)}_{j,k}\}$ between different sublattices and $\{\wT^{(r)}_{j,k}\}$ within the same sublattice.
These algebras are isomorphic to the algebras generated using imaginary hoppings, and this can be understood via a transformation $c_{j,\sigma} \to i c_{j,\sigma}$ for $\sigma \in \{\uparrow, \downarrow\}$ on one of the sublattices.
Under this transformation, $\{T^{(r)}_{j,k}\}_{\tnn} \rightarrow \{T^{(i)}_{j,k}\}_{\tnn}$ and $M_j$ remains invariant, hence the bond algebras of Eq.~(\ref{eq:Ars}) transform as $\mA_r \rightarrow \mA_i$ and $\mA_{r,h} \rightarrow \mA_{i,h}$, defined in Eqs.~(\ref{eq:spinfulimag}) and (\ref{eq:Aih}) respectively.
Note that $\mA_{i,h}$ is in turn isomorphic to $\mA_{c,\mu}$ of Eq.~(\ref{eq:spinlesscmu}), as discussed in App.~\ref{subsec:spinfulimagh}.
The isomorphism of the bond algebras also implies an isomorphism of the respective commutants $\mC_r$ and $\mC_{r,h}$ of $\mA_r$ and $\mA_{r,h}$ to $\mC_i$ and $\mC_{i,h}$ of $\mA_i$ and $\mA_{i,h}$.
While the transformation preserves the spin-$SU(2)$ generators $\{S^\alpha_{\tot}\}$, the $\ed_0$ and $\eta_0$ operators get mapped onto the more familiar $\eta^\dagger_{\pi}$ and $\eta_{\pi}$ studied in the context of the Hubbard model~\cite{yang1989eta, essler2005one, moudgalya2020eta}, defined in Eq.~(\ref{eq:etaalphadefns}).
The pseudospin $SU(2)$ is then generated by $\{\eta^\alpha_{\pi}\}$, and the singlets of $\mA_r$ and $\mA_{r,h}$ can be obtained using those of $\mA_{i}$ and $\mA_{i,h}$  with the aforementioned substitution.
Further, the operator $Q^\sigma_X$ maps to the operator $\tQ^\sigma_X$, where both are defined in Eq.~(\ref{eq:QXQXtdefnsspin}).
Hence the full commutants are given by 
\begin{equation}
    \mC_{r} = \lgen \{S^\alpha_{\tot}\}, \{\eta^\alpha_\pi\}, \{\tQ^\sigma_X\}\rgen,\;\;\;
    \mC_{r,h} = \lgen S^z_{\tot}, \{\eta^\alpha_\pi\}\rgen.
\end{equation}
We can obtain the centers $\mZ_r$ and $\mZ_{r,h}$ by applying the sublattice mapping to the centers of $\mZ_i$ and $\mZ_{i,h}$ respectively.
Hence we obtain $\mZ_r = \lgen \vec{S}^2_{\tot} + \vec{\eta}^2_\pi\rgen$ and $\mZ_{r,h} = \lgen S^z_{\tot}, \vec{\eta}^2_\pi\rgen$, where $\vec{\eta}^2_{\pi}$ is the Casimir operator the $SU(2)$ symmetry generated by $\{\eta^\alpha_{\pi}\}$. 

\section{Group and Algebra Partitionings of Fermionic Hilbert Spaces}
\label{app:partitioning}
In this Appendix, we provide details on the connection between group partitionings and algebra partitionings of the Hilbert space, discussed in Sec.~\ref{subsec:groupdecomp}.
Given a bond algebra $\mA$ that corresponds to a Lie group $G_1$ and a commutant algebra $\mC$ that corresponds to a Lie group $G_2$, the relations between the group and algebra decompositions depend on the relations between the Casimir algebras $\mZ_{G_1}$ and $\mZ_{G_2}$ with the center $\mZ \defn \mA \cap \mC$.
Here we discuss various situations that arise in the cases we study.
If the common center $\mZ$ of the algebras $\mA$ and $\mC$ is the same as the Casimir algebras of both groups, i.e., if $\mZ = \mZ_{G_1} = \mZ_{G_2}$, the algebra partitioning of Eq.~(\ref{eq:Hilbertdecomp}) is equivalent to the group partitioning in terms of irreps of $G_1 \times G_2$. 
The Hilbert space can be partitioned into sectors labelled by $(\{C^{G_1}_\alpha\}, \{C^{G_2}_\alpha\})$, and hence according to irreps of $G_1 \times G_2$.
Such a scenario occurs in cases \#1a-c in Tab.~\ref{tab:spinfulbondalgebra}, where the Hilbert space can be decomposed in terms of representations of $U(N) \times (SU(2) \times U(1))$ [or equivalently, in terms of $U(N) \times SU(2)$ within each sector labelled by a fixed $U(1)$ quantum number], which was also pointed out in \cite{pakrouski2020many} (also see App.~\ref{subsec:spincomplexmu}).
In such a case, since $\mZ_{G_1} = \mZ_{G_2}$, the Casimir elements of the groups of $G_1$ can be expressed in terms of $G_2$ and vice versa.
In Eq.~(\ref{eq:quadraticUN}), we show an example of such a relation between the quadratic Casimirs of $U(N)$ and $SU(2) \times U(1)$.
We expect similar relations between quadratic and higher order Casimirs in each of such cases in Tabs.~\ref{tab:spinlessbondalgebra} and \ref{tab:spinfulbondalgebra}, which can be directly computed for further analytical checks.
On the other hand, a  partitioning in terms of irreducible representations of $G_1 \times G_2$ is also possible if the common center $\mZ$ of $\mA$ and $\mC$ is a strict subalgebra of the Casimir algebra of $G_2$, i.e., if $\mZ = \mZ_{G_1} \subset \mZ_{G_2}$.
The situations we have in mind are where $G_2$ is a prominent ``subgroup'' of the commutant (like those listed in Tab.~\ref{tab:spinfulbondalgebra}) but does not generate the full commutant.
However, in this case, the group partitioning differs from the algebra partitioning of Eq.~(\ref{eq:Hilbertdecomp}), since there are operators in the Casimir algebra of $G_2$ that are not part of the center $\mZ$.
Nevertheless, since the Casimir algebra $\mZ_{G_2}$ is Abelian subalgebra of $\mC$ by definition, it should be (a subalgebra of) a maximal Abelian subalgebra of $\mC$.
Hence the group partitions are closed under the action of elements of $\mA$ (although the actions are reducible if $\mZ_{G_2}$ is not a maximal Abelian subalgebra of $\mC$), but are not closed under the action of certain elements of $\mC$ (see discussion in Sec.~\ref{subsec:Hilbertdecomp}).
The group partitioning thus misses some information on the degeneracy of the eigenvalues of operators in $\mA$, e.g., some of the singlets that are degenerate in $\mA$ (captured by belonging to the same block in the algebra partition) may belong to different blocks of the group partition.
%
%The Hilbert space can be partitioned according to labels $(\{C^{G_1}_\alpha\}, \{C^{G_2}_\alpha\}_{<})$, where $\{C^{G_2}_\alpha\}_{<}$ denotes a strict subset of Casimir elements of $G_2$.\footnote{Here we refer to any element of the Casimir algebra as a Casimir element. These need not correspond to the standard quadratic and higher order Casimir operators that are constructed in the literature.}
%
This scenario occurs in the cases \#2 and \#3a-b in Tab.~\ref{tab:spinfulbondalgebra} discussed in Apps.~\ref{subsec:spincomplex} and \ref{subsec:spinimaginary} respectively.
In case \#2, the center $\mZ$ can be shown to be strictly smaller than the Casimir algebra of $SU(2) \times U(1)$.
Hence, even though the Hilbert space can be partitioned according to irreps of $SU(N) \times (SU(2) \times U(1))$, the states $\ket{\Omega}$ and $\sket{\bar{\Omega}}$ transform under different irreps of the group, but transform under the same irrep of the algebra $\lgen \mA \cup \mC \rgen$ [i.e., they are part of the same  block in Eq.~(\ref{eq:Hilbertdecomp})].
Similarly, in case \#3a-b we can choose $G_2 = SU(2) \times SU(2)$, and partition the Hilbert space into blocks that transform under its irreps. 
Note that if $N$ is even, all the irreps of $SU(2) \times SU(2)$ that appear are also irreps of $SO(4)$, hence we can equivalently refer to $G_2$ as $SO(4)$ in that case~\cite{yang1990so, essler2005one}.\footnote{This is analogous to the correspondence between representations of $SU(2)$ and $SO(3)$ -- all odd dimensional irreps of $SU(2)$ are also irreps of $SO(3)$.}
In either case, the Casimir operators of $SU(2) \times SU(2)$ or $SO(4)$ are the same, and only one of the two independent Casimir operators is a part of the center $\mZ$.
Hence the states $\ket{\Omega}$ and $\ket{F}$ transform under different irreps of $G_2$ (i.e., $SU(2) \times SU(2)$ or $SO(4)$) while transforming under the same irrep of $\lgen \mA \cup \mC \rgen$.
Further, since $\mZ_{G_1} \subset \mZ_{G_2}$, the Casimir elements $\{C^{G_1}_\alpha\}$ of $G_1$ can be expressed in terms of Casimir elements $\{C^{G_2}_\alpha\}$ of $G_2$, but not all Casimir elements of $G_2$ can be expressed in terms of those of $G_1$.
In Eq.~(\ref{eq:quadraticSUN}) [resp.\ Eq.~(\ref{eq:quadraticSON})], we show such relations between the quadratic Casimirs of $SU(N)$ and $SU(2) \times U(1)$ [resp. $SO(N)$ and $SU(2) \times SU(2)$ or $SO(4)$], which were also derived in \cite{klebanov2018spectra, gaitan2020hagedorn}.
Apart from the cases we have discussed, we note that similar group decompositions can be identified in several cases in Tab.~\ref{tab:spinfulbondalgebra}.
Based on the centers derived in App.~\ref{app:spinfulfermion}, we find that the case \#4 belongs to the case where $\mZ = \mZ_{G_1} = \mZ_{G_2}$, whereas case \#5 $\mZ = \mZ_{G_1} \subset \mZ_{G_2}$, hence group decompositions (with the caveats discussed above) can be identified there too.
Similar ideas can also be applied to the spinless cases shown in Tab.~\ref{tab:spinlessbondalgebra}. With straightforward algebra it is easy to see that while case \#1 satisfies $\mZ = \mZ_{G_1} = \mZ_{G_2}$ (i.e., a $U(N) \times U(1)$ decomposition), cases \#2 and \#3 satisfy $\mZ = \mZ_{G_1} \subset \mZ_{G_2}$ (i.e, for $SU(N) \times U(1)$ and $SO(N) \times U(1)$ decompositions respectively), hence group decompositions also hold there.

\section{Some Proofs of Commutant Exhaustion}\label{app:commutantexhaustion}
In this appendix, we discuss some illustratory examples of bond algebras where we are able to \textit{prove} the expression for the commutant algebras we quote in Tabs.~\ref{tab:conventionalexamples} and \ref{tab:spinlessbondalgebra}.
We expect similar methods to also work for many other algebras we study in the main text. 
\subsection{Spinless Fermions with Pure Imaginary Hoppings}\label{subsec:spinlessimagproof}
For the spinless free-fermion bond algebra $\mA_i$ discussed in App.~\ref{subsec:spinlessimaginary}, we can prove the claimed commutant $\mC_i = \lgen N_{\tot}, Q_X \rgen$ as follows.
From the discussion in App.~\ref{subsec:spinlessimaginary} it is clear that $\lgen N_{\tot}, Q_X \rgen \subseteq \mC_i$, hence it is sufficient to show $\mC_i \subseteq \lgen N_{\tot}, Q_X \rgen$.
We begin with the identity
\begin{equation}
    e^{i x T^{(i)}_{j,k}} = 1 + i \sin(x) T^{(i)}_{j,k} - (1 - \cos(x)) \left(T^{(i)}_{j,k}\right)^2,\;\;T^{(i)}_{j,k} = i(\cd_j c_k - \cd_k c_j),\;\;\left(T^{(i)}_{j,k}\right)^2 = n_j + n_k - 2 n_j n_k,\;\;(T^{(i)}_{j,k})^3 = T^{(i)}_{j,k}.
\label{eq:usefulidentityTi}
\end{equation}
Then consider a unitary operator $U_{jk}$ with the following properties:
\begin{equation}
    U_{jk} \defn e^{i\frac{\pi}{2} T_{j,k}^{(i)}} = 1 + (-\cd_j c_k + \cd_k c_j) - (n_j + n_k - 2n_j n_k) ~, \quad 
    U_{jk}^\dagger U_{jk} = 1 ~, \quad 
    U_{jk}^2 = (-1)^{n_j + n_k},\quad U_{jk}^4 = 1.
\label{eq:Ujkproperties}
\end{equation}
Since $\{T^{(i)}_{j,k}\}$ are the generators of $SO(N)$, it is clear that the unitaries $\{U_{jk}\}$ generate a discrete subgroup of $SO(N)$.
Further, the generators $\{T^{(i)}_{j,k}\}$ can be expressed in terms of $U_{jk}$ using Eq.~(\ref{eq:Ujkproperties}) (i.e., $ T^{(i)}_{j,k} = i(U^3_{jk} - U_{jk})/2$).
Hence the group algebra (with complex coefficients) of the discrete subgroup generated by the unitaries $\{U_{jk}\}$ is equal to the bond algebra $\mA_i$.
Hence we can derive the full commutant $\mC_i$ by requiring that operators $\Gamma \in \mC_i$ commute with $U_{jk}$ for all $j$ and $k$.
We start by examining the conjugate action of $U_{jk}$ on on-site operators.
It is straightforward to verify that the conjugation by this swaps fermionic operators at sites $j$ and $k$ with a relative minus sign:
\begin{equation}
    U_{jk} c_j U_{jk}^{-1} = c_k ~, \quad 
    U_{jk} c_k U_{jk}^{-1} = -c_j ~, \quad 
    U_{jk} c_\ell U_{jk}^{-1} = c_\ell ~,~ \ell \neq j, k ~.
    \label{eq:Ujkaction}
\end{equation}
It follows that $U_{jk} n_j U_{jk}^{-1} = n_k$, $U_{jk} n_k U_{jk}^{-1} = n_j$, and also
\begin{equation}
    U_{jk}^2 c_j U_{jk}^{-2} = -c_j ~, \quad
    U_{jk}^2 c_k U_{jk}^{-2} = -c_k ~, \quad
    U_{jk}^2 n_j U_{jk}^{-2} = n_j ~, \quad
    U_{jk}^2 n_k U_{jk}^{-2} = n_k ~.
    \label{eq:Ujk2action}
\end{equation}
Hence, for any operator $O_{j,k}$ with support on sites $j$ and $k$ it is easy to see that $U^2_{jk} O_{j,k} U^{-2}_{jk} = O_{j,k}$ if $O_{j,k}$ is ``bosonic"/``even" (i.e., consists of even number of fermionic operators) and  $U^2_{jk} O_{j,k} U^{-2}_{jk} = -O_{j,k}$ if $O_{j,k}$ is ``fermionic"/``odd" (i.e., consists of odd number of fermionic operators).
Consider an operator $\Gamma$ from the commutant $\mC_i$, expanded in the complete operator basis consisting of all operator strings
\begin{equation}
    \Gamma = \sum_{a_1, a_2, \dots, a_N \in \{\id, n, +, - \}} \gamma_{a_1, a_2, \dots, a_N} \, \omega_1^{a_1} \omega_2^{a_2} \dots \omega_N^{a_N} ~, \quad\quad \{\omega_\ell^{\id}, \omega_\ell^{n}, \omega_\ell^{+}, \omega_\ell^{-} \} \defn \{\mathbb{1}_\ell, n_\ell, \cd_\ell, c_\ell \} ~.
\end{equation}
Note that the operator strings can contain on-site ``bosonic''/``even" (i.e., $\mathds{1}_\ell$ or $n_{\ell}$) or ``fermionic''/``odd" (i.e., $\cd_{\ell}$ or $c_{\ell}$) operators, and the operator ordering is important; in the above writing, we assume a fixed ordering of sites $1, 2, \dots, N$, but the specific choice is immaterial.
Since $\Gamma \in \mC_i$, it must commute with both $U_{jk}$ and $U_{jk}^2$.
Starting with $U_{jk}^2 \Gamma U_{jk}^{-2} = \Gamma$ and using Eq.~(\ref{eq:Ujk2action}), it follows that $\gamma_{\dots, a_j, \dots, a_k, \dots}$ can be non-zero only if $\omega^{a_j}_j \omega^{a_k}_k$ is even,  which happens when $\omega_j^{a_j}$ and $\omega_k^{a_k}$ have the same parity (i.e., $a_j, a_k \in \{\id, n \}$ or $a_j, a_k \in \{+, - \}$).
Requiring this condition for all pairs $j, k$, it follows that $\gamma_{a_1, a_2, \dots, a_N}$ can be non-zero only if all $\omega_{\ell}^{a_\ell}$ have the same parity (i.e., $a_\ell \in \{\id, n \}$ or $a_\ell \in \{+, - \}$).
We can collect the former ``bosonic"/``even" parts into  $\Gamma^{(b)}$ and the latter ``fermionic"/``odd" parts into $\Gamma^{(f)}$, and split $\Gamma$ as $\Gamma = \Gamma^{(b)} + \Gamma^{(f)}$.
We now consider the condition $U_{jk} \Gamma U_{jk}^{-1} = \Gamma$, which can be applied separately to the bosonic and fermionic parts  $\Gamma^{(b)}$ and $\Gamma^{(f)}$.
Using Eq.~(\ref{eq:Ujkaction}), it is easy to verify that $U_{jk} \omega_j^a U_{jk}^{-1} \cdots U_{jk} \omega_k^b U_{jk}^{-1} = \omega_j^b \cdots \omega_k^a$ for all cases when $\omega^a$ and $\omega^b$ have the same parity and for any operator string part between the sites $j$ and $k$ marked with ``$\cdots$''.
It then follows that for both $\Gamma^{(b)}$ and $\Gamma^{(f)}$ we require symmetry of $\gamma_{a_1, a_2, \cdots, a_N}$ under exchange of its indices at positions $j$ and $k$:
\begin{equation}
    \gamma_{\cdots, a_j, \cdots, a_k, \cdots} = \gamma_{\cdots, a_k, \cdots, a_j, \cdots} ~.
\label{eq:gammaexcondition}
\end{equation}
Since this is required for all pairs $j, k$, we conclude that the amplitudes are the same for all strings labelled by permutations of a fixed \textit{set} of $\{a_\ell\}$.
Hence, we can further subdivide both $\Gamma^{(b)}$ and $\Gamma^{(f)}$ into $N+1$ independent operators each, 
\begin{equation}
    \Gamma^{(b)} = \sum_{m_{\id} = 0}^N \Gamma^{(b)}_{m_\text{id}, m_n = N - m_\text{id}} ~, \quad\quad  
    \Gamma^{(f)} = \sum_{m_+ = 0}^N \Gamma^{(f)}_{m_+, m_- = N - m_+} ~,
\end{equation}
where $\Gamma^{(b)}_{m_{\id}, m_n}$ is an {\it equal-amplitude superposition} of all operator strings containing precisely $m_{\id}$ on-site operators $\mathbb{1}_\ell$ and $m_n = N - m_{\id}$ operators $n_\ell$, and similarly for $\Gamma^{(f)}_{m_+, m_-}$ in terms of operator strings containing precisely $m_+$ operators $\cd_\ell$ and $m_- = N - m_+$ operators $c_\ell$.
We know that such operators $\{\Gamma^{(b)}_{m_{\id}, m_n}, m_n = 0, 1, \dots, N \}$
form a complete basis in the span of $\{N_\text{tot}^m, m = 0, 1, \dots, N \}$ (see, e.g., discussion in Sec.~VII B of \cite{moudgalya2021hilbert}), hence $\Gamma^{(b)}_{m_{\id}, m_n} \in \lgen N_{\tot} \rgen$.
On the other hand, for the operators $\Gamma^{(f)}_{m_+, m_-}$, it is easy to verify that
\begin{equation}
 \Gamma^{(f)}_{m_+, m_-} \sim Q_X P_{N_\text{tot} = m_-},
\end{equation}
where $Q_X$ is from Eq.~(\ref{eq:QXQXtdefns}) and $P_{N_\text{tot} = m_-}$ is a projector into the sector with the total fermion number $N_\text{tot} = m_-$.
Since $P_{N_{\tot} = m_-} \in \lgen N_\text{tot} \rgen$, it follows that all $\Gamma^{(f)}_{m_+, m_-} \in \lgen N_\text{tot}, Q_X \rgen$.
This shows that the commutant $\mC_i \subseteq \lgen N_{\tot}, Q_X \rgen$, completing the proof that $\mC_i = \lgen N_{\tot}, Q_X \rgen$.
Note that this proof also shows that the linear dimension of the commutant $\mC_i$ is $2(N+1)$, the number of linearly independent operators $\Gamma$ that satisfy the required conditions.
The bond unitaries $U_{jk}$ can also be used to understand the singlets of the bond algebra.
Even though we already know all the singlets from arguments given earlier, we will start from scratch looking for states that are simultaneous eigenstates of all bond unitaries, $U_{jk} \ket{\Psi} = \lambda_{jk} \ket{\Psi}$, with no assumptions about $\lambda_{jk}$ but using $U_{jk} \ket{\Omega} = \ket{\Omega}$ (which already provides one singlet).
We can directly verify the following relations among the bond unitaries:
\begin{equation}
    U_{jk} U_{kl} = U_{kl} U_{lj} = U_{lj} U_{jk} ~, \quad
    (U_{jk} U_{kl})^3 = \mathbb{1} ~, \quad
    U_{jk}^4 = \mathbb{1}.
\label{eq:Ujkrelations}
\end{equation}
These relations imply that all $\lambda_{jk}$ are the same and satisfy $\lambda_{jk}^2 = 1$.
Hence $U_{jk}^2 \ket{\Psi} = \ket{\Psi}$ on any singlet $\ket{\Psi}$, which immediately implies that only configurations with $n_j = n_k$ can contribute to $\ket{\Psi}$.
It is easy to show that $\ket{\Psi}$ must then be in the span of $\ket{\Omega}$ and $\sket{\bar{\Omega}}$, which shows that $\ket{\Omega}$ and $\sket{\bar{\Omega}}$ are the only singlets of $\mA_i$. 
\subsection{Spinless Fermions with Complex Hoppings}\label{subsec:spinlesscomplexproof}
It is also interesting to see in this language how the commutant is ``reduced" from $\mC_i$ to $\mC_c$ when real hoppings are added to the bond algebra $\mA_i$ to obtain the algebra $\mA_c$, see Sec.~\ref{subsec:spinlesscomplex}.
We begin with the identity
\begin{equation}
    e^{i x T^{(r)}_{j,k}} = 1 + i \sin(x) T^{(r)}_{j,k} - (1 - \cos(x)) \left(T^{(r)}_{j,k}\right)^2,\;\;T^{(r)}_{j,k} =  \cd_j c_k + \cd_k c_j,\;\;\left(T^{(r)}_{j,k}\right)^2 = n_j + n_k - 2 n_j n_k,\;\;(T^{(r)}_{j,k})^3 = T^{(r)}_{j,k}.
\label{eq:usefulidentityTr}
\end{equation}
The corresponding bond unitary operators $\{V_{jk}\}$ are
\begin{equation}
    V_{jk} \defn e^{i\frac{\pi}{2} T_{jk}^{(r)}} = 1 + i(\cd_j c_k + \cd_k c_j) - (n_j + n_k - 2 n_j n_k) ~, \quad
    V_{jk}^\dagger V_{jk} = \mathbb{1} ~, \quad
    V_{jk}^2 = (-1)^{n_j + n_k} ~,
\label{eq:Vjkproperties}
\end{equation}
which act on the local operators as
\begin{equation}
    V_{jk} c_j V_{jk}^{-1} = -i c_k ~, \quad 
    V_{jk} c_k V_{jk}^{-1} = -i c_j ~, \quad 
    V_{jk} c_\ell V_{jk}^{-1} = c_\ell ~,~ \ell \neq j, k ~.
\label{eq:Vjkactions}
\end{equation}
We can then follow similar steps as discussed in App.~\ref{subsec:spinlessimagproof} to constrain the operator strings that appear in the expansion of any operator $\Gamma \in \mC_c$, requiring commutation with both the unitaries $V_{jk}$ and $U_{jk}$. 
The key difference from the action of $U_{jk}$ appears from the study of parts in $\Gamma^{(f)}$ in the steps that lead to Eq.~(\ref{eq:gammaexcondition}).
Here, $V_{jk} \omega_j^a V_{jk}^{-1} \dots V_{jk} \omega_k^b V_{jk}^{-1} = -\omega_j^b \dots \omega_k^a$ for $(a,b)=(+,-)$ or $(-,+)$, which has the opposite sign compared to the $U_{jk}$ action.
Hence combining the requirements $[\Gamma, U_{jk}] = 0$  and $[\Gamma, V_{jk}]=0$ imposes the constraint that $\gamma_{a_1, a_2, \dots, a_N} = 0$ if $(a_j, a_k) = (+,-)$ or $(-,+)$ for any $j \neq k$.\footnote{Another way of deriving this is to note that $U_{jk} V_{jk} c_j V_{jk}^{-1} U_{jk}^{-1} = i c_j$, $U_{jk} V_{jk} c_k V_{jk}^{-1} U_{jk}^{-1} = -i c_k$, so the conjugate action by $U_{jk} V_{jk}$ flips the signs of $c_j^\dagger c_k$ and $c_j c_k^\dagger$, and hence such terms are prohibited from $\Gamma \in \mC_c$.} 
Hence, only $\gamma_{+,+,\cdots,+}$ and $\gamma_{-,-,\cdots,-}$ parts remain, corresponding to operators $\ketbra{\bar{\Omega}}{\Omega}$ and $\ketbra{\Omega}{\bar{\Omega}}$.
This shows that the commutant $\mC_c = \lgen N_{\tot}, \ketbra{\bar{\Omega}}{\Omega}\rgen$.
\subsection{Regular \texorpdfstring{$\boldsymbol{SU(2)}$}{} Symmetry}\label{subsec:regularsu2proof}
Finally we show that similar ideas can be applied to the bond algebra $\mA_{SU(2)}$ that appears in the $SU(2)$ spin-1/2 models discussed in Sec.~\ref{subsec:regularSU2}. 
We start by noting that the two-site permutation operators $\{P_{jk}\}$ that generate $\mA_{SU(2)}$, defined as
\begin{equation}
 P_{jk} \defn 2 \vec{S}_j \cdot \vec{S}_k + \frac{1}{2}, \quad P_{jk} \ket{\sigma, \sigma'}_{jk} = \ket{\sigma', \sigma}_{jk} ~,
 \end{equation}
can be viewed as unitary operators analogous to $\{U_{jk}\}$ and $\{V_{jk}\}$ in the previous sections. 
The conjugate actions of this unitary on the local spin operators read
\begin{equation}
P_{jk} \vec{S}_j P_{jk}^{-1} = \vec{S}_k, \quad P_{jk} \vec{S}_k P_{jk}^{-1} = \vec{S}_j ~. 
\end{equation}
We can then expand operators $\Gamma \in \mC_{SU(2)}$ in the Pauli string basis and constrain the Pauli strings that appear in the expansion based on invariance under the conjugation by all unitaries $\{P_{jk}\}$. 
In particular, it is easy to show that any such $\Gamma$ can be divided into a sum of operators $\Gamma_{(m_0,m_x,m_y,m_z)}$ with $m_0 + m_x + m_y + m_z = N$, where $\Gamma_{(m_0,m_x,m_y,m_z)}$ is an equal weight superposition of all Pauli strings with precisely $m_\mu$ operators $S^\mu_\ell$ ($\mu \in \{0, x, y, z\}$), where we $S^0_\ell \defn \mathds{1}_{\ell}$.
As we have discussed in \cite{moudgalya2021hilbert} (see Eq.~(77) there), this is precisely the span of the algebra $\lgen S^x_{\tot}, S^y_{\tot}, S^z_{\tot} \rgen$, and hence this derives the expression for the commutant $\mC_{SU(2)}$. 
\section{Details on Hubbard Algebras}\label{app:Hubbard}
In this appendix, we provide more details on the Hubbard algebras discussed in Sec.~\ref{sec:Hubbard}. 
As we now show, the Hubbard algebras are different from any of the free-fermion algebras shown in Tab.~\ref{tab:spinfulbondalgebra}. 
\subsection{Imaginary hoppings}\label{subsec:Hubbardimag}
\subsubsection{Regular \texorpdfstring{$SU(2)$}{} symmetries}
We start with Hubbard algebras of Eq.~(\ref{eq:Ahubbard}) with imaginary hopping terms and Hubbard terms, given by
\begin{equation}
    \mA_{i,\hub} \defn \lgen \{T^{(i)}_{j,k}\}_{\tnn}, \{V_j\} \rgen. 
\label{eq:Aihub}
\end{equation}
By definition, this algebra clearly satisfies the relations $\mA_i \subseteq \mA_{i, \hub} \subseteq \mA_{c,\mu} \cap \mA_{i,h}$, where $\mA_{c,\mu}$, $\mA_{i,h}$, and $\mA_i$  are the free-fermion bond algebras corresponding to cases \#1a, \#1b, and \#3a in Tab.~\ref{tab:spinfulbondalgebra} respectively (discussed in App.~\ref{subsec:spincomplexmu}, \ref{subsec:spinfulimagh}, and \ref{subsec:spinimaginary} respectively). 
This implies the relations between the commutant of $\mA_{i,\hub}$, denoted by $\mC_{i,\hub}$, and the free-fermion commutants, i.e., $\lgen \mC_{c,\mu} \cup \mC_{i,h} \rgen \subseteq \mC_{i, \hub} \subseteq \mC_i$.\footnote{We would like to remind readers that we are using $\cup$ and $\cap$ in the sense of sets. Hence the intersection of two algebras is an algebra whereas the union of two algebras is not necessarily an algebra.}
Using the expressions for $\mC_{c, \mu}$ and $\mC_{i,h}$ in Tab.~\ref{tab:spinfulbondalgebra}, we obtain $\lgen \mC_{c, \mu} \cup \mC_{i,h}\rgen = \lgen \{S^\alpha_{\tot}\}, \{\eta^\alpha_0\}\rgen \subseteq \mC_{i,\hub}$.
Further, it is easy to verify that the Hubbard terms $\{V_j\}$ anti-commute with the ``Shiba transformation" operators $Q_X^\sigma \in \mC_i$, defined in Eq.~(\ref{eq:QXQXtdefnsspin}), hence we obtain $Q^\sigma_X \notin \mC_{i,\hub}$ and $\mC_{i,\hub} \subset \mC_i$. 
While $V_j$ does commute with $Q_X^\uparrow Q_X^\downarrow$, it is easy to show that $Q^\uparrow_X Q^\downarrow_X \in \lgen S_\tot^y + \eta_0^y \rgen \subset \lgen \{S_\tot^\alpha \}, \{\eta_0^\alpha \} \rgen$.
This leads us to the conjecture that 
\begin{equation}
    \mC_{i, \hub} = \lgen \{S^\alpha_{\tot}\}, \{\eta^\alpha_{0}\} \rgen ~.
\end{equation}
Indeed, this can also be verified numerically via methods we discuss in \cite{moudgalya2022numerical}. 
Since this commutant satisfies $\mC_{i,\hub} = \lgen \mC_{c,\mu} \cup \mC_{i,h} \rgen$, this conjecture also implies that the Hubbard algebra is the intersection of the free-fermion algebras $\mA_{c,\mu}$ and $\mA_{i,h}$, i.e., $\mA_{i,\hub} = \mA_{c,\mu} \cap \mA_{i,h}$.
Note that $\{S^\alpha_{\tot}\}$ and $\{\eta^\alpha_0\}$ are the generators of the spin and pseudospin $SU(2)$ groups and they commute with each other, hence $\mC_{i,\hub}$ has an $SU(2) \times SU(2)$ subgroup.
Moreover, the full commutant $\mC_{i,\hub}$ can be viewed as being generated by the full family of on-site unitaries correspoding to these $SU(2)$ groups, i.e., by $\exp(i \sum_\alpha{\theta_\alpha S^\alpha_{\tot}})$ and  $\exp(i \sum_\alpha{\theta_\alpha \eta^\alpha_{0}})$, a property that enables the application of the DCT via Lems.~\ref{lem:strloc} and \ref{lem:typeI}.
Hence the symmetry group for Hamiltonians within $\mA_{i,\hub}$ is referred to as $SU(2) \times SU(2)$. 
Further, as discussed in Sec.~\ref{subsec:groupdecomp}, when the number of sites $N$ is even, irreducible representations of $SU(2) \times SU(2)$ are also representations of $SO(4)$, which is the usual symmetry group of the Hubbard model~\cite{yang1990so, essler2005one}. 
Finally, the singlets of the Hubbard algebra $\mA_{i,\hub}$ consist of the towers of states annihilated by all the hopping terms, which are the ferromagnetic tower given by $\{(S^-_{\tot})^n \ket{F}\}$ and the ``eta-pairing" tower $\{(\eta^\dagger_0)^n\ket{\Omega}\}$, where $S^-_{\tot}$ and $\eta^\dagger_0$ are defined in Eqs.~(\ref{eq:Salphadefn}) and (\ref{eq:etaalphadefns}) respectively.  
However, the towers differ in their eigenvalues under the Hubbard terms $\{V_j\}$, hence $\mA_{i,\hub}$ has two sets of degenerate singlets that are not degenerate with each other.
\subsubsection{Dynamical \texorpdfstring{$SU(2)$}{} symmetries}\label{subsubsec:hubbardimagdyn}
We can expand the algebra of Eq.~(\ref{eq:Aihub}) to include extensive local terms that lead to dynamical symmetries in the commutants.
We consider the algebra
\begin{equation}
    \mA^{(\dyn \eta)}_{i,\hub} \defn \lgen \{T^{(i)}_{j,k}\}, \{V_j\}, N_{\tot} \rgen = \lgen \{T^{(i)}_{j,k}\}, \{V_j\}, \eta^z_0 \rgen,
\label{eq:Aidynhub}
\end{equation}
where we have used the fact that $N_{\tot}$ and $\eta^z_0$ are related according to Eq.~(\ref{eq:etaalphadefns}).
As discussed in Sec.~\ref{subsec:Hubbarddynamical}, this algebra is relevant for the study of the Hubbard model in the presence of a chemical potential.
Since $\eta^z_0$ is one of the generators of the pseudospin $SU(2)$ symmetry in the commutant $\mC_{i,\hub}$, the relation between $\mA_{i,\hub}$ and $\mA^{(\dyn \eta)}_{i,\hub}$
is analogous to that between $\mA_{SU(2)}$ and $\mA_{\dyn SU(2)}$ discussed in Sec.~\ref{sec:commutantexamples}. 
In particular, the addition of $\eta^z_0$ to the bond algebra $\mA_{i,\hub}$ ``breaks" the pseudospin $SU(2)$ down to its maximal Abelian subalgebra, i.e., $\lgen \{\eta^\alpha_0\} \rgen \rightarrow \lgen \vec{\eta}^2_0, \eta^z_0 \rgen$.
It is then easy to check that the commutant of $\mA^{(\dyn \eta)}_{i,\hub}$ reads
\begin{equation}
    \mC^{(\dyn \eta)}_{i,\hub} = \lgen \{S^\alpha_{\tot}\}, \vec{\eta}^2_0, \eta^z_0 \rgen.
\label{eq:Cidynhub}
\end{equation}
This commutant has a spin regular $SU(2)$ symmetry along with a pseudospin dynamical $SU(2)$ symmetry, similar to the case discussed in Sec.~\ref{subsec:dynamicalSU2}.
The ferromagnetic tower and the $\eta$-pairing tower are still the singlets of the local algebra $\mA^{(\dyn \eta)}_{i,\hub}$, but the latter states are now split in energy by the $N_\tot$ term and are hence non-degenerate singlets.
In addition to or instead of a pseudospin dynamical $SU(2)$ symmetry, we could similarly add a uniform magnetic field $S^z_{\tot}$ to the bond algbera $\mA_{i,\hub}$ to ``break" the regular spin $SU(2)$ symmetry down to a dynamical spin $SU(2)$ symmetry. 
Hence, we can also obtain the following pairs of local and commutant algebras:
\begin{gather}
    \mA^{(\dyn S)}_{i, \hub} \defn \lgen \{T^{(i)}_{j,k}\}, \{V_j\}, S^z_{\tot} \rgen,\;\;\;\mC^{(\dyn S)}_{i,\hub} = \lgen \vec{S}^2_{\tot}, S^z_{\tot}, \{\eta^\alpha_{0}\} \rgen\nn; \\
    \mA^{(\dyn S, \dyn \eta)}_{i, \hub} \defn \lgen \{T^{(i)}_{j,k}\}, \{V_j\}, S^z_{\tot}, N_{\tot} \rgen,\;\;\;\mC^{(\dyn S, \dyn \eta)}_{i,\hub} = \lgen \vec{S}^2_{\tot}, S^z_{\tot}, \vec{\eta}^2_0, \eta^z_0 \rgen.  
\end{gather}
As discussed in Sec.~\ref{subsec:Hubbarddynamical},  these algebras are relevant for the study of the Hubbard model in the presence of a uniform magnetic field or a uniform chemical potential (or both). 
Note that the algebras $\mA^{(\dyn S)}_{i,\hub}$ and $\mA^{(\dyn \eta)}_{i,\hub}$, and hence their commutants, are isomorphic to each other.
This can be deduced by conjugating the generators of $\mA^{(\dyn S)}_{i,\hub}$ by $Q^\uparrow_X$ or $Q^\downarrow_X$ defined in Eq.~(\ref{eq:QXQXtdefnsspin}) (also known as the Shiba transformation~\cite{essler2005one}). 
In particular this transfoms the generators as $S^z_{\tot} \rightarrow \eta^z_0$ (see Eq.~(\ref{eq:etaSrelation})), $T^{(i)}_{j,k} \rightarrow T^{(i)}_{j,k}$ (since $Q^\downarrow_X \in \mC_i$, see Eq.~(\ref{eq:Ciapp})), and $V_j \rightarrow -V_j$ (which can be verified by a direct application of Eq.~(\ref{eq:QXQXtdefns})). 
Hence this transformation exactly maps the algebra $\mA^{(\dyn S)}_{i,\hub} = \lgen \{T^{(i)}_{j,k}\}, \{V_j\}, S^z_{\tot} \rgen$ to $\lgen \{T^{(i)}_{j,k}\}, \{-V_j\}, \eta^z_{0} \rgen = \mA^{(\dyn \eta)}_{i,\hub}$.
\subsection{Real hoppings}\label{subsec:Hubbardreal}
We now replace the imaginary hopping terms $\{T^{(i)}_{j,k}\}_{\tnn}$ with real hoppings $\{T^{(r)}_{j,k}\}_{\tnn}$ in the algebra $\mA_{i, \hub}$ of Eq.~(\ref{eq:Aihub}), and study the algebra
\begin{equation}
    \mA_{r,\hub} \defn \lgen \{T^{(r)}_{j,k}\}_{\tnn}, \{V_j\} \rgen.
\label{eq:Arhub}
\end{equation}
This algebra contains the well-studied Hubbard model with real hoppings~\cite{essler2005one}. 
Similar to the case of spinful fermions with real hoppings discussed in App.~\ref{subsec:spinfulreal}, on a non-bipartite lattice, it is easy to see that all complex hoppings $\{T^{(c)}_{j,k}\}$ are eventually generated, and the algebra $\mA_{r,\hub}$ is equal to the $\mA_{c,\hub}$, the Hubbard algebra with complex hoppings, which we discuss in App.~\ref{subsec:Hubbardcomplex}.
However, on a bipartite lattice, a distinct closed algebra is generated, similar to the case of spinful fermions with real hoppings discussed in App.~\ref{subsec:spinfulreal}.
As discussed there, the Lie algebra generated by $\{T^{(r)}_{j,k}\}_{\tnn}$ consists of real hoppings $\{T^{(r)}_{j,k}\}$ between different sublattices and imaginary hoppings $\{T^{(i)}_{j,k}\}$ within the same sublattices.
This algebra is isomorphic to the Lie algebra generated from imaginary hoppings, and this can be understood via a transformation $c_{j,\sigma} \to i c_{j,\sigma}$ for $\sigma \in \{\uparrow, \downarrow\}$ on one of the sublattices.
Under this transformation, $\{T^{(r)}_{j,k}\}_{\tnn} \rightarrow \{T^{(i)}_{j,k}\}_{\tnn}$ while the Hubbard terms $\{V_j\}$ remain invariant, hence the bond algebras of Eq.~(\ref{eq:Arhub}) transforms as $\mA_{r,\hub} \rightarrow \mA_{i,\hub}$ defined in Eq.~(\ref{eq:Aihub}).
The singlets of $\mA_{r,\hub}$ can also be obtained using those of $\mA_{i, \hub}$ using this substitution.
The isomorphism of the bond algebras also implies an isomorphism of the respective commutants $\mC_{r,\hub}$ of $\mA_{r,\hub}$ to $\mC_{i,\hub}$ of $\mA_{i,\hub}$.
As discussed in App.~\ref{subsec:spinfulreal}, the sublattice transformation preserves the spin-$SU(2)$ generators $\{S^\alpha_{\tot}\}$, while the operators $\{\eta^\alpha_0\}$ get mapped onto $\{\eta^\alpha_{\pi}\}$, both defined in Eq.~(\ref{eq:etaalphadefns}).
Hence the full commutant with real bipartite hoppings is given by 
\begin{equation}
    \mC_{r, \hub} = \lgen \{S^\alpha_{\tot}\}, \{\eta^\alpha_\pi\}\rgen.
\end{equation}
Similar to $\mC_{i,\hub}$ discussed in the previous section, this conjecture implies that $\mA_{r,\hub} = \mA_{c,\mu} \cap \mA_{r,h}$.
Further, $\{S^\alpha_{\tot}\}$ and $\{\eta^\alpha_{\pi}\}$ generate the spin and pseudospin $SU(2)$ groups respectively, and moreover, the full commutant can be understood as being generated by the family on-site unitary operators corresponding to these groups. 
Hence the symmetry group for Hamiltonians within $\mA_{r,\hub}$ is referred to as $SU(2) \times SU(2)$. 
As discussed in Sec.~\ref{subsec:spinfulfermions} and App.~\ref{subsec:Hubbardimag}, if the number of sites $N$ is even, irreps of this group can also be viewed as irreps of $SO(4)$, which is usually referred to as the symmetry group of the Hubbard model. 
The sublattice transformation that results in the isomorphism between $\mA_{r,\hub}$ and $\mA_{i,\hub}$ can also be applied in the presence of a uniform chemical potential or uniform magnetic field, where the regular $SU(2)$ symmetries in the commutant are broken down to dynamical $SU(2)$ symmetries. 
Hence, performing this transformation on the algebras discussed in Sec.~\ref{subsubsec:hubbardimagdyn}, we obtain the following pairs of local and commutant algebras with real hoppings (on a bipartite lattice):
\begin{gather}
    \mA^{(\dyn \eta)}_{r,\hub} \defn \lgen \{T^{(r)}_{j,k}\}, \{V_j\}, N_{\tot} \rgen,\;\;\;\mC^{(\dyn \eta)}_{r,\hub} = \lgen \{S^\alpha_{\tot}\}, \vec{\eta}^2_\pi, \eta^z_\pi \rgen\nn; \\
    \mA^{(\dyn S)}_{r, \hub} \defn \lgen \{T^{(r)}_{j,k}\}, \{V_j\}, S^z_{\tot} \rgen,\;\;\;\mC^{(\dyn S)}_{r,\hub} = \lgen \vec{S}^2_{\tot}, S^z_{\tot}, \{\eta^\alpha_{\pi}\} \rgen\nn; \\
    \mA^{(\dyn S, \dyn \eta)}_{r, \hub} \defn \lgen \{T^{(r)}_{j,k}\}, \{V_j\}, S^z_{\tot}, N_{\tot} \rgen,\;\;\;\mC^{(\dyn S, \dyn \eta)}_{r,\hub} = \lgen \vec{S}^2_{\tot}, S^z_{\tot}, \vec{\eta}^2_\pi, \eta^z_\pi \rgen.  
\end{gather}
These algebras are relevant for the Hubbard model with real hoppings in the presence of a magnetic field or a chemical potential, which includes the standard form of the Hubbard model~\cite{essler2005one}. 
Similar to the algebras in App.~\ref{subsubsec:hubbardimagdyn}, the algebras of $\mA^{(\dyn S)}_{r,\hub}$ and $\mA^{(\dyn\eta)}_{r,\hub}$ are isomorphic to each other, which can be understood using the Shiba transformation $\tQ^\sigma_X$ of Eq.~(\ref{eq:QXQXtdefnsspin}).
\subsection{Complex hoppings}\label{subsec:Hubbardcomplex}
Finally, we study the Hubbard algebra generated in the presence of both real and imaginary hopping terms along with the on-site Hubbard terms: 
\begin{equation}
    \mA_{c,\hub} \defn \lgen \{T^{(c)}_{j,k}\}_{\tnn}, \{V_j\}\rgen = \lgen \{T^{(r)}_{j,k}\}_{\tnn}, \{T^{(i)}_{j,k}\}_{\tnn}, \{V_j\}\rgen.
\label{eq:Achub}
\end{equation}
By definition, this algebra clearly satisfies the relations $\mA_{c} \subset \mA_{c, \hub} \subseteq \mA_{c,\mu} \cap \mA_{c,h}$, where $\mA_c$, $\mA_{c,\mu}$, and $\mA_{c,h}$ are the free-fermion bond algebras corresponding to cases \#2, \#1a, and \#5 in Tab.~\ref{tab:spinfulbondalgebra} respectively (discussed in Apps.~\ref{subsec:spincomplex}, \ref{subsec:spincomplexmu}, and \ref{subsec:spincomplexmag} respectively). 
This implies the relations between the commutant of $\mA_{c,\hub}$, denoted by $\mC_{c,\hub}$, and the other commutants, i.e., $\lgen\mC_{c,\mu} \cup \mC_{c,h}\rgen \subseteq \mC_{c, \hub} \subset \mC_{c}$.
Using the expressions for the commutants from Tab.~\ref{tab:spinfulbondalgebra}, we conjecture
\begin{equation}
    \mC_{c,\hub} = \lgen \{S^\alpha_{\tot}\}, N_{\tot}, \ketbra{\bar{\Omega}}{\Omega} \rgen,
\label{eq:Cchub}
\end{equation}
where $\ket{\Omega}$ and $\sket{\bar{\Omega}}$ are the vacuum and the fully filled states respectively. 
Similar to the previous cases, this conjecture implies that $\mC_{c,\hub} = \lgen\mC_{c,\mu} \cup \mC_{c,h}\rgen$, and hence the Hubbard algebra can be expressed as $\mA_{c,\hub} = \mA_{c,\mu} \cap \mA_{c,h}$.
The singlets of $\mA_{c,\hub}$ can also be understood using the properties of singlets of $\mA_{c}$, listed in Tab.~\ref{tab:spinfulbondalgebra}.
The addition of the Hubbard terms to $\mA_c$ splits the degeneracy between the ferromagnetic tower $\{(S^-_{\tot})^n\ket{F}\}$ and $\ket{\Omega}$ but not between states $\ket{\Omega}$ and $\sket{\bar{\Omega}}$, hence we obtain two sets of degenerate singlets, $\{(S^-_{\tot})^n\ket{F}\}$ and $\{\ket{\Omega}, \sket{\bar{\Omega}}\}$.
Similar to discussions in the previous sections, a uniform chemical potential or a uniform magnetic field can be added to $\mA_{c,\hub}$. 
This gives us the following pairs of local and commutant algebras with complex hoppings:
\begin{gather}
    \mA^{(\dyn \eta)}_{c,\hub} \defn \lgen \{T^{(c)}_{j,k}\}, \{V_j\}, N_{\tot} \rgen = \mA_{c,\mu} ,\;\;\;\mC^{(N_{\tot})}_{c,\hub} = \lgen \{S^\alpha_{\tot}\}, N_{\tot} \rgen = C_{c,\mu}\nn; \\
    \mA^{(\dyn S)}_{c, \hub} \defn \lgen \{T^{(c)}_{j,k}\}, \{V_j\}, S^z_{\tot} \rgen,\;\;\;\mC^{(\dyn S)}_{c,\hub} = \lgen \vec{S}^2_{\tot}, S^z_{\tot}, N_{\tot}, \ketbra{\bar{\Omega}}{\Omega} \rgen\nn; \\
    \mA^{(\dyn S, \dyn \eta)}_{c, \hub} \defn \lgen \{T^{(c)}_{j,k}\}, \{V_j\}, S^z_{\tot}, N_{\tot} \rgen,\;\;\;\mC^{(\dyn S, \dyn \eta)}_{r,\hub} = \lgen \vec{S}^2_{\tot}, S^z_{\tot}, N_{\tot} \rgen.  
\end{gather}
Note that unlike the previous Hubbard cases, $\mC_{c,\hub}$ does not possess a pseudospin $SU(2)$ symmetry, and the addition of $N_{\tot}$ to the bond algebra only results in a breaking of the degeneracy between the states $\ket{\Omega}$ and $\sket{\bar{\Omega}}$.
Hence, in the first case we simply recover the free-fermion bond algebra with arbitrary hoppings and chemical potentials, while in the third case we obtain its dynamical-spin-$SU(2)$ deformation.
\end{document}